\documentclass[12pt]{article}
\pdfoutput=1
\usepackage{jheppub}
\usepackage{mathtools,amssymb,amsthm}
\usepackage[utf8]{inputenc}
\usepackage{graphics}

\def\be{\begin{equation}}
\def\ee{\end{equation}}
\def\ba#1\ea{\begin{align}#1\end{align}}
\def\bg#1\eg{\begin{gather}#1\end{gather}}
\def\bm#1\em{\begin{multline}#1\end{multline}}
\def\bmd#1\emd{\begin{multlined}#1\end{multlined}}

\def\vp{\chi}

\def\la{\label}

\def\er{\eqref}

\def\pa{\partial}

\def\wg{\wedge}

\def\no{\nonumber}

\def\({\left(}
\def\){\right)}
\def\[{\left[}
\def\]{\right]}
\def\<{\langle}
\def\>{\rangle}

\def\bea{\begin{eqnarray}}
\def\eea{\end{eqnarray}}

\newcommand{\tr}{\operatorname{tr}}
\newcommand{\Tr}{\operatorname{Tr}}

\newcommand{\zb}{{\bar z}}

\def\nn{\nonumber}

\begin{document}

\global\long\def\aad{(a\tilde{a}+a^{\dagger}\tilde{a}^{\dagger})}%

\global\long\def\ad{{\rm ad}}%

\global\long\def\bij{\langle ij\rangle}%

\global\long\def\df{\coloneqq}%

\global\long\def\bs{b_{\alpha}^{*}}%

\global\long\def\bra{\langle}%

\global\long\def\dd{{\rm d}}%

\global\long\def\dg{{\rm {\rm \dot{\gamma}}}}%

\global\long\def\ddt{\frac{{\rm d^{2}}}{{\rm d}t^{2}}}%

\global\long\def\ddg{\nabla_{\dot{\gamma}}}%

\global\long\def\del{\mathcal{\delta}}%

\global\long\def\Del{\Delta}%

\global\long\def\dtau{\frac{\dd^{2}}{\dd\tau^{2}}}%

\global\long\def\ul{U(\Lambda)}%

\global\long\def\udl{U^{\dagger}(\Lambda)}%

\global\long\def\dl{D(\Lambda)}%

\global\long\def\da{\dagger}%

\global\long\def\id{{\rm id}}%

\global\long\def\ml{\mathcal{L}}%

\global\long\def\mm{\mathcal{\mathcal{M}}}%

\global\long\def\mf{\mathcal{\mathcal{F}}}%

\global\long\def\ket{\rangle}%

\global\long\def\kpp{k^{\prime}}%

\global\long\def\lr{\leftrightarrow}%

\global\long\def\lf{\leftrightarrow}%

\global\long\def\ma{\mathcal{A}}%

\global\long\def\mb{\mathcal{B}}%

\global\long\def\md{\mathcal{D}}%

\global\long\def\mbr{\mathbb{R}}%

\global\long\def\mbz{\mathbb{Z}}%

\global\long\def\mh{\mathcal{\mathcal{H}}}%

\global\long\def\mi{\mathcal{\mathcal{I}}}%

\global\long\def\ms{\mathcal{\mathcal{\mathcal{S}}}}%

\global\long\def\mg{\mathcal{\mathcal{G}}}%

\global\long\def\mfa{\mathcal{\mathfrak{a}}}%

\global\long\def\mfb{\mathcal{\mathfrak{b}}}%

\global\long\def\mfb{\mathcal{\mathfrak{b}}}%

\global\long\def\mfg{\mathcal{\mathfrak{g}}}%

\global\long\def\mj{\mathcal{\mathcal{J}}}%

\global\long\def\mk{\mathcal{K}}%

\global\long\def\mmp{\mathcal{\mathcal{P}}}%

\global\long\def\mn{\mathcal{\mathcal{\mathcal{N}}}}%

\global\long\def\mq{\mathcal{\mathcal{Q}}}%

\global\long\def\mo{\mathcal{O}}%

\global\long\def\qq{\mathcal{\mathcal{\mathcal{\quad}}}}%

\global\long\def\ww{\wedge}%

\global\long\def\ka{\kappa}%

\global\long\def\nn{\nabla}%

\global\long\def\nb{\overline{\nabla}}%

\global\long\def\pathint{\langle x_{f},t_{f}|x_{i},t_{i}\rangle}%

\global\long\def\ppp{p^{\prime}}%

\global\long\def\qpp{q^{\prime}}%

\global\long\def\we{\wedge}%

\global\long\def\pp{\prime}%

\global\long\def\sq{\square}%

\global\long\def\vp{\varphi}%

\global\long\def\ti{\widetilde{}}%

\global\long\def\wg{\widetilde{g}}%

\global\long\def\te{\theta}%

\global\long\def\tr{{\rm Tr}}%

\global\long\def\ta{{\rm \widetilde{\alpha}}}%

\global\long\def\sh{{\rm {\rm sh}}}%

\global\long\def\ch{{\rm ch}}%

\global\long\def\Si{{\rm {\rm \Sigma}}}%

\global\long\def\sch{{\rm {\rm Sch}}}%

\global\long\def\vol{{\rm {\rm {\rm Vol}}}}%

\global\long\def\reg{{\rm {\rm reg}}}%

\global\long\def\zb{{\rm {\rm |0(\beta)\ket}}}%

\title{Islands in Non-Minimal Dilaton Gravity: Exploring Effective Theories for Black Hole Evaporation}
\author{Chih-Hung Wu and Jiuci Xu}
\affiliation{Department of Physics, University of California, Santa Barbara, CA 93106, USA}
\emailAdd{chih-hungwu@physics.ucsb.edu, Jiuci\_Xu@ucsb.edu}

\abstract{We start from $(3 + 1)$-dimensional Einstein gravity with minimally coupled massless scalar matter, through spherical dimensional reduction, the matter theory is non-minimally coupled with the dilaton in $(1 + 1)$-dimensions. Despite its simplicity, constructing a self-consistent one-loop effective theory for this model remains a challenge, partially due to a Weyl-invariant ambiguity in the effective action. With a universal splitting property for the one-loop action, the ambiguity can be identified with the state-dependent part of the covariant quantum stress tensor. By introducing on-shell equivalent auxiliary fields to construct minimal candidates of Weyl-invariant terms, we derive a one-parameter family of one-loop actions with unique, regular, and physical
stress tensors corresponding to the Boulware, Hartle-Hawking and Unruh states. We further study the back-reacted geometry and the corresponding quantum extremal islands that were inaccessible without a consistent one-loop theory. Along the way, we elaborate on the implications of our construction for the non-minimal dilaton gravity model.}

\maketitle


\section{Introduction}

The black hole information paradox \cite{Hawking:1976} has been recognized as one of the major mysteries whose resolution may lead us to a full understanding of quantum gravity. Recent progress in the gravitational path integral indicates that the quantum extremal islands \cite{Penington:2019npb,Almheiri:2019psf,Almheiri:2019hni} emerge from the replica wormhole saddles  \cite{Penington:2019kki, Almheiri:2019qdq}, and the fine-grained entropy of the Hawking radiation should include contributions from islands. As a result, we anticipate that after the Page time \cite{Page:1993wv, Page:2013dx}, information from the black hole interior will begin to leak out into Hawking radiation, as predicted by a unitary quantum theory. The island formula has been successfully applied to various scenarios \cite{Almheiri:2019psy, Gautason:2020tmk,Anegawa:2020ezn, Hartman:2020swn, Hashimoto:2020cas, Matsuo:2020ypv, Wang:2021mqq, Tian:2022pso, Gan:2022jay, Djordjevic:2022qdk, Yu:2022xlh, Guo:2023gfa, Krishnan:2020fer, Hartman:2020khs, Aguilar-Gutierrez:2021bns}, outside its original context in AdS/CFT \cite{Maldacena:1997re,Gubser:1998bc, Witten:1998qj}.

Although the ultimate goal is to understand black hole evaporation and islands in general dimensions, generic higher-dimensional models are known to be intricate due to the lack of conformal symmetry. Dimensional reduction has been shown to be a successful strategy, particularly with the introduction of the dilaton gravity models. In fact, most recent studies in understanding the gravitational path integral and island formula have focused on $(1+1)$-dimensional dilaton gravity, such as the Jackiw-Teitelboim (JT) \cite{Penington:2019kki, Almheiri:2019qdq, Goto:2020wnk} and the Callan-Giddings-Harvey-Strominger (CGHS) \cite{Hartman:2020swn} models. These models admit higher-dimensional interpretation. For example, JT gravity \cite{Jackiw:1984je, Teitelboim:1983ux} can be viewed as a dimensional reduction in the near-horizon limit of near-extremal Reissner-Nordstr\"om black hole where the spacetime factorizes into ${\rm AdS}_2\times S^{2}$. CGHS \cite{Callan:1992rs}, on the other hand, comes from a four-dimensional near-extremal magnetically charged dilaton black hole in the string frame. It becomes an exactly solvable model for $(1+1)$-dimensional asymptotically flat dilaton gravity theory by including a local counterterm in the one-loop action known as the Russo-Susskind-Thorlacius (RST) term \cite{Russo:1992ax}. These models are constructed with the nice property of exact solvability, but it is not clear that they are generic or represent black holes in our universe. 

Furthermore, the gravitational sector typically has a higher-dimensional origin, whereas the matter sector does not. This is because we generally do not apply dimensional reduction to the matter theory. It is preferable to study minimally coupled scalar matter field $f$ in two dimensions for its simplicity
\be \la{matter1}
S_{\text{matter}}=-\frac{1}{4 \pi} \int d^2 x \sqrt{-g} (\nabla f)^2.
\ee
In this case, one could properly account for the back-reaction problem by following the prescription of Christensen-Fulling \cite{Christensen}. That is, we first adopt the $(1+1)$-dimensional conformal anomaly 
\be \la{minanomaly}
\langle T \rangle=\frac{\hbar}{24 \pi }R,
\ee
which gives the trace of the stress tensor. Note that this is a universal and geometrical result of the matter theory, which is state-independent. The remaining components can be derived by integrating the $(1+1)$-dimensional conservation law
\be
\nabla^a \langle T_{ab} \rangle=0.
\ee
One can also construct a unique one-loop action that reproduces the conformal anomaly by functionally integrating the following defining equation
\be \la{Poldef}
-\frac{2}{\sqrt{-g}} g^{ab} \frac{\delta \Gamma_P}{\delta g^{ab}} \equiv \langle  T   \rangle,
\ee
where the final product is known as the non-local \textit{Polyakov action} \cite{Polyakov:1981rd}
\be \la{Polyakov}
\Gamma_P=-\frac{\hbar}{96 \pi} \int d^2 x \sqrt{-g} R \Box^{-1} R.
\ee
It is also clear that conformal anomaly results from the one-loop action where there is a non-vanishing trace due to broken conformal symmetry. In fact, the variation of the Polyakov action yields a quantum stress tensor consistent with the one constructed from the conservation law, beyond the trace. This is of course a special feature of the minimally coupled theory. 

Due to the limitation, we instead focus on a more general yet simple model. As previously stated, the gravitational sector has a clear higher-dimensional origin; however, the same does not hold true for classical matter theory and one-loop action. A more physical scenario is to consider $(3+1)$-dimensional Einstein gravity coupled with a scalar matter field, with a spherical dimensional reduction to $(1+1)$ dimensions. Focusing on the matter theory, we start with the matter action that is minimally coupled in four dimensions
\be
S^{(4)}_{\text{matter}}=-\frac{1}{8 \pi} \int d^4 x \sqrt{-g^{(4)}}(\nabla f)^2,
\ee
upon spherical dimensional reduction to two dimensions
\be \la{matter2}
S_{\text{matter}}=-\frac{1}{2}\int d^2x \sqrt{-g} e^{-2 \phi}(\nabla f)^2,
\ee
becomes non-minimally coupled with the dilaton $\phi$. The matter action has a four-dimensional origin, and it is important to note that the connection is not limited to four dimensions, but to general dimensions where similar dimensional reduction can be performed.\footnote{For a general connection between $D$-dimensional Einstein gravity with dimensionally reduced models, see \cite{Kummer_1999, Kummer_1999-1}.} This model, which was first considered in \cite{Mukhanov_1994}, is the simplest possible extension of the general spherical reduction gravity.

We expect the theory to capture the s-wave sector of generic higher-dimensional models. In addition, there are new features associated with this model that make it worth studying. The conformal anomaly and the corresponding one-loop theory are deformed due to the presence of dilaton coupling to the matter. As we will see shortly, an important Weyl-invariant ambiguity will arise in the one-loop action.

To study the back-reaction problem and the corresponding quantum extremal islands, we need to solve the semi-classical Einstein equations sourced by the quantum stress tensor. An expression of the quantum stress tensor should be obtained from the one-loop theory. In fact, various methods for finding a consistent one-loop action had been considered. To name a few, using auxiliary fields for the one-loop action to solve the appropriate boundary conditions \cite{Buric:1998xv, Balbinot_1999, Balbinot_1999_2, Buric:2000cj}, the effective action formalism based on perturbative heat kernel \cite{Mukhanov_1994, Lombardo:1998iw, Gusev:1999cv, Kummer_1999, Kummer_1999-1,  Hofmann:2004kk, Hofmann:2005yv}, canonical quantization by solving the field equations \cite{Balbinot_2001, Balbinot:2002bz}, and a generalized transformation law for the normal-ordered stress tensor \cite{Fabbri:2003vy}. 

For such a simple model, the surprising thing is that the results from these approaches are incompatible, and may even lead to unphysical predictions. Examples include untamable logarithmic divergence in the stress tensor at the horizon from heat kernel and canonical quantization; while using the auxiliary fields, one encounters thermal equilibrium in a thermal bath of negative energy, or black hole anti-evaporation, where the black hole is absorbing energy instead of evaporating. Different approaches have their advantages, but also suffer different weaknesses. Obtaining a self-consistent one-loop theory becomes a significant problem that hinders progress, and it is one of the main reasons the model has been overlooked for a while.

With lessons from previous studies, we will impose a few reasonable assumptions on the theory to obtain unique, regular, and physical quantum stress tensors. We first impose the dilaton-deformed conformal anomaly, and it allows us to fix the one-loop action up to Weyl-invariant terms as these terms do not contribute to the trace. Contrary to the minimal model, the anomaly equation and the conservation law fail to determine all the components of the stress tensor in a unique way due to dilaton coupled scalar matter. Therefore, the Weyl-invariant terms pose an ambiguity in the one-loop action, and it is the ambiguity that partially motivates the study of a consistent one-loop theory in the literature.

A crucial observation is that these Weyl-invariant terms are state-dependent. A general consequence in two dimensions is that we can decompose the covariant one-loop action into two non-covariant parts, where we can isolate the contribution of the conformal anomaly and absorb any Weyl-invariant terms into state-dependent quantities. It suggests that we can elevate the problem of state choice to the choice of effective action, where we shall construct minimal candidates of the Weyl-invariant terms for the purpose of describing the states of physical interest. To achieve this, we introduce on-shell equivalent auxiliary fields to the model and solve the corresponding constraint equations. It turns out that by imposing boundary conditions associated with different quantum states, we are able to find a one-parameter family of actions that produces a unique quantum stress tensor for each state. 

The quantum states we have considered include the Boulware state \cite{Boulware1975} describing vacuum polarization exterior to a static black hole; the Hartle-Hawking state \cite{Hartle:1976tp, Israel:1976ur}, describing a black hole in thermal equilibrium; and the $|in \rangle$ state \cite{Davies:1976ei, Hiscock:1980ze, Hiscock:1981xb, Fabbri:2005} describing a black hole formed from gravitational collapse where at late times it gives rise to the Unruh state \cite{Unruh:1976db} for an evaporating black hole. For the first time in the non-minimal dilaton gravity model, unique and completely regular stress tensors could be obtained for these physical states. The near-horizon and asymptotic behaviors are in accordance with the s-wave approximation from four dimensions. However, different states impose different constraints on the possible Weyl-invariant terms, leading to different physical interpretations. To examine the effect, we solve the back-reaction geometry under semi-classical Einstein equations. Straightforward application of the island prescription indicates a unitary Page curve, which is expected from a consistent study of the one-loop theory.

The plan of this paper is as follows. In Sec.~\ref{s2}, we give a precise definition of the non-minimal dilaton gravity model and show how the Weyl-invariant ambiguity arises in the one-loop action. We briefly discuss the earlier studies in the literature before moving on to our resolution, which is based on a universal splitting property of the effective action. In Sec.~\ref{s3}, we apply the formalism we have developed to the construction of effective theories of physical quantum states, including the Boulware, Hartle-Hawking and Unruh states. In Sec.~\ref{s4}, we study the back-reaction and island problems in the eternal and evaporating black hole scenarios. In Sec.~\ref{s5}, we summarize our findings and discuss a few subtitles relevant to the model. In Appendix.~\ref{sA}, we discuss the implication of general covariance and the splitting property we used in Sec.~\ref{s2.3}. We derive a generalized Virasoro anomaly that is crucial in interpreting our results. In Appendix.~\ref{sB}, we perform a non-perturbative analysis for the back-reaction problem associated with the Boulware state, where we show that the back-reaction leads to a no-horizon geometry that resembles a static quantum star. Appendix.~\ref{sC} and Appendix.~\ref{sD} are devoted to the details of the island calculations.

\section{A Non-Minimal Dilaton Gravity Model} \la{s2}

\subsection{Dimensional Reduction and the One-Loop Theory} \la{s2.1}

In this subsection, we introduce the non-minimal dilaton gravity model. Consider the $(3+1)$-dimensional Einstein-Hilbert action coupled with a scalar matter field $f$
\be
S^{(4)}=\frac{1}{16 \pi G^{(4)}_N}\int d^4 x \sqrt{-g^{(4)}} R^{(4)}-\frac{1}{8 \pi} \int d^4 x \sqrt{-g^{(4)}}(\nabla f)^2,
\ee
where we use a superscript $(4)$ to denote the $(3+1)$-dimensional quantities. Here $G^{(4)}_N$ represents the Newton's constant, $g^{(4)}_{\mu \nu}$ $(\mu, \nu=0,1,2,3)$ is the metric, and $R^{(4)}$ is the Ricci scalar. Under spherical dimensional reduction with the following ansatz
\be
ds^2_{(4)}=g_{ab}dx^a dx^b+\lambda^{-2} e^{-2 \phi} d \Omega^2,
\ee
where $a, b=0,1$ and the metric $g_{ab}$ will only depend on $x^{0,1}$. We omit any superscripts for the $(1+1)$-dimensional quantities. Here, a dilaton field $\phi$ is introduced for the radial coordinate $r=\lambda^{-1} e^{- \phi}$. By expressing our $(3+1)$-dimensional theory using the $(1+1)$-dimensional quantities, we arrive at the following action
\be \la{nonmin}
S=\frac{1}{4 G_N} \int d^2 x \sqrt{-g}[e^{-2 \phi} (R+2 (\nabla \phi)^2)+2 \lambda^2]-\frac{1}{2}\int d^2x \sqrt{-g} e^{-2 \phi}(\nabla f)^2,
\ee
where $G_N =\lambda^2 G^{(4)}_N$ and note that $\lambda^2$ term plays the role of a cosmological constant. From now on we set $\lambda=1$ for simplicity. We can generalize to $N$ massless scalar fields by including a factor of $N$ in the matter sector, but we only focus on the case of a single dilaton field where there is no kinetic term associated with the dilaton. A review on general dilaton gravity in two dimensions can be found in \cite{Grumiller:2002nm}. 

Now, both the gravity and the matter sectors have clear four-dimensional origins. In order to study the back-reaction problem, we need to construct a one-loop effective action for this model, and we expect new ingredients involving the dilaton field. 

Let us start with the anomaly equation. The conformal anomaly associated with this matter theory had been derived as \cite{Mukhanov_1994, Bousso, Mikovic_1998, Elizalde, Ichinose, Dowker_1998, katanaev1997generalized, Nojiri:1999vv}
\be \la{confana}
\langle T \rangle= \frac{\hbar}{24 \pi} (R-6(\nabla \phi)^2+6 \Box \phi),
\ee
where we can see explicitly new terms involving the dilaton $\phi$ compared with \er{minanomaly}. We will simply call \er{confana} the dilaton-deformed conformal anomaly, as the dilaton is not quantized, but is treated as an external field. Following a similar procedure as in the Polyakov action \er{Poldef}, we can obtain a one-loop action via the functional integral
\be \la{effe}
-\frac{2}{\sqrt{-g}} g^{ab} \frac{\delta \Gamma_{\text{eff}}}{\delta g^{ab}}=\langle T \rangle=\frac{\hbar}{24 \pi} (R-6 (\nabla \phi)^2+6 \Box \phi).
\ee
We can obtain the \textit{anomaly induced effective action} $\Gamma_{\text{anom}}$ \cite{Balbinot_1999} as a particular solution to \er{effe}, where
\be \la{anomm}
\Gamma_{\text{anom}}=-\frac{\hbar}{96 \pi} \int d^2 x \sqrt{-g} (  R \Box^{-1} R -12 (\nabla \phi)^2 \Box^{-1} R +12 \phi R ),
\ee
where the first term corresponds to the Polyakov action \er{Polyakov}. Being a particular solution, we may always add Weyl-invariant terms that would not affect the defining equation \er{effe}
\be \la{weyl}
\Gamma_{\text{eff}}=\Gamma_{\text{anom}}+ \text{Weyl-invariant terms},
\ee
where we refer $\Gamma_{\text{eff}}$ as the full one-loop effective action. Unlike the case of the Polyakov action, $\Gamma_{\text{anom}}$ does not have all the information about the quantum stress tensor. This is because the quantum conservation law is also modified to be
\be 
\nabla^a \langle T_{ab} \rangle - \frac{1}{\sqrt{-g}}\langle \frac{\delta \Gamma_{\text{eff}}}{\delta \phi} \rangle \nabla_b \phi=0,
\ee
which comes from the dimensional reduction of the four-dimensional conservation law $\nabla^a \langle T^{(4)}_{ab} \rangle=0$. Due to an unfixed degree of freedom, the Weyl-invariant ambiguity indicates that a more well-defined procedure is required for the one-loop action.

\subsection{Challenges in Constructing the One-Loop Effective Action} \la{s2.2}

The Weyl-invariant ambiguity and more generally on how to obtain the correct one-loop theory had been intensively investigated. So far, none of them lead to satisfactory results for Hawking evaporation. Let us briefly discuss the pros and cons of early approaches in the literature, as they addressed important aspects of the problem.\\

\noindent \textbf{Local Effective Action with Auxiliary Fields}

Given the non-local terms in $\Gamma_{\text{anom}}$, one can find a local expression for the action by introducing two auxiliary fields. For examples, in \cite{Buric:1998xv, Balbinot_1999_2, Buric:2000cj} (see also \cite{Balbinot_1999}), the authors introduced
\be
\Box \psi = R, \quad \Box \chi = (\nabla \phi)^2.
\ee
One can express the action in terms of these auxiliary fields, and the new action is on-shell equivalent to $\Gamma_{\text{anom}}$. With the action in local form, one can obtain the quantum stress tensor $\bra T_{ab}\ket$ by varying the effective action, and the state dependence will be encoded in the boundary conditions associated with the auxiliary fields. By taking the Schwarzschild metric as the background, the solutions of $\psi$ and $\chi$ will involve integration constants that encode such state dependence. 
 
However, the treatment can lead to unphysical results associated with different quantum states of black hole spacetime. For example, in the Hartle-Hawking state $| H \rangle$ \cite{Hartle:1976tp} describing a black hole in thermal equilibrium, the thermal bath is of \textit{negative energy} \cite{Buric:1998xv, Balbinot_1999_2} 
\be
\langle H | T_{uu} |H \rangle=\langle H | T_{vv} | H \rangle \to \frac{\hbar}{768 \pi M^2}(1-6),
\ee
where we explicitly keep the $-6$ factor, as it comes from the non-local dilaton term in \er{anomm}. Or similarly the $| in \rangle$ state describing evaporation of a black hole formed by gravitational collapse of a null shock wave \cite{Hiscock:1980ze, Hiscock:1981xb}. In this case, the black hole is \textit{anti-evaporating} \cite{Balbinot_1999_2, Buric:2000cj} 
\be
\langle in | T_{uu} | in \rangle \to \frac{\hbar}{768 \pi M^2} (1-6),
\ee
asymptotically.\footnote{On the other hand, according to \cite{Balbinot_1999}, one has the usual negative ingoing flux near the horizon $r \to 2 M$
\be
\langle in | T_{vv} | in \rangle \to \frac{-\hbar}{768 \pi M^2},
\ee
which makes the interpretation even more unfeasible.
} At late times, the $|in \rangle$ state reproduces the usual Unruh state $|U \rangle$ \cite{Unruh:1976db}, where the same anti-evaporation is found \cite{Balbinot_1999_2}. This means that the black hole is in fact absorbing energy from the vacuum. The results are not only unphysical, but also in violation of the weak energy condition in the asymptotic region \cite{Balbinot_1999_2, Hofmann:2004kk}.

A possibility is that such negative energy occurs because the spherical dimensional reduction only takes the s-wave mode into account. This argument, however, falls short because it does not explain why some models produce positive flux when using only the s-wave sector \cite{Buric:2000cj}. Furthermore, the inclusion of the angular modes should only change the flux's numerical factors rather than its sign.  This interpretation also ignores the non-local dilaton term, which is the source of the flux's negative component and is extremely sensitive to the boundary condition. At least, we expect the result to be corresponding to the flux dimensionally reduced from four dimensions. In four dimensions, we do expect a positive asymptotic flux.\footnote{Nevertheless, the authors in \cite{Buric:1998xv, Buric:2000cj} treated the negative flux as a feature of the model and computed the back-reacted geometry for the Hartle-Hawking and Unruh states. } 

We should stress that the above conclusion comes from the fact that only $\Gamma_{\text{anom}}$ is used as the input and the Weyl-invariant ambiguity in \er{weyl} is neglected. The ambiguity, however, is suggesting that we do not have the complete effective theory. Including more Weyl-invariant terms is a logical solution to this issue, as demonstrated in \cite{Mukhanov_1994}. The output is to remove the $-6$ factor coming from the non-local dilaton term. But as detailed by \cite{Balbinot_1999}, it is an \textit{ad hoc} approach that suffers other physical inconsistencies. For example, the theory does not satisfy Wald's axioms \cite{Wald1978}. \\

\noindent \textbf{Effective Action from Covariant Perturbation Theory}

The Schwinger-DeWitt expansion of the heat kernel is a standard technique for studying the one-loop action \cite{DeWitt:1964mxt}  (see also \cite{Shapiro:2008sf, parker_toms_2009}). In this formalism, we consider the effective action $W[g_{\mu \nu}]$ via the Euclidean path integral
\be
e^{i W[g_{\mu \nu}]}=\int D \phi e^{iS[\phi;g_{\mu \nu}]},
\ee
where we temporarily take $\phi$ to represent the set of all matter fields and $D \phi$ the covariant measure of the functional integration. The effective action of gravity admits a loop expansion in powers of $\hbar$. 
\be
W[g_{\mu \nu}]=S_{\text{vac}}[g_{\mu \nu}]+\hbar \Gamma_{\text{1-loop}}+\cdots.
\ee
For generic metric and potential $V(x)$ associated with a differential operator $F(\nabla, V)$ that depends on the theory, we define a heat kernel with a proper time parameter $\tau$ 
\be \la{kernel}
K(\tau|x,y)=e^{\tau F(\nabla, V)}\delta(x,y),
\ee
and then the one-loop action can be rewritten as
\be
\Gamma_{\text{1-loop}}= \frac{1}{2} \int_0^\infty \frac{d \tau}{\tau} \Tr K(\tau), \quad \Tr K(\tau) = \int dx K(\tau|x,x),
\ee
where the one-loop action is generally non-local (with generic positions $x$ and $y$ in \er{kernel}), as is evident from the structure of $\Gamma_{\text{anom}}$.

Under some mild assumptions about the quantum fields, we can assume there is an asymptotic curvature expansion in small $\tau$ for the heat kernel. This local Schwinger-DeWitt expansion in curvatures allows us to analyze the UV divergences of the theory. As a consequence, the conformal anomaly is a robust result coming from regularization in the UV, and it is regularization scheme independent.

However, we also need to mention the limits of the Schwinger-DeWitt expansion. It contains local covariant expressions with increasing powers of metric derivatives, where the general expressions are not available. In general, there is no way to compute the quantum effective action completely, because the expansion contains an infinite series in curvature tensor and its derivatives, which indicates an infinite amount of non-local insertions. There were early attempts that investigated various ways to resum the heat kernel expansion. However, it is a subtle issue and none of the approaches are satisfactory (see \cite{Vassilevich:2003xt} for a review).

The expansion also does not allow us to evaluate the finite part of the one-loop action as it requires a direct integration of the full $\tau$-range. Additionally, we are interested in studying non-local terms and IR divergences that may arise from the upper-limit of $\tau$. 

Covariant perturbation theory \cite{Barvinsky:1987uw, Barvinsky:1990up, Barvinsky:1990uq, Barvinsky:1993en, Barvinsky:1994ic} was developed as a powerful tool to approach these issues. The objective is to study the late-time asymptotic expansion of the heat kernel. This method corresponds to an infinite resummation of all possible terms with the potential and arbitrary derivatives acting on it. By finding an expansion in the infrared $\tau \to \infty$, one can successfully reproduce the Polyakov action \er{Polyakov} from covariant perturbation \cite{Barvinsky:1990up, Barvinsky:1994ic} for the minimal model. Note that the method requires $V(x)$ to be sufficiently small, which is a reflection of its perturbative nature.

The authors in \cite{Hofmann:2004kk} considered covariant perturbation up to second-order in curvatures for the non-minimal dilaton gravity model \er{nonmin}. The correct asymptotic flux of the s-wave contribution and the conformal anomaly \er{confana} are successfully reproduced from this method. However, an unavoidable logarithmic divergence in the stress tensor of Unruh state at the horizon occurs\footnote{There are several early efforts in deriving an effective action for the non-minimal dilaton gravity model based on heat kernel \cite{Mukhanov_1994, Lombardo:1998iw, Gusev:1999cv, Kummer_1999, Kummer_1999-1}. Especially in \cite{Kummer_1999, Kummer_1999-1}, a similar logarithmic divergent structure at the horizon was discovered. However, in \cite{Gusev:1999cv}, instead of expanding in curvatures, one can expand in orders of dilaton, where the divergence does not occur.}, and it persists when back-reaction is included. The divergence may be attributed to the IR divergent structure found using the covariant perturbation, and it indeed implies the effective action is intrinsically divergent. The issue is not yet fully understood as IR convergence is only guaranteed for $d \geq 3$ in the covariant perturbation theory \cite{Barvinsky:1990up}.

In order to determine whether the IR divergence is generic and whether it can be controlled, the authors further computed the effective action up to third-order in curvatures \cite{Hofmann:2005yv}. However, they discovered that the IR divergence persists in the third-order covariant perturbation. This led them to hypothesize that the IR divergences would appear in all orders. Note that the second-order divergence can be eliminated by renormalizing the theory with a counterterm, which is a coupling to an external field. However, it is unclear whether this can always be achieved if there is IR divergence at each order. Nevertheless, with some simplifying assumptions, the authors asserted that the renormalization can remove the IR divergences to all orders with a resummation based on the third-order structure. This is not conclusive, as whether the divergence is generic remains an open question. Also, as pointed out by the authors, the counterterm is not conformally invariant and may lead to other contributions to the conformal anomaly.

To recap, due to the perturbative nature, no complete closed form could be obtained for the effective action unless a well-defined resummation method is found. It is an open question whether the divergences can be resolved via some other non-perturbative improvement for calculating the heat kernel, such as the formalism developed in \cite{Barvinsky:2002uf, Barvinsky:2003rx, Barvinsky:2004he}.\\\\\\

\noindent \textbf{Canonical Quantization}

It is conceivable that the problem lies in the perturbative scheme of action formalism. Another standard method is based on canonical quantization, where it starts by finding a complete set of solutions associated with the dilaton coupled equation of motion. Approximate analytic expressions of $\langle T_{ab} \rangle$ for the Boulware and Hartle-Hawking states can be obtained by the point-splitting regularization and a WKB approximation of the normal modes. For details, we refer to \cite{Balbinot_2001} (see also \cite{Balbinot:2002bz} for an action formalism). A similar logarithmic divergent behavior at the horizon was discovered. The authors attributed this divergence as an artifact of the WKB approximation, which should not be applicable to some near-horizon low-frequency modes. One can hence argue that the stress tensor for the Hartle-Hawking state is regular at the horizon. Unfortunately, the applicability of the canonical quantization approach is still limited as the calculations of the normal modes are rather involved. There is no analytical expression that interpolates between the regular near-horizon behavior and the approximate WKB result far from the horizon.

Furthermore, although the results are consistent with the conformal anomaly, the asymptotic stress tensor does not correspond to the s-wave approximation. Hence it is not in agreement with the action formalism from covariant perturbation. This fact may have something to do with the \textit{dimensional reduction anomaly} \cite{Frolov_1999, Cognola:2000wd, Cognola:2000xp}, which states that the quantization procedure does not commute with dimensional reduction. To be more precise, the s-wave contribution to the renormalized stress tensor of the four-dimensional theory does not coincide with the renormalized stress tensor of the two-dimensional reduced theory. The reason behind this is that the sum over the higher angular modes will in general be divergent, although each of them is finite. Therefore, the higher dimensional theory would require more counterterms and counterterms of different types \cite{Grumiller:2002nm}. 

In light of these unsatisfactory results, the remainder of the paper seeks to address the non-minimal dilaton gravity model appropriately and come up with a consistent solution for the stress tensors describing physical quantum states. In the next subsection, we will further elaborate the role of the Weyl-invariant ambiguity, which is essential to our resolution.

\subsection{Universal Spliting of $\Gamma_{{\rm eff}}$ and the Role of Weyl Ambiguity} \la{s2.3}

In previous subsections, we have described how the anomaly equation \er{confana} determines the action up to a Weyl-invariant term, namely
\begin{equation}
\Gamma_{{\rm eff}}=\Gamma_{\text{anom}}+\Gamma_{W},
\end{equation}
and how this fact leads to an ambiguity in the effective action and several related problems in achieving a workable form of the stress tensors. Now we present a resolution toward deriving unique, regular, and physical stress tensors for the non-minimal dilaton gravity model. This is based on a universal way of splitting the effective action into a local part and a  Weyl-invariant part \cite{Karakhanian:1994gs, Jackiw:1995qh, Navarro-Salas:1995lmi, Fabbri:2005} that holds generally in two dimensions. That is 
\begin{equation}\label{split1}
\Gamma_{{\rm eff}}=\Gamma_{{\rm loc}}+\Gamma_{W},
\end{equation}
where $\Gamma_{{ W}}$ is a generally non-local Weyl-invariant action in metric and any potential matter contents. On the other hand, $\Gamma_{{\rm loc}}$ is a local action that captures the geometrical contribution to the stress tensor, especially the conformal anomaly.

We shall elaborate on the roles played by the two parts of the action in the following. We start by commenting on certain features and indications of \er{split1}:
\begin{itemize}
\item $\Gamma_{{\rm loc}}$ is always local despite $\Gamma_{{\rm eff}}$ being non-local in general.
This indicates that the effect of non-locality can all be  attributed to the Weyl-invariant part $\Gamma_{W}$. 
\item The stress tensor defined by $\Gamma_{{\rm loc}}$ captures the dilaton-deformed conformal anomaly \er{confana}.
This means if we define
\begin{equation}
\langle T^{\text{geo}}_{ab}\rangle=\frac{-2}{\sqrt{-g}}\frac{\del \Gamma_{{\rm loc}}}{\del g^{ab}},
\end{equation}
then the trace is given by
\begin{equation} \label{tgeo}
\langle T^{\text{geo}}\rangle=\frac{\hbar}{24\pi}\left(R-6\left(\nn\phi\right)^{2}+6\sq\phi\right).
\end{equation}
Here the superscript "geo" means that the contribution to the stress tensor comes from the geometry of the background together with the dilaton profile. In other words, the saddle breaks the Weyl invariance of the theory, leading to a non-vanishing stress tensor that sources the back-reaction. 
\item There is a canonical expression for $\Gamma_{\text{loc}}$ in terms of the local quantities consisting of the metric and dilaton field given by
\begin{equation}
  \begin{split} \label{def-split}
    \Gamma_{{\rm loc}}& =\frac{\hbar}{96\pi}\int\dd^{2}x \big( \sqrt{-g} \log\sqrt{-g}\sq\log\sqrt{-g}\\
     &+\log\sqrt{-g}\left(2R-12\left(\nn\phi\right)^{2}+12\sq\phi\right)  \big).
  \end{split}
\end{equation}
\item The above choice of $\Gamma_{{\rm loc}}$ is universal and state-independent.\footnote{Here the state dependence means the definition of the vacuum state from where $\Gamma_{{\rm eff}}$ describes the local excitations. By claiming that $\Gamma_{{\rm loc}}$ is state-independent, we mean the form of \er{def-split} as a functional of metric and dilaton field does not depend on the choice of vacuum. } We expect that any
$\Gamma_{{\rm eff}}$ which serves as the solution to the anomaly equation \er{confana} would produce the same $\Gamma_{{\rm loc}}.$ 
 
\end{itemize}

The essence of the splitting is to find a universal part of the effective
action that captures the conformal anomaly. Let us first comment on the universality of $\Gamma_{{\rm loc}}$
and then verify that it satisfies the properties mentioned above.

Since $\Gamma_{{\rm loc}}$ captures the anomaly, this means
the difference between $\Gamma_{{\rm eff}}$ and $\Gamma_{{\rm loc}}$ is Weyl-invariant. Therefore, we can write 
\begin{equation}\label{split2}
\Gamma_{{\rm eff}}\left(g_{ab},\psi\right)=\Gamma_{W}\left(g_{ab},\psi\right)+\Gamma_{loc}\left(g_{ab},\psi\right),
\end{equation}
where $g_{ab}$ is the metric and $\psi$ represents generic field content
excluding $g_{ab}$. Note that the above equation holds for off-shell
configurations of $g_{ab}$ as well, by plugging in a Weyl-invariant combination $\sqrt{-g}^{-1}g_{ab}$ 
we find 
\begin{equation}
\Gamma_{{\rm eff}}\left(\sqrt{-g}^{-1}g_{ab},\psi\right)=\Gamma_{W}\left(g_{ab},\psi\right)+\Gamma_{{\rm loc}}\left(\sqrt{-g}^{-1}g_{ab},\psi\right),
\end{equation}
where we have used the fact that $\Gamma_{W}$ is Weyl-invariant. Note that $\Gamma_{{\rm eff}}\left(\sqrt{-g}^{-1}g_{ab},\psi\right)$ as a functional of the metric is already invariant under Weyl transformation, which means it should be equal to the Weyl-invariant part of itself. This indicates
\begin{equation}\label{zero1}
    \Gamma_{{\rm loc}}\left(\sqrt{-g}^{-1}g_{ab},\psi\right)  =0.
\end{equation}
Combined with equation \er{split2}, we find the expression for $\Gamma_{{\rm loc}}$ in terms of $\Gamma_{{\rm eff}}$
\begin{equation}\label{def-split2}
\Gamma_{{\rm loc}}\left(g_{ab},\psi\right) =\Gamma_{{\rm eff}}\left(g_{ab},\psi\right)-\Gamma_{{\rm eff}}\left(\sqrt{g}^{-1}g_{ab},\psi\right).
\end{equation}
Note that the canonical $\Gamma_{{\rm loc}}$ in \er{def-split} already satisfies \er{zero1}. In fact, it vanishes for any metric configurations with a unit determinant.

Note that \er{def-split2} gives a concrete construction
for $\Gamma_{{\rm loc}}$ if one knows the form of $\Gamma_{{\rm eff}}$. The above derivation for $\Gamma_{{\rm loc}}$ has not specified any matter content $\psi$ in $\Gamma_{{\rm eff}} $,
which means it is applicable for
arbitrary field content with the corresponding conformal anomaly. To illustrate this point, we start with
a concrete and sufficient example 
\begin{equation}
\begin{split} \label{model}
\Gamma_{{\rm eff}}^{\pp} & =\Gamma_{\chi_{1}}+\Gamma_{\chi_{2}}+\Gamma_{\phi},\\
\Gamma_{\chi_{1}} & =\hbar\int\sqrt{-g}\left(\frac{1}{2}\left(\nn\chi_{1}\right)^{2}+\chi_{1}\left(\lambda_{1}R+\lambda_{2}\left(\nn\phi\right)^{2}\right)\right),\\
\Gamma_{\chi_{2}} & =\hbar\int\sqrt{-g}\left(-\frac{1}{2}\left(\nn\chi_{2}\right)^{2}+\chi_{2}\left(\mu_{1}R+\mu_{2}\left(\nn\phi\right)^{2}\right)\right),\\
\Gamma_{\phi} & =\frac{\hbar}{8\pi}\int\sqrt{-g}\phi R,
\end{split}
\end{equation}
where $\chi_{1}$ and $\chi_{2}$ are local fields with appropriate boundary conditions, $\phi$ is the dilaton field, and $(\lambda_{i}, \mu_{i})$ $(i=1,2)$
are arbitrary coupling constants. The new action in terms of the auxiliary fields is on-shell equivalent to the original one. It is straightforward to verify that 
\begin{equation}
    \begin{split}
\Gamma_{\chi_{1}}\left(g_{ab},\chi_{1}\right) -  \Gamma_{\chi_{1}}\left(\sqrt{-g}^{-1}g_{ab},\chi_{1}\right) = 
  &-\hbar\int\sqrt{-g}\Big(\frac{\lambda_{1}^{2}}{2}\log\sqrt{-g}\sq\log\sqrt{-g} \\ &+\lambda_{1}\left(\lambda_{1}R+\lambda_{2}(\nn\phi)^{2}\right)\log\sqrt{-g}\Big),
     \end{split}
\end{equation}
\begin{equation}
\Gamma_{\phi}\left(g_{ab},\phi\right) - \Gamma_{\phi}\left(\sqrt{-g}^{-1}g_{ab},\phi\right)  =-\frac{\hbar}{8\pi}\int\sqrt{-g}\phi\sq\log\sqrt{-g}.
\end{equation}

The corresponding $\Gamma_{{\rm loc}}$ is then given by 
\begin{equation}
  \begin{split}
    \Gamma_{{\rm eff}}^{\pp}\left(g_{ab},\psi\right)- & \Gamma_{{\rm eff}}^{\pp}\left(\sqrt{-g}^{-1}g_{ab},\psi\right)  =\frac{\hbar}{2}\bigg[\int\sqrt{-g}\left(\lambda_{1}^{2}-\mu_{1}^{2}\right)\log\sqrt{-g}\sq\log\sqrt{-g}\\  \label{difference}
     & +\int\sqrt{-g}\log\sqrt{-g}\left(\lambda_{1}^{2}-\mu_{1}^{2}\right)R\\ 
     & +\sqrt{-g}\log\sqrt{-g}\left(\left(\lambda_{1}\lambda_{2}-\mu_{1}\mu_{2}\right)\left(\nn\phi\right)^{2}+\frac{1}{8\pi}\sq\phi\right)\bigg].
    \end{split}
\end{equation}
The coupling constants $\left(\lambda_{1},\lambda_{2}\right)$ and
$\left(\mu_{1},\mu_{2}\right)$ are not completely independent. They
are constrained by the requirement that $\Gamma_{{\rm eff}}^{\pp}$ satisfies
the anomaly equation. Working out the trace of stress tensor for $\Gamma_{{\rm eff}}^{\pp}$,
we find the constraints on the parameters are precisely
\begin{equation}
\lambda_{1}^{2}-\mu_{1}^{2}  =\frac{1}{48\pi}, \quad \lambda_{1}\lambda_{2}-\mu_{1}\mu_{2}  =-\frac{1}{8\pi}.
\end{equation}
Therefore, we see that \er{difference} indeed
produces the canonical $\Gamma_{{\rm loc}}$. Note that there are still
free parameters in the family of the theory. They produce the same
$\Gamma_{{\rm loc}}$ and differ only by a Weyl-invariant term. 

Now let us take a closer look at the stress tensor defined by $\Gamma_{{\rm loc}}$
\begin{equation}
\bra T^{{\rm geo}}_{ab} \ket=\frac{-2}{\sqrt{-g}}\frac{\del \Gamma_{{\rm loc}}}{\del g^{ab}}.
\end{equation}
In a conformal gauge with $\dd s^{2}=-e^{2\rho}\dd x^+\dd x^-$, this yields
the following components of the stress tensor 
\be\label{trace2}
\bra T^{{\rm geo}}_{+-}\ket=-\frac{\hbar}{12\pi}\partial_{+}\partial_{-}\rho+\frac{\hbar}{4\pi}\left(-\partial_{+}\phi\partial_{-}\phi+\partial_{+}\partial_{-}\phi\right),
\ee
\be \label{tpp}
\bra T^{{\rm geo}}_{\pm\pm} \ket =\frac{\hbar}{12\pi}\left(\partial_{\pm}^{2}\rho-\left(\partial_{\pm}\rho\right)^{2}\right)+\frac{\hbar}{2\pi}\left(\rho\left(\partial_{\pm}\phi\right)^{2}+\partial_{\pm}\rho\partial_{\pm}\phi\right).
\ee
By expanding in component value of $(x^+,x^-)$, \er{trace2} precisely reproduces the dilaton-deformed conformal anomaly \er{confana}.
While for \er{tpp}, we show in Appendix.~\ref{sA} that the first term corresponds to a Schwarzian derivative, and the second term can be viewed as a deformed part 
 of the transformation
law in the presence of dilaton. One may also notice that the above result is not invariant under a constant shift of $\rho$, which should not cause any actual physical effect. We leave the detailed discussion on how to resolve this ambiguity to appendix \ref{sA}.

If
one requires general covariance to be maintained in the effective
action $\Gamma_{{\rm eff}}$ and in particular the existence of a
covariant quantum stress tensor $\bra T_{ab}\ket$, then there must be a way to incorporate the difference between
$\bra T_{ab}\ket$ and the normal-ordered part $\bra:T_{ab}:\ket$. The latter is not covariant in general because the normal-ordering breaks the general covariance by subtracting divergent parts in a specific coordinate.  Indeed, the breaking is captured by $\bra T^{{\rm geo}}_{ab}\ket$. 
In Appendix.~\ref{sA}, we show explicitly that the combination
\begin{equation}
\bra T_{ab}\ket=\bra:T_{ab}:\ket+\bra T_{ab}^{{\rm geo}}\ket,
\end{equation}
is covariant and unaffected by the value that one assigns
with the normal-ordered part in any specific coordinate.

Given that $\bra T_{ab}^{{\rm geo}}\ket $ is completely determined by the universal part $\Gamma_{{\rm loc}}$, the normal-ordered part can only come from $\Gamma_{W}$, which was previously viewed as an ambiguity in solving the anomaly equation. This makes sense because the normal-ordered part of stress tensor encodes the definition of the state, and can not be fixed by the state-independent anomaly.\footnote{Note that if we are not restricting to the s-wave approximation, even the trace of the stress tensor itself can be state-dependent. This implies we no longer have the canonical choice for $\Gamma_{\text{loc}}$. See, however, a critique for exploiting the state dependence of the four-dimensional effective action \cite{Bardeen:2018gca} for different scenarios.} For instance, in Minkowski vacuum one requires $\bra : T_{ab} : \ket=0$. Therefore, instead of interpreting $\Gamma_W$ as an ambiguity, we can now view it as an alternative definition of the state, and is incorporated into the specification of the effective action $\Gamma_{{\rm eff}}$.\footnote{This includes specifying the field contents and interactions in $\Gamma_{{\rm eff}}$, together with suitable boundary conditions that leads to correct asymptotic behavior of the quantum stress tensor.} In the next section, we shall explore the construction of $\Gamma_{{\rm eff}}$ for various states and show how it produces a covariant stress tensor compatible with the definition of the state.

\section{Effective Theories for Physical Quantum States} \la{s3}

As an application of the formalism developed in Sec.~\ref{s2.3}, we construct the one-loop effective theories for physical quantum states. An important lesson drawn from the role of the Weyl-invariant ambiguity is that, the ambiguity is associated with the state-dependent part $\langle : T_{ab} : \rangle$ that requires knowledge beyond the geometrical conformal anomaly. We adopt the viewpoint that $\langle : T_{ab} : \rangle$ is part of the definition of the theory, which should be determined by physical requirements.

Furthermore, from the discussion in Sec.~\ref{s2.2}, we believe the following conditions must be satisfied:
\begin{itemize}
    \item The state-independent dilaton-deformed conformal anomaly\footnote{A remark is that in the CGHS model, to get semi-classical exact solutions, the RST local term is added to the one-loop action by hand
    \be
    \Gamma_{\text{RST}}=-\frac{\hbar}{48 \pi} \int d^2 x \sqrt{-g} \phi R.
    \ee
For the island computation in asymptotically flat spacetime based on this model, see \cite{Gautason:2020tmk, Anegawa:2020ezn, Hartman:2020swn}. Note that the RST term is not the only choice, for example, one can have a different local term such as the Bose-Parker-Peleg term \cite{Bose:1995pz, Bose:1995bk} 
\be
\Gamma_{\text{BPP}}=\frac{\hbar}{24 \pi} \int d^2 x \sqrt{-g} [(\nabla \phi)^2-\phi R].
\ee
In fact, there are several other proposals for recovering the solvability \cite{Fabbri:1995bz, Cruz:1995zt, Zaslavsky:1998ca}. See also \cite{Wang:2021mqq, Tian:2022pso,  Yu:2022xlh} for island computations based on some of these models. These additional terms would change the conformal anomaly, but we would like to take the conformal anomaly as one of the first principles.}: from the defining equation        
    \be
    -\frac{2}{\sqrt{-g}} g^{ab} \frac{\delta \Gamma_{\text{eff}}}{\delta g^{ab}}=\langle T \rangle= \frac{ \hbar}{24 \pi} (R-6 (\nabla \phi)^2+6 \Box \phi).
    \ee
    This allows to fix the one-loop action up to Weyl-invariant terms
    \be 
    \Gamma_{\text{anom}}=-\frac{\hbar}{96 \pi} \int d^2 x \sqrt{-g} \bigg(  R \Box^{-1} R -12 (\nabla \phi)^2 \Box^{-1} R +12 \phi R \bigg),
    \ee
    \be 
    \Gamma_{\text{eff}}=\Gamma_{\text{anom}}+ \text{Weyl-invariant terms}.
    \ee
    \item Dilaton-deformed conservation law for the quantum stress tensor: 
    \be \la{conser}
    \nabla^a \langle T_{ab} \rangle - \frac{1}{\sqrt{-g}}\langle \frac{\delta \Gamma_{\text{eff}}}{\delta \phi} \rangle \nabla_b \phi=0.
    \ee
    This equation comes from the dimensional reduction of the four-dimensional conservation law $\nabla^\mu \langle T^{(4)}_{\mu \nu} \rangle=0$. It is a consequence of general covariance that is true in any dimension for any dilaton gravity theory whose effective action is of the form $\Gamma_{\text{eff}}=\Gamma_{\text{eff}}[g_{ab}, \phi]$ \cite{Balbinot_1999, Hofmann:2004kk}. Hence we do not need to impose it by hand once a covariant effective action is at hand \cite{parker_toms_2009}.

    \item Boundary conditions associated with the state: we impose appropriate boundary conditions associated with Boulware, Hartle-Hawking, and $|in \rangle$ states. It typically involves requiring the quantum stress tensor to be regular asymptotically or at the horizon.
    \item The quantum stress tensor must exhibit near-horizon and asymptotic behaviors that are consistent with s-wave approximation from four dimensions.
\end{itemize}
We will be explicit about the final two conditions in the following subsections. We will see that by imposing these physical conditions, one can fix the one-loop theory uniquely and determine completely regular stress tensors associated with different quantum states. 

\subsection{Building the Effective Theory for Boulware State} \la{s3.1}

In this subsection, we consider the simple vacuum state annihilated by operators using plane wave modes associated with the Eddington-Finkelstein coordinates $(u,v)$. The quantum state is called the Boulware state $|B \rangle$ \cite{Boulware1975}, and is considered to be describing the vacuum polarization of the spacetime outside a static black hole. The Boulware vacuum is simple in a sense that it reduces to the conventional Minkowski vacuum when the mass $M \to 0$.

Following Sec.~\ref{s2.3}, the starting point is to consider the one-loop theory constructed from the following actions with auxiliary fields $\chi_1$ and $\chi_2$ 
\bea
\Gamma_{\chi_1}&=& \hbar \int d^2 x \sqrt{-g} \bigg[ \frac{1}{2} (\nabla \chi_1)^2+ \chi_1 (\lambda_1 R + \lambda_2 (\nabla \phi)^2) \bigg],
\\
\Gamma_{\chi_2}&=& \hbar \int d^2 x \sqrt{-g} \bigg[ -\frac{1}{2} (\nabla \chi_2)^2+ \chi_2 (\mu_1 R + \mu_2 (\nabla \phi)^2) \bigg],
\\
\Gamma_{\phi} & =&-\frac{\hbar}{8\pi}\int\sqrt{-g}\phi R.
\eea
The introduction of the auxiliary fields should be perceived as merely a consistent method of dealing with the non-local feature of the effective action. Some comments about this setup are in order:
\begin{itemize}
    \item The effective action with auxiliary fields should be understood as on-shell equivalent to the full effective action describing the Boulware state. The state dependence will be encoded in the solutions of the equations of motion for the auxiliary fields on a background with the appropriate choice of boundary conditions. Hence, the coefficients $(\lambda_i, \mu_i)$ and any integrations constants that might arise are to be determined by the physical constraints associated with the Boulware state.
    \item Note that other candidate terms could exist as long as they do not contribute to the anomaly equation. There is no \textit{a priori} reason to say the state dependence is encoded in a single type or fixed types of terms off-shell. However, we are attempting to construct minimal candidates of possible Weyl-invariant terms. As we have demonstrated in Sec.~\ref{s2.3}, it is sufficient that by solving the constraints, we will get a one-parameter family of effective action with $\lambda_2$.
\end{itemize}

The quantum stress tensors associated with the auxiliary fields $\chi_1$, $\chi_2$ and the dilaton $\phi$ are given by 
\bea
\frac{-2}{\sqrt{-g}} \frac{\delta \Gamma_{\chi_1}}{\delta g_{ab}} =\langle T^{(\chi_1)}_{ab} \rangle &=& \hbar \bigg[ -\nabla_a \chi_1 \nabla_b \chi_1 +\frac{1}{2} g_{ab} (\nabla \chi_1)^2  + 2\lambda_1 ( \nabla_a \nabla_b \chi_1- g_{ab} \Box \chi_1) 
\no\\
&\quad&
  - 2 \lambda_2 \chi_1 \bigg( \nabla_a \phi \nabla_b \phi-\frac{1}{2} g_{ab} (\nabla \phi)^2 \bigg) \bigg],
\\
\frac{-2}{\sqrt{-g}} \frac{\delta \Gamma_{\chi_2}}{\delta g_{ab}} =\langle T^{(\chi_2)}_{ab} \rangle &=&  \hbar \bigg[ \nabla_a \chi_2 \nabla_b \chi_2 -\frac{1}{2} g_{ab} (\nabla \chi_2)^2  + 2\mu_1 ( \nabla_a \nabla_b \chi_2- g_{ab} \Box \chi_2) 
\no\\
&\quad&
- 2 \mu_2 \chi_2 \bigg( \nabla_a \phi \nabla_b \phi-\frac{1}{2} g_{ab} (\nabla \phi)^2 \bigg) \bigg],
 \\
\frac{-2}{\sqrt{-g}} \frac{\delta \Gamma_\phi}{\delta g_{ab}} =\langle T^{(\phi)}_{ab} \rangle &=& -\frac{\hbar}{4 \pi} (\nabla_a \nabla_b \phi- g_{ab} \Box \phi).
\eea
The full quantum stress tensor under consideration is the sum of the three terms
\be \la{stressT}
\langle T_{ab} \rangle = \langle T^{(\chi_1)}_{ab} \rangle+ \langle T^{(\chi_2)}_{ab} \rangle+\langle T^{(\phi)}_{ab} \rangle.
\ee
We will also have to solve equations of motion for the auxiliary fields $\chi_1$ and $\chi_2$ given by
\be
\Box \chi_1 =(\lambda_1 R + \lambda_2 (\nabla \phi)^2),
\ee
\be
\Box \chi_2=-(\mu_1 R + \mu_2 (\nabla \phi)^2),
\ee
with the classical background in the Eddington-Finkelstein coordinates
\be
ds^2=-\bigg( 1-\frac{r_0}{r} \bigg)du dv
\ee
where we have set $r_0=2M$. We can express the equations of motion as 
\be
\Box \chi_1=\left(\lambda_{1}-\frac{\lambda_{2}}{2}\right)\frac{8r_0}{r^{3}}+\frac{4\lambda_{2}}{r^{2}},
\ee
\be
\Box \chi_2 =-\left(\mu_{1}-\frac{\mu_{2}}{2}\right)\frac{8 r_0}{r^{3}}-\frac{4\mu_{2}}{r^{2}}.
\ee
The solutions are then given by
\begin{equation} \label{chi1}
    \begin{split}
        \chi_1=&-\lambda_{1}\log\left(1-\frac{r_0}{r}\right)-\frac{\lambda_{2}}{2}\left[\log\left(\frac{r}{r_0}-1\right)+\log\left(\frac{r}{r_0}\right)\right]\\
        &+C_{1}\left[\frac{r}{r_0}+\log\left(\frac{r}{r_0}-1\right)\right]+C_{2},
    \end{split}
\end{equation}
\begin{equation}\label{chi2}
    \begin{split}
       \chi_2&=\mu_{1}\log\left(1-\frac{r_0}{r}\right)+\frac{\mu_{2}}{2}\left[\log\left(\frac{r}{r_0}-1\right)+\log\left( \frac{r}{r_0}\right) \right]\\
       &+C_{3}\left[\frac{r}{r_0}+\log\left(\frac{r}{r_0}-1\right)\right]+C_{4}, 
    \end{split}
\end{equation}
with four integration constants $C_i$, $(i=1 \sim 4)$ that parametrizes the zero modes of the d'Alembertian. These constants are also to be determined by the physical conditions imposed in the theory. Note that the dilaton is given by $\phi=-\ln{r}$ from dimensional reduction.

Following a similar discussion in Sec.~\ref{s2.3}, in order to restore $\Gamma_{\text{anom}}$, the following requirements must be satisfied 
\be \la{constraint1}
\lambda_1^2 -\mu_1^2=-\frac{1}{48 \pi}, \quad \lambda_1 \lambda_2-\mu_1 \mu_2=\frac{1}{8 \pi}, \quad \lambda_2^2 -\mu_2^2=0. 
\ee
The last constraint requires that there is no additional Weyl-invariant term $\left(\nn\phi\right)^{2}\frac{1}{\sq}\left(\nn\phi\right)^{2}$ in the action. The set of constraints \er{constraint1} allows us to express the stress tensor \er{stressT} in terms of only $\lambda_2$ by the following two sets of solutions
\be \la{sol1}
\{\lambda_1=\frac{1}{16 \pi \lambda_2}-\frac{\lambda_2}{12}, \quad \mu_1=\frac{-1}{16 \pi \lambda_2}-\frac{\lambda_2}{12}, \quad \lambda_2=\mu_2\},
\ee
or
\be \la{sol2}
\{\lambda_1=\frac{1}{16 \pi \lambda_2}-\frac{\lambda_2}{12}, \quad \mu_1=\frac{1}{16 \pi \lambda_2}+\frac{\lambda_2}{12}, \quad \lambda_2=-\mu_2\}.
\ee
Let us examine whether $\Gamma_{\text{anom}}$ is sufficient to reproduce the correct physics associated with the Boulware state. 

The Boulware state is required to reduce to the Minkowski vacuum as $M \to 0$. This imposes
\be \la{BConditions}
\lim_{M \to 0} \langle B| T_{uu}| B \rangle=\lim_{M \to 0} \langle B| T_{vv}| B \rangle=0, \quad \lim_{M \to 0} \langle B| T_{uv}| B \rangle=0.
\ee
If we use the first set of solution \er{sol1}, we get 
\be
\lim_{M\to0}\bra B|T_{uu}|B\ket = \lim_{M\to0}\bra B|T_{vv}|B\ket =C_1^2-C_3^2+\frac{2(C_2+C_4)\lambda_2}{r^2}=0,
\ee
which implies $C_1=\pm C_3$ and $C_2=-C_4$;
If we use the second set of solution \er{sol2}, we get
\be
\lim_{M\to0}\bra B|T_{uu}|B\ket = \lim_{M\to0}\bra B|T_{vv}|B\ket=C_1^2-C_3^2+\frac{2(C_2-C_4)\lambda_2}{r^2}=0,
\ee
which implies $C_1=\pm C_3$ and $C_2=C_4$. Note that $\lim_{M \to 0} \langle B| T_{uv}| B \rangle$ imposes no constraint on the parameters, which is expected as $\langle T_{uv} \rangle$ must reproduce the anomaly equation \er{confana}. 

Hence, we find that there is a unique solution to the theory corresponding to the state, which fixes the Weyl-invariant ambiguity sourced by $\lambda_{2}$. The components of the stress tensor in this case read
\be \la{BoulwareT}
\langle B| T_{uu} | B \rangle= \langle B| T_{vv}| B \rangle= \frac{\hbar}{24 \pi } \bigg(\frac{3r^2_0}{8 r^4}-\frac{r_0}{2r^3} \bigg) +\frac{\hbar}{16 \pi} \frac{(r-r_0)^2 \ln{(1-\frac{r_0}{r})}}{r^4},
\ee
\be \la{BoulwareTuv}
\langle B| T_{uv} | B \rangle= \frac{\hbar r_0}{24 \pi r^3} \bigg(1-\frac{r_0}{r} \bigg).
\ee
Note that these results are in agreement with \cite{Balbinot_1999}, which is expected as the authors were also adopting $\Gamma_{\text{anom}}$. One can immediately verify that the stress tensor vanishes at asymptotic infinity $r \to \infty$, which is also a physical property of the Boulware state such that it should always reduce to the Minkowski vacuum asymptotically. An interesting observation is that the first piece in the $\langle B| T_{uu} | B \rangle$ or $\langle B| T_{vv}| B \rangle$ is exactly the stress tensor one would get had we chosen the minimal model \er{matter1} in the matter sector; thus the second piece can be viewed as originating from the non-minimal dilaton coupling \er{matter2}. Note that $\langle B| T_{uv} | B \rangle$ indeed agrees with the anomaly equation \er{confana}.

By transforming to local regular coordinates such as the Kruskal coordinates $(U,V)$, the stress tensor of the Boulware state is divergent at the horizon. This is a generic feature in the Boulware state. The interpretation is that the physical portion of the Schwarzschild black hole that the Boulware state is describing does not contain the horizon. However, for an intriguing back-reaction calculation that alters such an interpretation for the Boulware state, we refer to Apppendix.~\ref{sB}.

The self-consistent analysis above implies that we do not need to add any additional Weyl-invariant terms  to describe the Boulware state. It indicates $\Gamma_{\text{anom}}$ is the natural action that incorporates the state and does not suffer the Weyl-invariant ambiguity as described in Sec.~\ref{s2.2}. 

However, what if we want to do so? If it is possible to include additional Weyl-invariant terms to describe the Boulware state, then the results we found seem to be non-unique. A simple consistency check is to relax the final constraint in \er{constraint1} to be some constant $L$
\be
\lambda_2^2- \mu_2^2=L.
\ee
By imposing again the physical conditions in \er{BConditions} with this new constraint, a straightforward calculation shows that the constant must be zero. This enforces our initial condition, which confirms the uniqueness of our discovery. 

However, as we commented earlier, there could be terms that do not contribute to the anomaly equation (see, for example, \cite{Mukhanov_1994} with terms arising from the heat kernel expansion) that can still capture Boulware-like states. But by incorporating these terms, we will encounter other issues as discussed in \cite{Balbinot_1999}, such as the violation of Wald's axioms. We will return to the implication for $\langle : T_{ab}: \rangle$ in our approach and Wald's axioms in Sec.~\ref{s5}. 

\subsection{Building the Effective Theory for Hartle-Hawking State} \la{s3.2}

The next physical scenario we want to consider is an eternal black hole where it stays in thermal equilibrium with the environment. An equal amount of radiation from past null infinity balances the thermal radiation emitted from the black hole. The quantum state corresponds to the Hartle-Hawking state $|H \rangle$ \cite{Hartle:1976tp, Israel:1976ur}, which is annihilated by operators defined with respect to the plane wave modes using the Kruskal-type coordinates $(U,V)$. This case is worth studying as we will be able to demonstrate how our formalism in Sec.~\ref{s2.3} works and how additional Weyl-invariant terms would explicitly appear. We will also be able to see whether issues mentioned in Sec.~\ref{s2.2}, such as thermal equilibrium with a negative energy bath or logarithmic divergence, would occur. 

Following a similar construction as in Sec.~\ref{s3.1}, we consider the same effective actions with two auxiliary fields $\chi_1$ and $\chi_2$, where the solutions for $\chi_1$ and $\chi_2$ again given by \er{chi1} and \er{chi2}. We then examine the conditions for the Hartle-Hawking state. Here we impose the following physical conditions: 
\begin{itemize}
    \item Regularity conditions: regularity in both the future and past horizons can be achieved by imposing
    \be \la{regu}
    \lim_{r \to r_0} \frac{|\langle H | T_{uu} | H \rangle|}{(1-\frac{r_0}{r})^2} = \lim_{r \to r_0} \frac{|\langle H | T_{vv} | H \rangle|}{(1-\frac{r_0}{r})^2} <\infty , \quad \lim_{r \to r_0} \frac{|\langle H | T_{uv} | H \rangle|}{(1-\frac{r_0}{r})} <\infty.
    \ee
    Note that these are the same conditions as the regularity conditions discussed in Appendix.~\ref{sA} for the Boulware state.
    \item Asymptotic behaviors: for the thermal equilibrium with a thermal bath, we expect a balanced radiation and incoming flux asymptotically
    \be \la{asym}
    \lim_{r \to \infty} \langle H| T_{uu} | H \rangle= \lim_{r \to \infty} \langle H| T_{vv}| 
    H \rangle=\frac{\hbar}{192 \pi r^2_0}.
    \ee
    We take the value given by the s-wave result from four dimensions. We will only need the asymptotic value to fix the stress tensor completely, even though the full four-dimensional answer is unknown.
\end{itemize}

Can we use solely the $\Gamma_{\text{anom}}$ to capture the Hartle-Hawking state? That is, we require no additional Weyl-invariant terms to appear, then we just have the same constraints as in \er{constraint1}. The answer turns out to be no. The solutions in \er{sol1} or \er{sol2} are incompatible with the two conditions \er{regu} and \er{asym}. This means that we should relax the final constraint in \er{constraint1} to be with some constant $L$
\be \la{cons}
\lambda_2^2- \mu_2^2=L.
\ee
Without loss of generality, we can set $\lambda_1=0$ in \er{constraint1} and solve $(\mu_1, \mu_2, \lambda_2)$. We have four roots from \er{constraint1} combined with \er{cons}
\be \la{root1}
\bigg\{ \mu_1 = -\frac{1}{4 \sqrt{3 \pi}}, \quad \mu_2 = \sqrt{\frac{3}{4\pi}},  \quad \lambda_2= \mp \sqrt{L+\frac{3}{4 \pi}} \bigg\}, 
\ee
and
\be \la{root2}
\bigg\{ \mu_1 = \frac{1}{4 \sqrt{3 \pi}}, \quad \mu_2 = -\sqrt{\frac{3}{4\pi}},  \quad \lambda_2= \mp \sqrt{L+\frac{3}{4 \pi}} \bigg\}.
\ee
We substitute each of the four roots into either the $\langle H| T_{uu}| H \rangle$ or $\langle H| T_{vv}| H \rangle$ by expanding around the horizon $r= r_0+x$. With the regularity conditions \er{regu}, we require terms proportional to $\frac{1}{x}$ and $\ln{x}$ must vanish under $x \to 0$. This procedure will give two constraint equations for each of the four roots. In combination of the asymptotic behaviors \er{asym} that gives the following constraint for the four roots
\be
\frac{1}{4} (C^2_3-C^2_1)=\frac{\hbar}{192 \pi r_0^2}.
\ee
All these roots lead to a unique choice of $L$ being
\be \la{Leq}
L= -\frac{1}{2 \pi}.
\ee
With this choice, we can immediately write down the one-loop effective action for the Hartle-Hawking state as
\be \la{HHaction}
\begin{split}
    \Gamma_{\text{HH}}=&-\frac{\hbar}{96 \pi} \int d^2 x \sqrt{-g} \bigg( R \Box^{-1} R -12 (\nabla \phi)^2 \Box^{-1} R +12 \phi R 
\\
&+24 (\nabla \phi)^2 \Box^{-1} (\nabla \phi)^2\bigg)
\end{split}
\ee
which differs with $\Gamma_{\text{anom}}$ by an additional Weyl-invariant term $(\nabla \phi)^2 \frac{1}{\Box} (\nabla \phi)^2$.

Let us continue to solve the remaining constraints. For the choice \er{root1}, we have
\be \la{solu1}
\bigg\{C_1=0, \quad  C_3=- \frac{1}{4 \sqrt{3 \pi} r_0} \bigg\} \text{ or } \bigg\{C_1= \mp \frac{1}{4 \sqrt{\pi} r_0 }, \quad  C_3=- \frac{1}{2 \sqrt{3 \pi}r_0} \bigg\},
\ee 
their predicted $\langle H| T_{uu}| H \rangle$ or $\langle H| T_{vv}| H \rangle$ will be the same once we set $C_2=0$. For \er{root2}, we have 
\be \la{solu2}
\bigg\{C_1=0, \quad  C_3=\frac{1}{4 \sqrt{3 \pi}r_0} \bigg\} \text{ or } \bigg\{C_1= \mp \frac{1}{4 \sqrt{\pi} r_0 }, \quad  C_3= \frac{1}{2 \sqrt{3 \pi} r_0} \bigg\},
\ee 
again, $\langle H| T_{uu}| H \rangle$ or $\langle H| T_{vv}| H \rangle$ will be the same as we set $C_2=0$. On the other hand, when comparing between the two cases \er{solu1} and \er{solu2}, we have $C_4=0$. Therefore, we have the following unique quantum stress tensor corresponding to the Hartle-Hawking state compatible with \er{regu} and \er{asym}
\be \la{HHTensor1}
 \langle H| T_{uu} | H \rangle= \langle H| T_{vv} | H \rangle=\frac{\hbar}{192 \pi r_0^2} \bigg(1-\frac{r_0}{r} \bigg)^2 \bigg[1+\frac{2r_0}{r}+\frac{9 r_0^2}{r^2} \bigg(1-4 \ln{\frac{r}{\ell}} \bigg) \bigg],
\ee
\be \la{HHTensor2}
 \langle H| T_{uv} | H \rangle=  \frac{\hbar r_0}{24 \pi r^3} \bigg(1-\frac{r_0}{r} \bigg).
\ee
where $\ell$ is an arbitrary length scale that we may set $\ell = r_0$. We can clearly see that the stress tensor is regular, and no logarithmic divergence at the horizon like the ones predicted in \cite{Balbinot_2001, Balbinot:2002bz, Hofmann:2004kk, Hofmann:2005yv} is observed. Again, the $uv$-component is in agreement with the anomaly equation \er{confana}.

As we have pointed out in Sec.~\ref{s3.1}, since the constant $L$ is not zero, the physical spectrum of the effective action \er{HHaction} does not contain the Boulware state.

\subsection{Gravitational Collapse and $|in \rangle$ State} \la{s3.3}

Now we consider the case of an evaporating black hole in the non-minimal dilaton gravity model. We expect more dynamics to enter into the calculations of an evaporating black hole. In particular, we are interested in understanding whether the same problems, such as the anti-evaporation or logarithmic divergence at the horizon we mentioned in Sec.~\ref{s2.2}, occur.

We start from the construction of the $|in \rangle$ vacuum state that corresponds to a dynamical black hole formed from the gravitational collapse of a spherical null shell \cite{Hiscock:1980ze, Hiscock:1981xb, Fabbri:2005} (see also \cite{Davies:1976ei} for a time-like shell). The $|in \rangle$ state is defined such that it corresponds to the Minkowski vacuum on past null infinity. This vacuum state corresponds to the Unruh state $|U \rangle$ \cite{Unruh:1976db} in the late-time limit, which is annihilated by the operators defined by plane waves with respect to $(U,v)$. 

\begin{figure} 
\centering
    \includegraphics[width=0.30\textwidth]{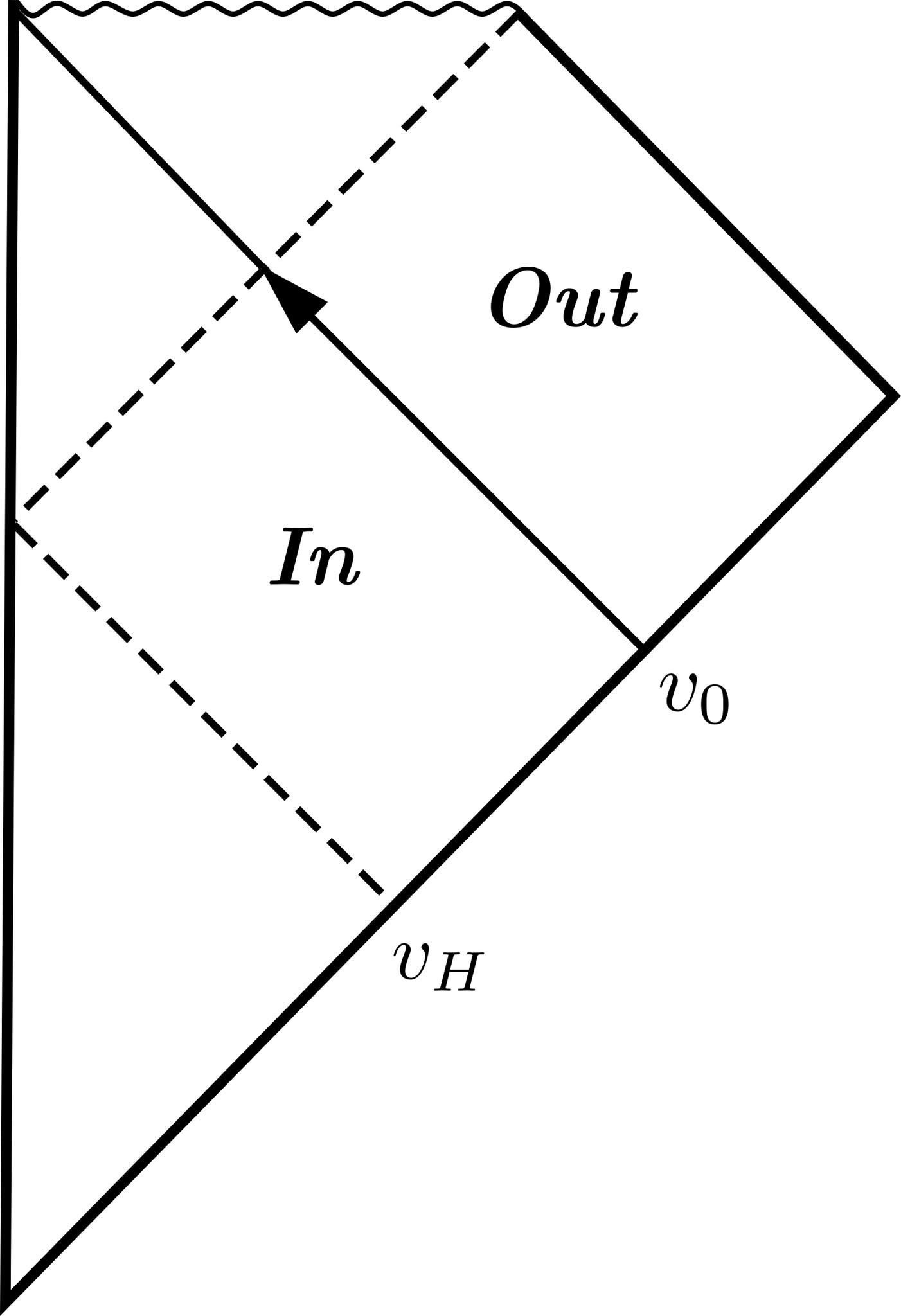}
    \caption{Gravitational collapse with a null shockwave at $v=v_0$. The in region with $v< v_0$ is given by a flat spacetime; while the out region is given by a back-reacted black hole geometry. The two spacetimes are junctioned at $v=v_0$}
    \label{Collapse}
\end{figure}

This case is more tricky as we need to construct the geometry and the stress tensor more carefully due to its dynamical nature. Let us follow a similar construction as in \cite{Hiscock:1980ze, Hiscock:1981xb} by considering the gravitational collapse of a spherical null shell with the back-reaction geometry.

Consider a null shockwave at $v=v_0$ that forms the black hole. In the "in" region $v< v_0$, the spacetime is flat (see Figure~\ref{Collapse})
\be
ds^2_{in}=-du_{in}dv,
\ee
For $v> v_0$, the "out" geometry is a "back-reacted" black hole geometry
\be 
ds^2_{out}=-F(r,v)e^{2 \epsilon \varphi(r, v)}du_{out} dv,
\ee
where
\be
F(r,v)=1-\frac{r_0}{r}+ \frac{\epsilon m(r,v)}{r}.
\ee
Note that both the functions $m(r,v)$ and $\varphi(r,v)$ are generally time-dependent. With this metric ansatz, we have the following nice tortoise coordinate
\be
\frac{dr^\ast}{dr}=F^{-1}(r,v)e^{- \epsilon \varphi(r,v)},
\ee
where $r^\ast$ is also generally time-dependent. Note that $d u_{out}=dv-2 dr^\ast$, which allows us to transform the metric into the following ingoing Vaidya form, which is describing the geometry outside the ingoing null shell
\be \la{Vaidya}
ds^2=-F(r,v)e^{2 \epsilon \varphi(r,v)} dv^2+2 e^{\epsilon \varphi (r,v)} dv dr.
\ee
We will always write the radius of the back-reacted event horizon as $r_H$. Now we consider the junction condition at $v=v_0$. Requiring the metric on the shockwave to be continuous on both sides implies
\be
r(u_{in},v_0)=r(u_{out},v_0),
\ee
where
\be
r(u_{in},v_0)=\frac{v_0-u_{in}}{2}, \quad r^\ast(u_{out},v_0)=\frac{v_0-u_{out}}{2}.
\ee
We can solve the following relation approximately
\be \la{junc}
u_{out} \approx \frac{u_{in}+(2 \kappa r_H-1) v_0}{2 \kappa r_H}-\frac{1}{\kappa} \ln{ \bigg(\frac{v_0-u_{in}-2 r_H}{2 r_H} \bigg)}.
\ee
Now we solve $u_{in}$ as
\be
u_{in} =v_0-2r_H -2r_H W[e^{-1+\kappa v_0- \kappa u_{out}}],
\ee
where $W(x)$ is the Lambert W function. At late times $u_{out} \to \infty$
\be \la{latetimeu}
u_{in} \simeq v_0-2 r_H -2r_H e^{-\kappa u_{out}}  \approx v_H+U,
\ee
where $v_H= v_0 -2r_H$. Here the second equality is also an approximation as $U\equiv-\frac{1}{\kappa} e^{-\kappa u_{out}}$. We are taking their difference to be of $O (\epsilon)$ since $\kappa \approx \frac{1}{2 r_0}+O(\frac{\epsilon}{r_0})$ and $r_H=r_0+O(\epsilon)$. A final remark is that $(u_{in},v) $ covers the entire spacetime, while $(u_{out},v)$ does not cover the black hole region. The vacuum state $|in \rangle$ is therefore defined to be annihilated by the annihilation operators defined with respect to the modes $(u_{in}, v)$. As we can see from \er{latetimeu}, at late times, it is captured by $U$. We take the Unruh state $| U \rangle$ to be defined with respect to $(U,v)$. The Unruh state can be understood as the quantum state that describes the gravitational collapse in the late-time near-horizon limit.

Now we focus on finding the regular stress tensor associated with the Unruh state $| U \rangle$ for the non-minimal dilaton gravity model. We will demonstrate that the covariant stress tensor can be obtained by a Weyl transformation from $(u_{out},v)$ to $(u_{in},v)$ without assuming any \textit{a priori} knowledge about $\Gamma_W$. In a general local conformal gauge $ds^2=-e^{2 \rho} dx^+ dx^-$, the stress tensors associated with $\Gamma_{\text{loc}}$ and $\Gamma_W$ can be written in the form 
\be \la{Tloc}
\langle T^{{\text{geo}}}_{\pm \pm} \rangle=-\frac{ \hbar}{12 \pi} [(\pa_\pm \rho)^2-\pa_\pm^2 \rho ]+\frac{\hbar}{ 2 \pi} [\pa_\pm \rho \pa_\pm \phi + \rho (\pa_\pm \phi)^2].
\ee
\be
\langle T_{\pm \pm}^{W} \rangle= \langle \Psi| : T_{\pm \pm}:| \Psi \rangle.
\ee
The two terms are non-covariant if considered separately, but the sum of these two pieces must form a covariant stress tensor, see Appendix.~\ref{sA}.

As we have stressed in the beginning, the normal-ordered piece is determined by physical conditions associated with the quantum state. We will see that by imposing the following vanishing normal-ordering part of the stress tensor for $| in \rangle$ state
\begin{equation}
    \bra in | : T_{u_{in} u_{in}} : |in \ket =0, \quad \langle in | : T_{vv}: | in \rangle=0,
\end{equation}
which means we only need to take into account $\langle in|T^{\text{geo}}_{u_{in} u_{in}} | in \rangle$ in the covariant stress tensor $\langle in | T_{u_{in} u_{in}} | in \rangle$, and similarly for the $vv$ component, we will be able to derive a workable form of stress tensor for the Unruh state $| U \rangle$.

Let us be explicit, the coordinate $u_{{\rm in }}$ is related to $u_{\text{out}} \equiv u$ via the following junction condition to leading order in $\epsilon$ from \er{junc}
\begin{equation}
    u=u_{in}-2r_0\ln{\bigg(\frac{v_0-u_{in}-2r_0}{2r_0} \bigg)},
\end{equation}
and the conformal factor $\rho^\prime$ is related to $\rho$ defined in $(u,v)$ through
\be
ds^2=-e^{2 \rho} du dv=-e^{2 \rho} u' du_{in} dv=-e^{2 \rho'} du_{in} dv,
\ee
with $\rho'=\rho+\frac{1}{2}\ln{ u'}$ and $u'=du/du_{in}$. Hence the $uu$ component of \er{Tloc} is transformed to be
\bea \la{Tloc1}
\langle in | T_{u_{in} u_{in}} | in \rangle&\equiv& \langle in | T^{(1)}_{u_{in} u_{in}} | in \rangle+\langle in | T^{(2)}_{u_{in} u_{in}} | in \rangle
\no\\
&=&-\frac{\hbar}{12 \pi} [ (\pa_{u_{in}} \rho')^2-\pa^2_{u_{in}} \rho']+\frac{\hbar}{ 2 \pi} [\pa_{u_{in}} \rho' \pa_{u_{in}} \phi+\rho' (\pa_{u_{in}} \phi)^2].
\eea
Let us analyze the two pieces separately. For the first term with
\be
\langle in | T^{(1)}_{u_{in} u_{in}} | in \rangle=-\frac{\hbar}{12 \pi} [ (\pa_{u_{in}} \rho')^2-\pa^2_{u_{in}} \rho'],
\ee
let us work out
\be
\pa_{u_{in}} \rho'=\pa_{u_{in}} \rho+\frac{u''}{2 u'}, \quad \pa^2_{u_{in}} \rho'=\pa^2_{u_{in}} \rho+\frac{u'''}{2 u'}-\frac{u''^2}{u'^2}.
\ee
Then we can organize the equation into the following form
\be \la{Tloc2}
-\frac{\hbar}{12 \pi} [ (\pa_{u_{in}} \rho')^2-\pa^2_{u_{in}} \rho']=-\frac{\hbar}{12 \pi} \bigg[ (\pa_{u_{in}} \rho)^2-\pa_{u_{in}}^2 \rho+\pa_{u_{in}} \rho \frac{u''}{u'} \bigg]+\frac{\hbar}{24 \pi} \{u, u_{in} \},
\ee
where we notice the second piece is the Schwarzian derivative with $\{u, u_{in} \}=\frac{u'''}{u'}-\frac{3u''^2}{2u'^2}$. 

We can explicitly evaluate \er{Tloc2} by plugging the background values (omitting $O(\epsilon)$ pieces) of the conformal factor $\rho$ and the junction condition \er{junc} 
\be
\rho=\frac{1}{2} \ln{\bigg( 1-\frac{r_0}{r} \bigg)}, \quad u=u_{in}-2r_0\ln{\bigg(\frac{v_0-u_{in}-2r_0}{2r_0} \bigg)},
\ee
and the result is given by
\begin{equation} 
    \begin{split}
        -\frac{\hbar}{12 \pi} [ (\pa_{u_{in}} \rho')^2-\pa^2_{u_{in}} \rho']=&-\frac{\hbar}{12 \pi} \bigg[ (\pa_{u_{in}} \rho)^2-\pa_{u_{in}}^2 \rho+\pa_{u_{in}} \rho \frac{u''}{u'} \bigg]+\frac{\hbar}{24 \pi} \{u, u_{in} \}\\
        =&-\frac{\hbar}{12 \pi} \bigg( -\frac{3 r^2_0}{16 r^4}+\frac{r_0}{4r^3}\bigg) \frac{(u_{in}-v_0)^2}{(4M+u_{in}-v_0)^2}\\
        &-\frac{\hbar}{24 \pi} \frac{6 r^2_0 +4 r_0(u_{in}-v_0)}{(u_{in}-v_0)^2 (2 r_0+u_{in}-v_0)^2},
    \end{split}
\end{equation}
where the last line is the explicit evaluation of the Schwarzian term. Let us apply a coordinate transformation back to the coordinate $u$ in the out region and look at the $uu$-component
\be \la{Tuu1}
\begin{split}
    \langle in |T^{(1)}_{u u} | in \rangle&=\frac{du_{in}}{du} \frac{du_{in}}{du} \langle in |T^{(1)}_{u_{in} u_{in}} | in \rangle\\
    &=\frac{(u_{in}-v_0)^2}{(4M+u_{in}-v_0)^2} \langle in |T^{(1)}_{u_{in} u_{in}} | in \rangle\\
    &=\frac{\hbar}{24 \pi } \bigg(\frac{3r^2_0}{8 r^4}-\frac{r_0}{2r^3} \bigg)-\frac{\hbar}{ 24 \pi} \{u_{in},u \},
\end{split}
\ee
where for the Schwarzian term, we have applied the following inverse transformation law
\be
\{u,u_{in} \}=-\{u_{in},u \} \bigg(\frac{du}{du_{in}} \bigg)^2.
\ee
We can see clearly the first term in \er{Tuu1} captures the stress tensor of the Boulware state had we used the minimal model \er{matter1}, as we commented in Sec.~\ref{s3.1}. Hence the expression \er{Tuu1} is indeed the correct covariant stress tensor for the $|in \rangle$ state in the minimal model related by a Schwarzian derivative \cite{Hiscock:1980ze, Hiscock:1981xb, Fabbri:2005}.

Similarly, for the second term in \er{Tloc1} with 
\be
\langle in | T^{(2)}_{u_{in} u_{in}} | in \rangle=\frac{\hbar}{ 2 \pi} [\pa_{u_{in}} \rho' \pa_{u_{in}} \phi+\rho' (\pa_{u_{in}} \phi)^2],
\ee
since the dilaton $\phi$ is still given by $\phi=-\ln{r}$ in the out region, we work out
\be
\pa_{u_{in}} \phi=\frac{1}{2r}\bigg(1-\frac{r_0}{r} \bigg) \frac{u_{in}-v_0}{2r_0+u_{in}-v_0}.
\ee
After some algebra, we have
\bea
\frac{\hbar}{ 2 \pi} [\pa_{u_{in}} \rho' \pa_{u_{in}} \phi+\rho' (\pa_{u_{in}} \phi)^2]&=&\frac{\hbar}{2 \pi} \bigg[\frac{r_0 }{2r}\bigg( 1-\frac{r_0}{r} \bigg)+(u_{in}-v_0)^2 \bigg(\bigg(-\frac{r_0}{8r^3} \bigg) \bigg( 1-\frac{r_0}{r} \bigg) 
\no\\
&\quad &+\frac{1}{8 r^2} \bigg( 1-\frac{r_0}{r} \bigg)^2 \ln{\bigg\{ \bigg( 1-\frac{r_0}{r} \bigg) \bigg( \frac{u_{in}-v_0}{2r_0+u_{in}-v_0} \bigg) \bigg\} } \bigg) 
\no\\
&\quad& \bigg] \bigg/ (2 r_0+u_{in}-v_0)^2.
\eea
Again, we perform a coordinate transformation back to the $uu$-component
\be
\begin{split}
    \langle in | T^{(2)}_{u u }| in \rangle&=\frac{du_{in}}{du}  \frac{du_{in}}{du} \langle in | T^{(2)}_{u_{in} u_{in}} | in \rangle \\
    &=\frac{\hbar}{2 \pi}  \bigg[-\frac{r_0}{8 r^3} \bigg( 1-\frac{r_0}{r} \bigg)+\frac{r_0}{2r} \bigg( 1-\frac{r_0}{r} \bigg) \frac{1}{(u_{in}-v_0)^2} \\
    &\quad +\frac{1}{8 r^2} \bigg(1-\frac{r_0}{r}\bigg)^2 \ln{\bigg\{ \bigg( 1-\frac{r_0}{r} \bigg) \bigg( \frac{u_{in}-v_0}{2r_0+u_{in}-v_0} \bigg) \bigg\}   } \bigg].
\end{split}
\ee
This is the new result that corresponds to the non-minimal dilaton coupling. The full covariant stress tensor for the $| in \rangle$ state can be written as 
\be \la{fullT}
\begin{split}
    \langle T_{uu} \rangle&=\langle in | T^{(1)}_{u u} | in \rangle+\langle in | T^{(2)}_{u u}  | in \rangle\\
    &=\frac{\hbar}{24 \pi } \bigg(\frac{3r^2_0}{8 r^4}-\frac{r_0}{2r^3} \bigg)-\frac{\hbar}{ 24 \pi} \{u_{in},u \} \\
    &\quad +\frac{\hbar}{2 \pi}  \bigg[-\frac{r_0}{8 r^3} \bigg( 1-\frac{r_0}{r} \bigg)+\frac{r_0}{2r} \bigg( 1-\frac{r_0}{r} \bigg) \frac{1}{(u_{in}-v_0)^2} \\
    &\quad +\frac{1}{8 r^2} \bigg(1-\frac{r_0}{r}\bigg)^2 \ln{\bigg\{ \bigg( 1-\frac{r_0}{r} \bigg) \bigg( \frac{u_{in}-v_0}{2r_0+u_{in}-v_0} \bigg) \bigg\}   } \bigg],\\
\langle T_{vv} \rangle&=\frac{\hbar}{24 \pi } \bigg(\frac{15 r^2_0}{8 r^4}-\frac{2r_0}{r^3} \bigg)+\frac{\hbar}{16 \pi} \frac{(r-r_0)^2 \ln{(1-\frac{r_0}{r})}}{r^4}\\
&\quad+\frac{\hbar }{16 \pi r^2 }\bigg(1-\frac{r_0}{r}\bigg)^2\ln{\bigg(\frac{u_{in}-v_0}{2r_0+u_{in}-v_0}\bigg)},\\
\langle T_{uv} \rangle&=\frac{\hbar r_0}{24 \pi r^3} \bigg(1-\frac{r_0}{r} \bigg),
\end{split}
\ee
where the $vv$ component is derived by following exactly the same procedure as the $uu$ component, while with $\pa_v \rho'=\pa_v \rho$. The $uv$ component again corresponds to the dilaton-deformed conformal anomaly.

To examine whether the covariant stress tensor we obtained makes sense, let us perform a few sanity checks that should be satisfied by the $|in \rangle$ state. Now we consider whether the following boundary conditions for the $|in \rangle$ state can be satisfied
\begin{itemize}
    \item At early times where we take $u_{in} \sim u \to -\infty$, $\langle T_{uu} \rangle$ and $\langle T_{vv} \rangle$ should reproduce the Boulware-type terms. This is because the $|in \rangle$ state is defined such that it reduces to the Minkowski vacuum on past null infinity. This can be easily verified as
    \be \la{EarlyT}
    \lim_{u_{in} \to -\infty} \langle T_{uu} \rangle =\lim_{u_{in} \to -\infty} \langle T_{vv} \rangle =\frac{\hbar}{24 \pi } \bigg(\frac{15 r^2_0}{8 r^4}-\frac{2r_0}{r^3} \bigg) +\frac{\hbar}{16 \pi} \frac{(r-r_0)^2 \ln{(1-\frac{r_0}{r})}}{r^4},
    \ee
    which actually does not coincide with \er{BoulwareT}. The reason has to do with the fact that we implicitly assumed $\langle B|: T_{ab}:| B \rangle \neq 0$ in our calculation leading to \er{BoulwareT}, we will address this issue in Sec.~\ref{s5}. Note that this corresponds to pure vacuum polarization, which goes to zero asymptotically.
    \item  Regularity conditions at the future horizon. Since in our configuration, there is no past horizon, we only require the following conditions to hold 
    \be \la{reguU}
    \lim_{r \to r_0} \frac{|\langle in | T_{uu} | in \rangle|}{(1-\frac{r_0}{r})^2} <\infty , \quad \lim_{r \to r_0} \frac{|\langle in | T_{uv} | in \rangle|}{(1-\frac{r_0}{r})} <\infty, \quad \lim_{r \to r_0} |\langle in | T_{vv} | in \rangle| <\infty.
    \ee
    These are clearly satisfied with \er{fullT}.
    \item At late times ($u \to \infty$), from \er{latetimeu}
    \be
    u_{in} \simeq v_0-2 r_H -2r_H e^{-\kappa u} \approx v_0-2 r_H,
    \ee
    we require
    \be \la{uuasym}
    \langle in | T_{uu} |in \rangle \to \frac{\hbar}{192 \pi r_0^2} \quad \text{as } r \to \infty,
    \ee
    \be \la{vvhorizon}
    \langle in| T_{vv} |in \rangle \to \frac{-\hbar}{192 \pi r_0^2} \quad \text{as } r \to r_0
    \ee
    The condition on $\langle in| T_{uu} |in \rangle$ represents the positive outgoing flux of Hawking radiation at future null infinity, where the value should be given again by the s-wave result from four dimensions. This is clearly satisfied since the contribution from the dilaton part in $\langle in| T^{(2)}_{uu} |in \rangle$ vanishes.
    
    The condition on $\langle in| T_{vv} |in \rangle$ comes from the fact that there must be a negative influx of energy that makes the black hole shrink while compensating for the positive outgoing flux. This condition is also satisfied as $\langle in| T_{vv} |in \rangle$ is not affected by the dilaton contribution. 

    We should emphasize that our analytic results in \er{fullT} naturally lead to \er{uuasym} and \er{vvhorizon}, which corresponds to the s-wave approximation for a minimally coupled matter theory in two dimensions. However, in the dilaton-coupled matter theory, one does expect to get the following result with a grey-body factor even in s-wave
    \be
    \langle in | T_{uu} |in \rangle \to \frac{\hbar}{2\pi}\int_0^\infty \frac{w dw}{e^{4 \pi r_0 w}-1} \Gamma_{w, l=0},
    \ee
    where the s-wave grey-body factor $\Gamma_{w, l=0}$ comes from the transmission coefficient of the corresponding potential barrier of the matter theory. It is only when we take $\Gamma_{w, l=0}=1$ by ignoring the backscattering effect that we recover \er{uuasym} asymptotically. A similar statement also applies to \er{vvhorizon}. Therefore, our expressions in \er{fullT} can be considered as a useful approximation that captures the high-frequency limit where the backscattering is negligible.

\end{itemize}

\section{The Back-Reaction Geometry and Quantum Extremal Islands} \la{s4}

Having presented a self-consistent method to treat the non-minimal dilaton gravity model, this section is devoted to the back-reaction and island problems of this model. We will consider the case of eternal and evaporating black holes, corresponding to the Hartle-Hawking and Unruh states, respectively. The goal is to show that one can successfully reproduce the Page curve of black hole evaporation, which was unavailable until we have a consistent one-loop theory.

\subsection{Setup of the Back-Reaction Problem and Island Formula} \la{s4.1}

We consider the generic (1+1)-dimensional dilaton gravity model dimensionally reduced from (3+1)-dimensional Einstein-Hilbert action with a single massless scalar matter field, as presented in Sec.~\ref{s2.1}. For clarity, let us reproduce the classical action of our theory here
\bea
S_{\text{cl}} &=& S_{\text{grav}}+S_{\text{matter}} 
\\
&=&\frac{1}{4 G_N} \int d^2 x \sqrt{-g}[e^{-2 \phi} (R+2 (\nabla \phi)^2)+2 ]-\frac{1}{2}\int d^2x \sqrt{-g} e^{-2 \phi}(\nabla f)^2.
\eea
The classical equations of motion for the metric $g_{ab}$, the dilaton $\phi$, and the scalar field $f$ are given respectively by 
\be
e^{-2 \phi} \{2 \nabla_a \nabla_b \phi-2 \nabla_a \phi \nabla_b \phi+g_{ab}[3 (\nabla \phi)^2-2 \Box \phi] \} -g_{ab}=2G_N T^{(g)}_{ab},
\ee
\be \la{dilatonEOM}
e^{-2 \phi}\bigg[(\nabla \phi)^2-\Box \phi -\frac{R}{2} \bigg] =- G_N \frac{\delta S_{\text{matter}}}{\delta \phi},
\ee
\be
e^{-2 \phi} (\Box f-2 \nabla_a \phi \nabla^a f)=0,
\ee
and the classical stress tensor is given by 
\be
T^{(g)}_{ab}=\frac{-2}{\sqrt{-g}}\frac{\delta S_{\text{matter}}}{\delta g^{ab}}=e^{-2 \phi} \bigg[ \nabla_a f \nabla_b f- \frac{1}{2} g_{ab} (\nabla f)^2 \bigg].
\ee
Since the model is not exactly solvable, we take the Schwarzschild metric as the background solution for the following perturbative back-reaction problem.

We consider the quantum back-reaction problem by adopting a unitless perturbative parameter $\epsilon$ in orders of $G_N \hbar/\ell^2$, where $\ell$ is some length scale of the quantum fields that we omit by setting it to unity. Specifically, we define $\epsilon = \frac{G_N \hbar}{24 \pi}$. We then solve the semi-classical Einstein equations perturbatively with the back-reaction on top of the background solution. We take the classical stress tensor $T_{ab}$ to be vanishing outside the classical radius $r_0=2M$. The stress tensor for $r > r_0$ is given by $\langle T_{ab} \rangle$, which comes from the one-loop effective action as
\be
\langle T_{ab} \rangle=\frac{-2}{\sqrt{-g}}\frac{\delta \Gamma_{\text{eff}}}{\delta g^{ab}}.
\ee
Thus, we will be able to solve the back-reacted geometry consistently up to $O(\epsilon)$.

Once we have the back-reacted geometry, we apply the quantum extremal surface prescription \cite{Engelhardt:2014gca} that leads to the so-called island formula \cite{Penington:2019npb,Almheiri:2019psf,Almheiri:2019hni}
\be \la{islandfor}
S_\text{gen} (R)=\text{min}_{I} \bigg\{ \text{ext}_{I} \bigg[ \frac{\text{Area}(\partial I)}{4 G_N}+S_{\text{matter}}(I \cup R) \bigg] \bigg\}. 
\ee
For an intuitive discussion on how the island formula is derived and how it leads to a unitary Page curve \cite{Page:1993wv,Page:2013dx} once the quantum extremal island is found, we refer to the review \cite{Almheiri:2020cfm}. The island formula can be viewed as the correct prescription that computes the fine-grained entropy of Hawking radiation, and was derived from the replica wormhole saddles using techniques of Euclidean path integral in the context of JT gravity \cite{Penington:2019kki,Almheiri:2019qdq}.

The entropy is given by extremizing a generalized entropy-like functional over the islands $I$ followed by minimization over all extrema. Note that the area term here refers to the boundary of the island region $\pa I$. As should be obvious from the context, we will slightly abuse the term "island" as referring to $\pa I$. The $S_{\text{matter}}(I \cup R)$ term should be understood as the semi-classical entanglement entropy of the quantum fields with support on the combined radiation and the island systems $I \cup R$. We should emphasize that we are not assuming that such an extremal surface could always be found in the non-minimal dilaton gravity model, instead we will show it is the case.

\begin{figure} 
\centering
    \includegraphics[width=0.30\textwidth]{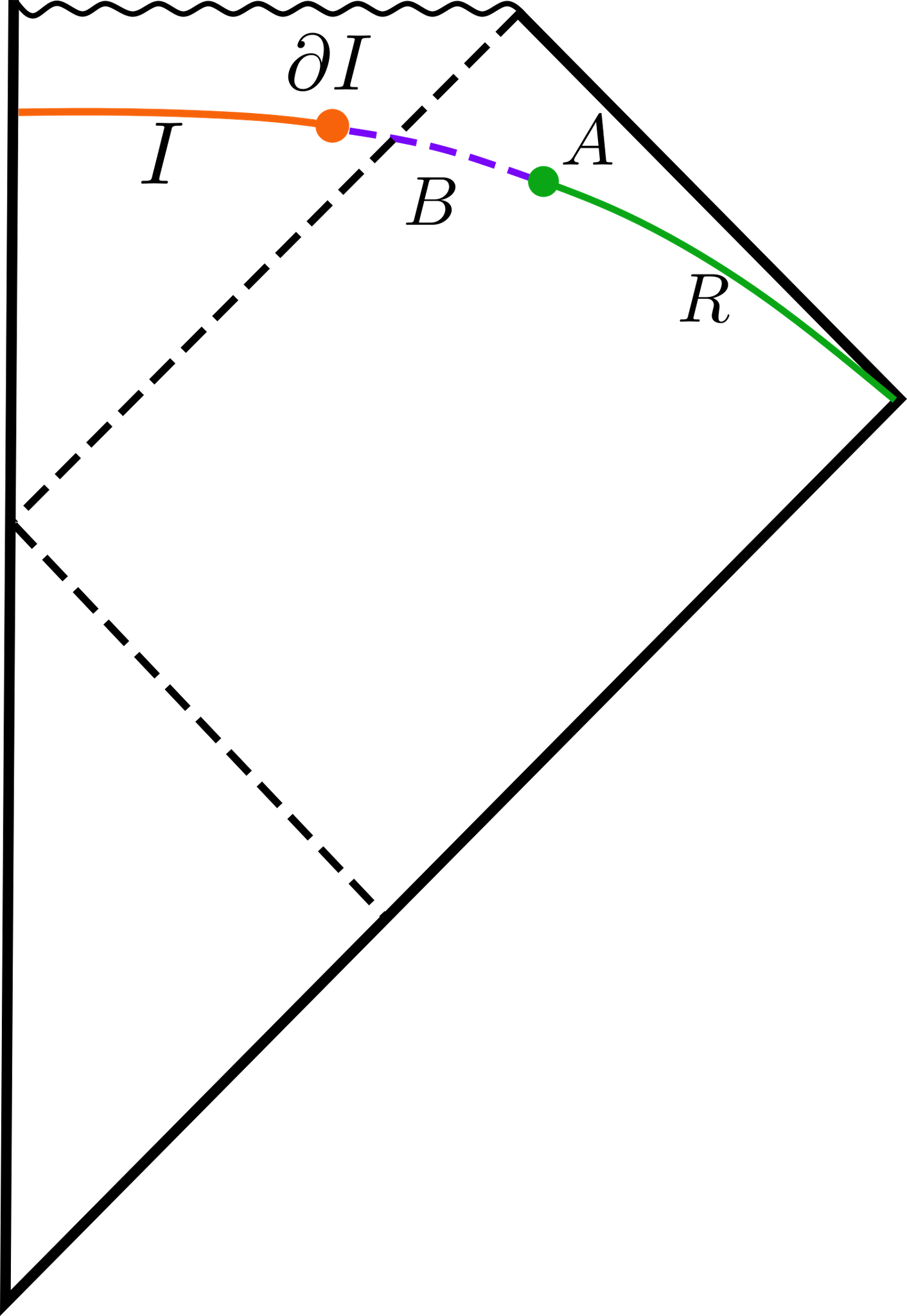}
    \caption{We begin with a pure vacuum state on the Cauchy slice $I \cup B \cup R$, and by complementarity property, we have $S_{\text{matter}}(I \cup R)=S_{\text{matter}}(B)$. In this figure, we take $I$ to represent the island region, where $\pa I$ is a quantum extremal surface. A cut-off surface is taken to be the boundary of the radiation region $A=\pa R$.}
    \label{Island}
\end{figure}

In practice, we need to compute $S_{\text{matter}}(I \cup R)$ in a curved background. We assume that $I \cup B \cup R$ is a Cauchy slice in Figure~\ref {Island} where we have a pure vacuum state. By the complementarity property, we have
\be
S_{\text{matter}}(I \cup R)=S_{\text{matter}}(B),
\ee
then we can adopt the single interval entropy formula that can significantly simplify the calculation. Given the metric $ds^2=-e^{2 \rho(x^+,x^-)} dx^+ dx^-$ with conformal factor $\rho(x^+,x^-)$, and let us take the cut-off surface to be $A=\pa R$ in Figure~\ref{Island}, the general formula that works in the curved background for a single interval is given by \cite{Fiola:1994ir}
\be \la{matterentropy}
S_{\text{matter}} (B)=\frac{c}{12}\ln \frac{(x^+(A)-x^+(I))^2 (x^-(A)-x^-(I))^2}{ \delta^4 e^{-2 \rho_A }e^{-2 \rho_I}},
\ee
and similarly a formula for two disjoint intervals as \er{twoint}. Note that $\delta$ is the UV cut-off and we treat the central charge $c$ as a constant here. For the Hartle-Hawking and Unruh states we are going to consider in the next two subsections, we detail the calculations involving this formula in Appendix.~\ref{sC} and Appendix.~\ref{sD}. We will discuss the applicability of this formula in Sec.~\ref{s5} for our construction. For now, we take \er{matterentropy} to be approximately true in our model.

\subsection{Eternal Black Hole Scenario} \la{s4.2}

We start from the quantum stress tensor for the Hartle-Hawking state in Sec.~\ref{s3.2} from the one-loop theory and solve the back-reacted geometry to $O(\epsilon)$. Consider the metric ansatz with two functions $m(r)$ and $\varphi(r)$
\be \la{ansatz1}
ds^2=-F(r) e^{2 \epsilon \varphi(r)}dt^2 + \frac{dr^2}{F(r)},
\ee
where we define 
\be
F(r)\equiv 1-\frac{r_0}{r}+\frac{\epsilon m(r)}{r}=F_0(r)+\frac{\epsilon m (r)}{r}.
\ee
With the Kruskal-type coordinates $(U,V)$ defined as
\be
 U \equiv -\frac{1}{\kappa}e^{-\kappa u}, \quad V \equiv \frac{1}{\kappa} e^{\kappa v},
\ee
then 
\be
ds^2=-F(r)e^{2 \epsilon \varphi(r)-2 \kappa r^\ast} dU dV =-e^{2 \rho (U,V)} dU dV,
\ee
with
\be
\rho(U,V)=\frac{1}{2} \ln h+\epsilon \varphi-\kappa r^\ast.
\ee
Note that the $(u,v)$ are the  Eddington-Finkelstein coordinates with 
\be
u \equiv t-r^\ast, \quad v \equiv t+r^\ast,
\ee
where the tortoise coordinate $r^\ast$ is defined as
\be
r^\ast =\int_r^\infty \frac{1}{F(r')e^{\epsilon \varphi(r')}} dr'.
\ee
In this case, the quantum corrected horizon position $r_H$ is determined by solving $g^{rr}(r_H)=0$, which is given by
\begin{equation}
    r_H = r_0-\epsilon m(r_0) + O(\epsilon^2).
\end{equation}
The surface gravity $\kappa$ at the horizon can be calculated from $\kappa^2=-\frac{1}{2}\nabla^a \chi^b \nabla_a \chi_b|_{H}$ where $\chi^a$ is the Killing vector for the stationary metric. Therefore
\be
\kappa = \frac{1}{2 r_0} \bigg[ 1+ \epsilon \bigg(\varphi (r_0)+ m'(r_0)+\frac{m(r_0)}{r_0} \bigg) \bigg] + O(\epsilon^2).
\ee
The remaining task is to solve the functions $m(r)$ and $\varphi(r)$ to $O(\epsilon)$ and determine  the geometry.

With the metric ansatz \er{ansatz1}, we have the following back-reaction equations
\be \la{back-reaction1}
- \epsilon F_0 (r) m'(r)= 2G_N \langle T_{tt} \rangle,
\ee
\be \la{back-reaction2}
2 \epsilon r F_0(r) \varphi'(r)= 2G_N \bigg( F_0(r) \langle T_{rr} \rangle +\frac{\langle T_{tt} \rangle}{F_0(r)} \bigg).
\ee
Starting from the regular stress tensor for the Hartle-Hawking state in Sec.~\ref{s3.2}, we can transform to $(t,r)$ coordinates by  
\bea
\langle T_{rr} \rangle &=&\frac{\pa x^a}{\pa r} \frac{\pa x^b}{\pa r} \langle T_{ab} \rangle
\no\\
&=& \frac{\hbar}{96 \pi r^2_0} \bigg[ 1+\frac{2r_0}{r}+\frac{9 r^2_0}{r^2} \bigg(1-4\ln{\frac{r}{r_0}} \bigg) \bigg] -\frac{\hbar r_0}{12 \pi r^3}\bigg( 1-\frac{r_0}{r} \bigg)^{-1},
\\
\langle T_{tt} \rangle &=&\frac{\pa x^a}{\pa t} \frac{\pa x^b}{\pa t} \langle T_{ab} \rangle
\no\\
&=& \frac{\hbar }{96 \pi r_0^2} \bigg( 1-\frac{r_0}{r} \bigg)^2 \bigg[1+\frac{2r_0}{r}+\frac{9 r^2_0}{r^2} \bigg(1-4 \ln{\frac{r}{r_0}} \bigg) \bigg] +\frac{\hbar r_0}{12 \pi r^3}\bigg(1-\frac{r_0}{r} \bigg).
\eea
By substituting into the back-reaction equations, we get the following solutions for $m(r)$ and $\varphi(r)$ 
\be
m(r)=-\frac{1}{2 r_0^2} \bigg[r+ r_0 \ln{\frac{r}{r_0}}+\frac{r_0^2}{r}\bigg(2 +9 \ln{\frac{r}{r_0}}  \bigg)-\frac{r_0^3}{r^2} \bigg(\frac{7}{4}+\frac{9}{2} \ln{\frac{r}{r_0}} \bigg) \bigg]+C_1,
\ee
\be
\varphi(r)=\frac{1}{2 r_0^2} \bigg[\ln{\frac{r}{r_0}}-\frac{2 r_0}{r}-\frac{r_0^2}{r^2}\bigg(\frac{9}{4}-\frac{9}{2}\ln{\frac{r}{r_0}} \bigg) \bigg]+C_2,
\ee
Now we need to determine the integration constants $C_1$ and $C_2$. Note that both functions behave regularly at the horizon under $r \to r_0$. However, both functions diverge asymptotically as $r \to \infty$, which indicates the theory is not asymptotically flat. Note that it is not a problem for the function $m(r)$, because it is $\frac{m(r)}{r}$ that appears in the metric and
\be
\lim_{r \to \infty} \frac{m(r)}{r}=\text{finite}.
\ee
Although there is no bearing for further analysis, we can still determine $C_1$ and $C_2$ by a cut-off. Let us introduce a cut-off at $r \to L$ and fix the integration constants by the following conditions 
\be
\lim_{L \to \infty} (\lim_{r \to L} m(r)) =0, \quad  \lim_{L \to \infty} (\lim_{r \to L} \varphi(r)) =0,
\ee
which allows us to set
\be
C_1=\frac{ 1}{2 r^2_0} \bigg(L+r_0 \ln{\frac{L}{r_0}} \bigg), \quad  C_2=-\frac{1}{2 r^2_0} \ln{\frac{L}{r_0}}.
\ee
Therefore, we have the following quantum-corrected horizon position and surface gravity to $O(\epsilon)$ as
\be
r_H \approx r_0-\epsilon m(r_0)=r_0+\frac{\epsilon }{2 r_0}\bigg( \frac{5}{4}-\frac{L}{r_0}-\ln{\frac{L}{r_0}}\bigg),
\ee
\bea
\kappa &\approx&\frac{1}{2 r_0} \bigg[ 1+ \epsilon \bigg(\varphi (r_0)+ m'(r_0)+\frac{m(r_0)}{r_0} \bigg) \bigg]
\no\\
&=&\frac{1}{2 r_0} \bigg[ 1- \frac{\epsilon }{2 r_0^2}\bigg(\frac{27}{2}-\frac{L}{r_0} \bigg) \bigg].
\eea

Given the back-reaction geometry, the island computation is straightforward, we refer to Appendix.~\ref{sC} for a self-contained treatment. Here we briefly recap the major results. 

We consider the no-island and island phases at late times. Without island, we can see clearly that $S_{\text{matter}}$ grows monotonically with time
\be
S_{{\rm matter}}\simeq\frac{1}{3}\ka t+{\rm const},
\ee
which is a general result in agreement with Hawking's prediction \cite{Hawking:1975, Hawking:1976}.

With island, we instead need to apply the $S_{\text{matter}}$ formula with two disjoint intervals as \er{twoint}. By extremizing $S_{\text{gen}}$, we have the following equation 
\be \la{islandeq}
\left[a+2\epsilon\rho^{\pp}\left(a\right)\right]F(a)e^{\epsilon\varphi(a)}=4\epsilon\frac{\kappa}{e^{\kappa\left(b^\ast-a^\ast\right)}-1},
\ee
where $b^\ast$ represents the cut-off surface and we take the island position $a$ to be outside the horizon with \cite{Almheiri:2019yqk}
\be
a =r_H+x, \quad x \ll r_H.
\ee
In Appendix.~\ref{sC}, we have considered two cases for $e^{\kappa a^\ast}$ on the right hand side of \er{islandeq}, where
\begin{itemize}
    \item The leading order piece of $e^{\kappa a^\ast}$ is an $O(1)$ constant, which means that the island position is at a small fixed location away from the horizon. The leading correction from $x$ is then $O(\epsilon)$
    \be
    x = \frac{2 \epsilon}{(r_H+2 \epsilon \rho'|_H)[e^{\kappa (b^\ast-a^\ast)}-1]} \approx \frac{2 \epsilon}{r_0 [e^{\kappa (b^\ast-a^\ast)}-1]}+O(\epsilon^2).
    \ee 
    \item The leading order piece of $e^{\kappa a^\ast}$ is $O(x)$, which means the island is extremely close to the horizon and they are nearly identical. The leading correction from $x$ is then $O(\epsilon^2)$
    \bea
    x&=& \frac{1}{r_H} \frac{ (\frac{2 \epsilon }{r_H} )^2 e^{1-2 \kappa b^\ast+\epsilon \alpha(r_H)}}{[1+ \frac{2 \epsilon }{r_H} (\rho'|_H -\frac{1}{r_H}e^{1-2\kappa b^\ast+\epsilon \alpha(r_H)} ) ]^2}
    \no\\
    &\approx&\frac{4 \epsilon^2}{r^3_h}e^{1-2 \kappa b^\ast}+O(\epsilon^3).
    \eea
\end{itemize}
These two cases indeed are different as we can see from $S_{\text{gen}}$
\bea
S_{\text{gen}}(a)&=&S_{\text{gen}}(r_H)+S_{\text{gen}}'(r_H)x + O(x^2)
\no\\
&\approx&S_{\text{gen}}(r_H)+\frac{4 \pi r_H}{G_N \hbar}x.
\eea
If $x \sim O(\epsilon)$, the correction can be $O(1)$ in $\epsilon$. However, the correction is essentially negligible if $x \sim O(\epsilon^2)$. Therefore, if we keep only up to the $O\left(1\right)$ terms of the
entropy, we can approximately think of the island location as the position of the back-reacted horizon. In either scenario, the fine-grained
entropy at late times is given by 
\be
S_{\text{FG}}=\text{min} \bigg \{S_{\text{gen,no-island}}, S_{\text{gen,island}} \bigg\}=\text{min} \bigg \{\frac{1}{3} \kappa t, S_{\text{gen}}(a) \bigg\}.
\ee
Hence the Page time can be determined approximately to be the transition time where 
\be
\frac{1}{3}\kappa t_P \approx S_{\text{gen}}(a) \implies t_P=3 \kappa S_{\text{gen}}(a).
\ee
Note that quantities such as $r_H$ and $\kappa$ are given by the back-reacted geometry of the non-minimal dilaton gravity model.

\subsection{Evaporating Black Hole Scenario} \la{s4.3}

Having obtained a consistent stress tensor for the $|in \rangle$ state in Sec.~\ref{s3.3}, we consider the back-reaction problem by adopting the following perturbative ingoing Vaidya metric describing the geometry outside an ingoing null shell as in \er{Vaidya}
\be
ds^2=-F(r,v)e^{2 \epsilon \varphi(r,v)} dv^2+2 e^{\epsilon \varphi (r,v)} dv dr,
\ee
where
\be
F(r,v)=1-\frac{r_0}{r}+ \frac{\epsilon m(r,v)}{r}.
\ee
As we have noted, we take the time dependence to be in the functions $m(r,v)$ and $\varphi(r,v)$, which is at $O(\epsilon)$. This is consistent with our understanding of quantum back-reaction where the stress tensor is time-dependent at the one-loop order, and it also corresponds to the case where the evaporation is quasi-static. 

The equations of motion for this metric up to $O(\epsilon)$ are given by 
\be
\epsilon  \pa_r \varphi(r) =\frac{G_N \langle T_{rr} \rangle}{r},
\ee
\be
\epsilon  \pa_r m(r,v)=2G_N \langle T_{rv} \rangle
\ee
\be
\epsilon  \pa_v m(r,v)= -2G_N [F_{0}(r) \langle T_{rv} \rangle+\langle T_{vv} \rangle]
\ee
With the full covariant stress tensor corresponding to the $|in \rangle$ state in Sec.~\ref{s3.3}, we perform a coordinate transformation with $v=v$, $u=v-2 r^\ast$. Then
\be
\begin{split}
    \langle T_{rr} \rangle&=4 \bigg(\frac{r}{r-2M} \bigg)^2 \langle T_{uu} \rangle,\\
    \langle T_{rv} \rangle&=\frac{-2r}{r-2M} (\langle T_{uu} \rangle+ \langle T_{uv }  \rangle),\\ 
    \langle T_{vv} \rangle&=\langle T_{uu} \rangle+2 \langle T_{uv }\rangle +\langle T_{vv} \rangle.
\end{split}
\ee
We should emphasize that the stress tensor for the $|in \rangle$ state describing the evaporation of the black hole is valid for $v>v_0$. 

However, it is difficult to work with a time-dependent stress tensor. Therefore, we will consider the problem at late times and near-horizon regime, corresponding to the Unruh state $|U \rangle$. This is achieved as we have noted around \er{latetimeu}, $u_{in}$ is related to the Kruskal $U$. Additionally, we expect the stress tensor to be regular near the horizon, and for simplicity, we take the near-horizon expansion $r=r_0+x$ by expanding in $x$. We have the following back-reaction equations 
\be
\epsilon  \pa_r\varphi(r)=\frac{3 \epsilon}{r_0^3} \bigg[3-e^{2-\frac{v}{r_0}} -4e^{1-\frac{v}{2r_0}}+ \frac{v}{r_0} \bigg]+O(\epsilon x),
\ee
\be
\epsilon \pa_r m(r,v)=-\frac{4 \epsilon}{r^2_0}+O(\epsilon x),
\ee
\be
\epsilon \pa_v m(r,v)=\frac{\epsilon}{4 r^2_0}+O(\epsilon x).
\ee
By omitting higher-order terms, we have the following solutions 
\be
\varphi(r,v )=\frac{9r}{r^3_0 } -\frac{3 r}{r^3_0 } e^{2-\frac{v}{r_0}}-\frac{12 r}{r^3_0} e^{1-\frac{v}{2 r_0}}+\frac{3 r v}{r^4_0 }+C_1,
\ee
\be
m(r,v)=-\frac{4r}{r^2_0 }+\frac{v}{4 r^2_0 } +C_2 .
\ee
Now we determine $C_1$ and $C_2$ by requiring the following relations to hold at some cut-off surface at $v=L$ while $L \to \infty$
\be
\lim_{L \to \infty} \bigg( \lim_{v \to L} m(r,v) \bigg) =0, \quad \lim_{L \to \infty} (\lim_{v \to L} \varphi(r,v))=0
\ee
then 
\be
C_1=-\frac{3 r L}{ r^4_0}, \quad C_2=-\frac{L}{4 r^2_0 }.
\ee
We present the island computation given the back-reacted geometry at hand. We find the no-island case is similar to the Hartle-Hawking case, while there is a slight difference in the island location, which is given by 
\begin{equation} \label{result-island}
    r_a=r_{H a}-\frac{\epsilon c\left(2+\kappa_a\left(v_a-v_b\right)\right)}{\kappa_a r_{H a}\left(v_a-v_b\right)}+O\left(\epsilon^2\right).
\end{equation}
We refer to Appendix.~\ref{sD} for a self-contained treatment and the notations used here. Now we comment on some interesting features of the result. Unlike the Hartle-Hawking state, the endpoint of the island can either locate inside or outside of the horizon, depending on the absolute value of $\ka_{a}(v_a - v_b) $. Since $\ka^{-1}_a$ can be effectively viewed as the horizon scale, this quantity is essentially the ratio of the (casual) length of the extremal surface to the horizon scale.     With $v_a - v_b <0$, we find that if $\ka_a\left|v_a-v_b\right|>2 $  the island is located inside the horizon, whereas if $\ka_a\left|v_a-v_b\right|<2 $, it extends beyond the horizon.  Note that when the ratio is significantly large, the location of the island is similar to that in the Hartle Hawking state. This can be understood as the difference in back-reaction between the two states being less important when the extremal surface is large. 

One may notice that there is no explicit dependence in time derivatives of either $r_H$ or $\ka$. This indicates that the effect of back-reaction can all be absorbed to quantum corrections of $\ka$ and $r_H$. Hence, the formula \er{result-island} is applicable whenever we are given the back-reacted surface gravity and horizon radius. 

On the other hand, when the length scale of the surface is comparable to the Planck scale $|v_a -v_b| \simeq \epsilon^{1/2}$, the island extends significantly outside the horizon. However, it is a regime where the formula \er{result-island} is no longer reliable as it is indicating a breakdown of perturbation in $\epsilon$.

\section{Discussion} \la{s5}

This paper begins with a simple physical model of $(3+1)$-dimensional Einstein gravity plus minimally coupled massless scalar matter. Through spherical dimensional reduction to $(1+1)$ dimensions, the new ingredient is that the dilaton field is non-minimally coupled with the scalar field. The model captures the s-wave sector of its higher-dimensional cousin. Despite its simplicity, a regular and consistent stress tensor was previously inaccessible, partially due to the issue of Weyl-invariant ambiguity in the effective action.  In response to the issue, we motivate the study with several reasonable assumptions on the one-loop action, including the dilaton-deformed conformal anomaly, conservation law, and boundary conditions associated with the state. From a universal splitting property we discussed in Sec.~\ref{s2.3}, the Weyl-invariant ambiguity in the one-loop action corresponds to the state-dependent part of the stress tensor; hence we can introduce on-shell equivalent auxiliary fields to the model and require that the resulting theory reproduces the same conformal anomaly. By constructing
minimal candidates of Weyl-invariant terms with the auxiliary fields, we derived a one-parameter family of self-consistent one-loop actions with unique and well-behaved stress tensors corresponding to the Boulware, Hartle-Hawking, and Unruh states. Their near-horizon and asymptotic behaviors are in accordance with the s-wave approximation from four dimensions.

As an application, we study the back-reaction problem under the semi-classical Einstein equations for the three quantum states describing different physical scenarios. Given the unique and regular stress tensors, we are able to determine the one-loop geometry without suffering any issues encountered in the literature summarized in Sec.~\ref{s2.2}. A straightforward application of the island formula indicates a unitary Page curve in each case, as expected from a consistent study of the one-loop theory.

We comment on the implications and relevant subtitles from our construction of the non-minimal dilaton gravity model: 
\begin{itemize}
    \item \textbf{Implications of one-loop effective theories:} As we noted in Sec.~\ref{s3}, we have found that different states enforce different conditions on the potential Weyl-invariant terms. For instance, Boulware state requires the absence of the Weyl-invariant term
    \be
    \int d^2 x \sqrt{-g} (\nabla \phi)^2 \Box^{-1} (\nabla \phi)^2,
    \ee
     while the Hartle-Hawking state cannot be captured without this term. In the case of Unruh state, we need that the normal-ordered part to be zero.
     
     As we have shown in Sec.~\ref{s2.3}, the Weyl-invariant terms can be identified with the normal-ordered stress tensor that is state-dependent. It is crucial that we do not assume the normal-ordered part has to be zero, instead it is determined by physical conditions. We believe that local excitations that do not change the boundary conditions should also be captured by the same one-loop theory.

     Our formalism does not rule out the possibility of describing these physical states by including more Weyl-invariant terms. However, our construction only assumes minimal candidate terms, which is consistent with Occam's razor. We will also comment on Wald's axioms later that were used to invalidate the inclusion of additional Weyl-invariant terms \cite{Balbinot_1999}.

     For other types of dilaton gravity theories with different anomaly equations \er{confana}, we expect that the Weyl-invariant ambiguity should generically appear and can be fixed with similar construction established in this work.

     From a Wilsonian renormalization group perspective, our model serves as a low energy effective theory that captures the conformal anomaly \er{confana}. However, the exact UV theory can be quite different from the low energy theory and one should not take the extrapolation too seriously, especially in view of the dimensional reduction anomaly \cite{Frolov_1999, Cognola:2000wd, Cognola:2000xp}.\footnote{We have assumed that the non-minimal dilaton gravity model should reproduce the s-wave contribution of the four-dimensional model. As we mentioned in Sec.~\ref{s2.2}, this is not guaranteed due to the dimensional reduction anomaly.}

    \item \textbf{Bridging the gaps between our results and the existing approaches:} We have presented a brief overview of approaches in tackling this model in Sec.~\ref{s2.2}. A few things are noteworthy. First of all, these approaches lead to unphysical stress tensors such as anti-evaporation or logarithmic divergence at the horizon. Second, the stress tensors obtained from these methods are in conflict with one another. Lastly, our construction in this paper differs from these approaches in terms of the regular and physical stress tensors that are predicted.
    
    Now we understand why the first approach in Sec.~\ref{s2.2} with local auxiliary fields fails. Since by adopting only $\Gamma_{\text{anom}}$, one missed other possible terms arising from the Weyl-invariant ambiguity \er{weyl}. However, our approach has no implications for the other two approaches, especially given the dimensional reduction anomaly \cite{Frolov_1999, Cognola:2000wd, Cognola:2000xp}. On the other hand, the other two approaches have inherent issues with their formulations. It is possible to obtain the same regular and physical stress tensors by modifying these approaches accordingly\footnote{For example, one could consider non-perturbative improvement for calculating the heat kernel beyond the covariant perturbation method, such as the formalism developed in \cite{Barvinsky:2002uf, Barvinsky:2003rx, Barvinsky:2004he}. It would be interesting to see whether one could obtain the one-loop effective action compatible with ours based on the new formalism.}, but how to bridge the gaps seems to be highly non-trivial.

    \item \textbf{A generalized Virasoro anomaly:} An approach that we did not discuss in Sec.~\ref{s2.2} to address this model was based on an anomalous transformation law for the normal-ordered quantum stress tensor derived in \cite{Fabbri:2003vy}. We have independently demonstrated in Appendix.~\ref{sA} that covariance is maintained by adding to the normal-ordered stress tensor a geometrical part $\bra T^{{\rm geo}}_{ab} \ket $.
    
    Virasoro anomaly in two dimensions is referring to the fact that by performing the conformal transformation $x^\pm \to y^{\pm}(x^\pm)$, the normal-ordered stress tensor $\langle :T_{\pm \pm}: \rangle$ would break general covariance and pick up a Schwarzian derivative
    \be \la{Trans1}
    \langle : T_{y^\pm y^\pm}: \rangle=\bigg(\frac{dx^\pm}{dy^\pm} \bigg)^2 \langle : T_{x^\pm x^\pm}: \rangle-\frac{1}{24 \pi} \{x^\pm(y^\pm),y^\pm \}.
    \ee
    The result holds for a free massless scalar field. For a more general theory with two-dimensional conformal invariance, we can multiply the Schwarzian term by a central charge $c$ corresponding to the particular theory.

    However, things have changed for the non-minimal dilaton gravity model \er{nonmin}. We do not expect the theory to follow the transformation law dictated in \er{Trans1}. Following standard OPE analysis with point-splitting regularization, it turns out that we have additional terms depending on the derivatives of the dilaton $\phi$ \cite{Fabbri:2003vy} (see also a self-contained derivation in Appendix.~\ref{sA})
    \bea \la{GVirasoro}
    \langle : T_{y^\pm y^\pm}: \rangle&=&\bigg(\frac{dx^\pm}{dy^\pm} \bigg)^2 \langle : T_{x^\pm x^\pm}: \rangle-\frac{1}{24 \pi} \{x^\pm(y^\pm),y^\pm \}
    \no \\
    &\quad&-\frac{1}{4\pi}\bigg[\frac{d^2 x^\pm}{dy^{\pm 2}}\bigg(\frac{dx^\pm}{dy^\pm} \bigg)^{-1}\frac{\partial \phi}{\partial y^\pm} +\ln \bigg(\frac{dx^+}{dy^+}\frac{dx^-}{dy^-} \bigg)\bigg(\frac{\partial \phi}{\partial y^\pm} \bigg)^2  \bigg],
    \eea
    which can be viewed as a generalization of the Virasoro-type anomaly. Note that the conformal symmetry can be recovered whenever $\pa_\pm \phi \to 0$.

    We can see clearly that \er{GVirasoro} breaks general covariance, and the motivation in \cite{Fabbri:2003vy} is to further impose the conservation law. The authors in \cite{Fabbri:2003vy} started by assuming the following conservation law
    \be \la{cons1}
    \pa_{\mp} \langle :T_{\pm \pm}: \rangle+ \pa_\pm \phi \langle \frac{\delta \Gamma}{\delta \phi} \rangle=0,
    \ee
    for the normal-ordered stress tensor. If $\langle :T_{\pm \pm}: \rangle$ transforms according to \er{GVirasoro} and we assume there is an associated transformation for $ \langle \frac{\delta \Gamma}{\delta \phi} \rangle$, then \er{cons1} is compatible with \er{GVirasoro} only if
    \be
    \Box \phi = (\nabla \phi)^2,
    \ee
    which may not hold in general. If the above relation is not true, then \er{cons1} must be modified to be
    \be \la{cons2}
    \pa_{\mp} \langle :T_{\pm \pm}: \rangle+\pa_\pm \langle T_{+-} \rangle+ \pa_\pm \phi \langle \frac{\delta \Gamma}{\delta \phi} \rangle=0,
    \ee
    where there is an extra trace term to be
    \be
    \langle T_{+-} \rangle=-\frac{1}{4 \pi} (\pa_+ \phi \pa_- \phi-\pa_+ \pa_- \phi),
    \ee
    which is consistent with the anomaly equation \er{confana} in flat spacetime. For general curved background, we may enforce general covariance with the quantum conservation law \er{conser}. It turns out that one could obtain the covariant stress tensor in agreement with \er{confana} as well as the one obtained from $\Gamma_{\text{loc}}$, which is apparent from our discussion in Appendix.~\ref{sA}. 

    We shall compare the difference of stress tensor following for Boulware state from \cite{Fabbri:2003vy} and our results in Sec.~\ref{s3.1}. According to \cite{Fabbri:2003vy}, the $\pm \pm$-components of the covariant stress tensor is given by
    \be \la{CovVirasoro}
    \langle \Psi | T_{\pm \pm} | \Psi \rangle= \langle \Psi |:T_{\pm \pm}: | \Psi \rangle-\frac{ \hbar}{12 \pi} [(\pa_\pm \rho)^2-\pa_\pm^2 \rho ]+\frac{\hbar}{ 2 \pi} [\pa_\pm \rho \pa_\pm \phi + \rho (\pa_\pm \phi)^2].
    \ee
    We notice that the last two pieces are in agreement with the variation of $\Gamma_{\text{loc}}$ in Sec.~\ref{s2.3}. The authors considered the case with $\langle B |:T_{\pm \pm}: | B \rangle=0$, which yields 
    \be \la{BoulwareT2}
    \langle B| T_{uu} | B \rangle= \langle B| T_{vv}| B \rangle= \frac{\hbar}{24 \pi } \bigg(\frac{15 M^2}{2 r^4}-\frac{4M}{r^3} \bigg) +\frac{\hbar}{16 \pi} \frac{(r-2M)^2 \ln{(1-\frac{2M}{r})}}{r^4}.
    \ee
    It is clearly different from our \er{BoulwareT}, but does agree with our \er{EarlyT}. We have checked that one cannot use an effective action by fine-tuning the parameters we had in Sec.~\ref{s3.1} to generate the same stress tensor as in \er{BoulwareT2}.

    The difference lies in the assumption of $\langle B |:T_{\pm \pm}: | B \rangle$. In our approach, we do not assume \textit{a priori} that $\langle B |:T_{\pm \pm}: | B \rangle=0$. Instead, we determine it by the physical requirements of the Boulware state. However, it is evident that \er{BoulwareT2} is also consistent with the boundary conditions we imposed in Sec.~\ref{s3.1}. Hence \er{BoulwareT2} is physical, at least under the criteria we discussed in Sec.~\ref{s3.1}. 

    The difference between \er{BoulwareT} and \er{BoulwareT2} is given by the following on-shell value in the Schwarzshcild background
    \be \la{addpiece}
    \frac{\hbar}{24 \pi} \bigg(\frac{6 M^2}{r^4}-\frac{3M}{r^3} \bigg).
    \ee
    As we have noted in Sec.~\ref{s2.3}, the normal-ordered part is given by $\Gamma_W$ where it involves the Weyl-invariant ambiguity. This means that when evaluated on-shell in the Boulware state under our construction in Sec.~\ref{s3.1}, we have implicitly imposed
    \be \la{Bnormal}
    \langle B |:T_{\pm \pm}: | B \rangle =-\frac{\hbar}{24 \pi} \bigg(\frac{6 M^2}{r^4}-\frac{3M}{r^3} \bigg),
    \ee
    to cancel the piece \er{addpiece}. Note that a similar difference should also appear in $\frac{\delta \Gamma_W}{\delta \phi}$.

     Presumably, there is no way to judge which approach for the Boulware state is more reasonable according to the criteria in Sec.~\ref{s3.1}. Our findings, however, can be understood as an alternative state that satisfies all boundary conditions for the Boulware state. In addition, as we have demonstrated in Sec.~\ref{s3.1}, our results are the direct consequence of $\Gamma_{\text{anom}}$ such that we do not require any additional Weyl-invariant terms. As a result, general covariance is automatically encoded in our approach. Further supports come from the fact that our results are in agreement with \cite{Balbinot_1999, Balbinot_1999_2}.

Despite the approach used in \cite{Fabbri:2003vy} being successful in describing the Boulware state with the correct boundary values\footnote{Note that a non-perturbative back-reaction analysis similar to our Appendix.~\ref{sB} based on the stress tensor for the Boulware state derived in \cite{Fabbri:2003vy} was already carried out in \cite{Ho:2017joh}. Again, it results in the absence of horizon structure.}, the difficulty lies in finding workable forms of stress tensors for the Hartle-Hawking and Unruh states beyond only the asymptotic and near-horizon values \cite{Fabbri:2003vy}. 

Similarly, we could work out the implicit assumptions made in $\langle H| : T_{ab}: | H \rangle$ for the Hartle-Hawking state by subtracting the state-independent geometrical contributions from \er{HHTensor1}, according to \er{CovVirasoro}
\bea
\langle H| : T_{\pm \pm}: | H \rangle&=&\langle H|  T_{\pm \pm}| H \rangle- \langle T^{\text{geo}}_{\pm \pm} \rangle
\no\\
&=&\frac{\hbar}{192 \pi r^4 r^2_0} \bigg\{ r^4+6 r^2 r^2_0 -6r^4_0
\no\\
&\quad&-12 (r-r_0)^2 r^2_0 \bigg[3 \ln{\frac{r}{r_0}} +\ln{ \bigg(1-\frac{r_0}{r} \bigg)} \bigg] \bigg\}.
\eea

    From the discussion so far, we also want to comment on Wald's axioms that were used to argue against including Weyl-invariant terms \cite{Balbinot_1999}. Wald's axioms \cite{Wald1978} are conditions that a reasonable four-dimensional quantum stress tensor should satisfy. These conditions include the conformal anomaly, conservation law, and the fact that semi-classical Einstein equations vanish for the Minkowski vacuum where $\langle T^{(4)}_{\mu \nu} \rangle= \langle : T^{(4)}_{\mu \nu} : \rangle$=0. Applying to our two-dimensional model, the first two conditions are clearly satisfied. Whether our model satisfies the third condition is slightly more tricky, where we have \er{Bnormal}, under the Minkowski limit, \er{Bnormal} vanishes and we do get back to the Minkowski vacuum. 
    
    \item \textbf{On the applicability of the island formula:} We need to discuss the applicability of $S_{\text{matter}}$ that we used in \er{matterentropy} for the island formula. The origin of the formula is that for general free field in four-dimensional flat spacetime, we have \cite{Casini:2005zv, Casini:2009sr}
    \be \la{areaformula}
    S_{\text{matter}}=-\kappa c \frac{\text{Area}}{L^{2}},
    \ee
    where for massless field $\kappa$ becomes a constant.\footnote{We can also take this formula to be approximately true in curved space when the distance between $\pa I$ and $A$ is small compared to the length scale of the curvature. For applications of this in the island computation, see \cite{Hashimoto:2020cas, Matsuo:2020ypv, Gan:2022jay}.}
    It reduces to the entropy formula for two-dimensional free massless fields in flat spacetime \cite{Holzhey:1994we, Calabrese:2009qy}
    \be \la{flatent}
    S_{\text{matter}}=\frac{c}{3} \ln {(d(\pa I,A))},
    \ee
    under dimensional reduction where we can use the s-wave approximation for a distance much larger than the correlation length of the massive modes. Note that $d(\pa I,A)$ is the distance between $\pa I$ and $A$. Under Weyl transformation, it gives the formula in general curved space as in \er{matterentropy}.

    Therefore,  $S_{\text{matter}}$ is in general state-dependent, and \er{matterentropy} is applicable in curved spacetime under the choice where we set the state-dependent normal-ordered part of the stress tensor $\langle :T_{ab}: \rangle=0$. If it is not zero, there shall be an additional state-dependent piece appearing in \er{matterentropy}. This additional piece may follow what is dictated in the new Virasoro anomaly that we commented.

    Our matter sector comes from the dimensional reduction of a four-dimensional free massless scalar field, where in flat spacetime we do have $\langle : T_{ab} : \rangle=0$ such that \er{flatent} holds. However, if we do not require that the normal-ordered stress tensor for quantum states in a curved background is zero, the matter entropy formula may acquire an additional piece that remains to be determined under our construction. We have assumed that \er{matterentropy} holds approximately true in our model as we believe the additional piece should not alter our conclusion for the unitarity of the Page curve.

    Even if we can bypass the above issue, there are still other open questions. A more fundamental debate was raised in \cite{Geng:2020fxl, Geng:2021hlu}, where it is argued that the fine-grained entropy should be a constant since there is no diffeomorphism-invariant way to split a radiation subregion. The entropy calculation in the main text can then be understood as certain coarse-grained entropy from that point of view. The argument applies to our case but is not crucial since our central point is to show that with consistent one-loop actions for physical states, we can always apply the island formula to the corresponding back-reacted geometry. Also, we can explore the extremal black hole solutions in the non-minimal dilaton gravity and how the island formula is applied \cite{Karananas:2020fwx, Ahn:2021chg}. Furthermore, it is interesting to see whether we can derive the island formula \er{islandfor} similar to the replica wormhole calculation in JT gravity \cite{Penington:2019kki, Almheiri:2019qdq} (see also \cite{Hartman:2020swn} for a case in CGHS model) for the non-minimal dilaton gravity model.

    \end{itemize}

    There are also a few fascinating future directions that are worth pointing out regarding the non-minimal dilaton gravity model. To name a few:
    \begin{itemize}
    \item \textbf{Generalizations of dilaton coupled theory:} The non-trivial ingredient in the model is the dilaton coupling in the matter sector. This can be generalized to more general gravity and matter sectors coupled with dilaton that may have physical origins from higher dimensions. It would be interesting to see whether exactly solvable models could arise in these scenarios, and whether similar Weyl-invariant ambiguities can be resolved in a consistent way.

    \item \textbf{Connections/Implications for holography:} We were working in asymptotically flat spacetime. However, it would be interesting to solve this type of dilaton coupling in the context of AdS/CFT. A particularly inspiring scenario is JT gravity in NAdS$_2$/NCFT$_1$ \cite{Almheiri:2014cka, Maldacena:2016upp, Engelsoy:2016xyb} (See also the review \cite{Mertens:2022irh}), where N means "nearly" as the boundary conformal symmetry is broken by the dilaton. By coupling our matter sector with JT gravity, an immediate consequence is that the variation of the dilaton does not necessarily enforce a local AdS$_2$ geometry. Instead, it  depends on the matter content. 
    
    Therefore, a more careful analysis of the configuration space of classical solutions, as well as the possibility of obtaining a solution with an asymptotic AdS boundary, is required. Intuitively, we need to impose suitable boundary conditions for the dilaton coupled matter theory consistent with asymptotic AdS boundary conditions. It is also interesting to explore such non-minimal matter coupling along the line of braneworld construction \cite{Geng:2022slq, Geng:2022tfc}.

    Furthermore, in view of \er{GVirasoro} regarding a new Virasoro anomaly, it would be interesting to see how it enters into the analysis of the boundary theory of JT gravity coupled with the non-minimal matter.

    \item \textbf{Implications for non-perturbative effects:} The analysis with JT gravity has recently uncovered new structures involving spacetime wormholes, where JT gravity under non-perturbative genus expansion is shown to be dual to a double-scaled matrix integral \cite{Saad:2019lba}. The analysis was mainly carried out in pure JT gravity or JT gravity with a minimally-coupled massive scalar field \cite{Jafferis:2022wez, Jafferis:2022uhu}. It would be interesting to see whether a similar study can be carried over to the case of dilaton coupling in the matter sector.

\end{itemize}

\acknowledgments

We thank Xi Dong, Steven Giddings, and Donald Marolf for their useful comments and encouragement of this work, and special thanks to Gary Horowitz for valuable feedback on the draft. We also thank Amirhossein Tajdini, Sean McBride, Diandian Wang, Ziyue Wang, and Wayne Weng for helpful discussions. C-H.W. was supported in part by the U.S. Department of Energy under Grant No. DE-SC0023275, and the Ministry of Education, Taiwan. J.X. was on the MURI grant and was supported in part by the U.S. Department of Energy under Grant No. DE-SC0023275. This material is based upon work supported by the Air Force Office of Scientific Research under award number FA9550-19-1-0360.

\begin{appendix}

\section{Implication of General Covariance in Quantum Stress Tensor} \la{sA}
We have argued that for a well-defined back-reaction problem, we need
the expectation value of stress tensor $\bra T_{ab}\ket$ to transform
covariantly under coordinate transformations. While it is clear that
the covariance is manifest once $\Gamma_{{\rm eff}}$ is covariant
and $\bra T_{ab}\ket$ is defined as a functional variation with respect to
the metric, the point here is that in general one does not know the concrete form of effective action. In addition, the non-local
nature of the effective action for gravity theory (as shown in \er{anomm})
makes it hard to verify that the general covariance still holds true
at the quantum level. 

In addition, it is shown in Section.~\ref{s2.3} that there is a universal
splitting of the effective action such that a local but non-covariant
part $\Gamma_{{\rm loc}}$ captures the geometrical contribution to $\langle T^{\text{geo}}_{ab} \rangle$.
As we shall see in the following, such a stress tensor is not covariant
as it corresponds to a specific gauge that fixes part of the diffeomorphism. Since covariance is a desirable property, the general expectation is that we can establish covariance by adding the normal-ordered part. We shall elaborate on this point
here by working out the detailed transformation law for both parts of
the stress tensor, and then show that the combination is indeed covariant.

Let us look at $\Gamma_{{\rm loc}}$ \er{def-split}, which is given by 
\begin{equation}
\Gamma_{{\rm loc}}=\frac{\hbar}{96\pi}\int\sqrt{-g}\left(\log\sqrt{-g}\sq\log\sqrt{-g}+\log\sqrt{-g}\left(2R-12\left(\nn\phi\right)^{2}+12\sq\phi\right)\right),
\end{equation}
we define the geometrical part of the stress tensor $\bra T_{ab}^{\text{geo}} \ket$ by funcitonal derivative of $\Gamma_{\rm{loc}}$ with respect to the metric
\begin{equation}
\bra T_{ab}^{\text{geo}}\ket =\frac{-2}{\sqrt{-g}}\frac{\del \Gamma_{{\rm loc}}}{\del g^{ab}},
\end{equation}
with some algebra, one can decompose the stress tensor into three parts
\begin{equation}
\bra T_{ab}^{\text{geo}}\ket =\bra T_{ab}^{\left(g\right)} \ket+\bra T_{ab}^{\left(\phi\right)} \ket+\bra T^{\text{tr}}\ket g_{ab} ,
\end{equation}
where 
\begin{equation}
\begin{split} \la{component}
\bra T_{ab}^{\left(g\right)} \ket & =\frac{\hbar}{48\pi}\left(\nn_{a}\log\sqrt{-g}\nn_{b}\log\sqrt{-g}-\frac{1}{2}g_{ab}\left(\nn\log\sqrt{-g}\right)^{2}\right)\\
 & +\frac{\hbar}{24\pi}\left(\nn_{a}\nn_{b}\log\sqrt{-g}-\frac{1}{2}g_{ab}\sq\log\sqrt{-g}\right),\\
\bra T_{ab}^{\left(\phi\right)}  \ket& =\frac{\hbar}{4\pi}\left(\nn_{(a}\log\sqrt{-g}\nn_{b)}\phi-\frac{1}{2}g_{ab}g^{cd}\nn_{c}\log\sqrt{-g}\nn_{d}\phi\right)\\
 & +\frac{\hbar}{4\pi}\log\sqrt{-g}\left(\nn_{a}\phi\nn_{b}\phi-\frac{1}{2}g_{ab}\left(\nn\phi\right)^{2}\right),\\
\bra T^{\text{tr}} \ket& =\frac{\hbar}{48\pi}\left(R-6\left(\nn\phi\right)^{2}+6\sq\phi\right),
\end{split}
\end{equation}
where we have split the stress tensor into pure gravity, dilaton, and trace parts. The above stress tensor is not covariant due to the manifest
dependence on $\sqrt{-g}$; however, it remains local, and every term in it follows certain transformation rules under diffeomorphism.
Therefore, one should expect that the stress tensor itself follows certain
transformation rule under diffeomorphism as well, and we are particularly interested in the deviation from the covariant transformation.

We will choose the conformal gauge $\dd s^{2}=-e^{2\rho}\dd u\dd v$
in the following discussion. For simplicity, let us look at the $uu$-component, which is evaluated as
\begin{equation} \label{tuu}
\bra T_{uu}^{{\rm geo}}\ket =\frac{\hbar}{12\pi}\left(\partial_{u}^{2}\rho-\left(\partial_{u}\rho\right)^{2}+6\partial_{u}\rho\partial_{u}\phi+6\rho\left(\partial_{u}\phi\right)^{2}\right).
\end{equation}
There is an issue in the value of $\log\sqrt{-g}$ we used in the evaluation
of the stress tensor component, resulting from the non-covariance
under diffeomorphism. We used $\log\sqrt{-g}=2\rho$ instead of
$2\rho+\log2$. We emphasize that this can be viewed as part of the definition of the state 
that one associates with the stress tensor components, in the same sense described in Sec.~\ref{s2.3}. 
Namely, the ambiguity of shifting $\rho$ by a constant can be viewed as a contribution to the stress tensor from the following Weyl-invariant term of the action
\begin{equation}
\Gamma_{\text{ct}}=\frac{\lambda\hbar}{8\pi}\int\sqrt{-g}\left(\nn\phi\right)^{2},
\end{equation}
where $\lambda$ is chosen to cancel the constant part in
$\rho$ to the stress tensor, and $\Gamma_{ \rm ct}$ is the counterterm that is added to cancel the ambiguity in $\rho$.  According to Sec.~\ref{s2.3}, we know $\Gamma_{{\rm ct}}$ belongs to $\Gamma_{{ W}}$ and therefore
the stress tensor given by $\Gamma_{{\rm ct}}$ actually contribute to
the normal-ordered part $\bra:T_{ab}:\ket$. By demanding $\log\sqrt{-g}=2\rho$ in the stress tensor, we are actually imposing the condition that $\bra  :T_{ab}: \ket$ vanishes for Minkowski vacuum, where $\rho$ is a constant. In conclusion,
adding $\Gamma_{{\rm ct}}$ to cancel the constant shift of $\rho$ in
the stress tensor amounts to setting the zero-point energy for the
theory. 

Now let us consider how $\bra T_{uu}^{\text{geo}} \ket$ transforms in the transformation $u\to U= U\left(u\right)$. This is a residual gauge symmetry to the conformal gauge that we chose, and under the transformation we find 
\begin{equation}
\dd s^{2}=-e^{2\rho}\dd u\dd v=-e^{2\widetilde{\rho}}\dd U\dd v ,
\end{equation}
where in the new coordinate, the conformal factor becomes
\begin{equation}
2\widetilde{\rho}=2\rho+\log u^{\pp}.
\end{equation}
By separating the covariant and non-covariant parts in $\bra T^{\text{geo}}_{uu} \ket$ under the transformation, it is easy to verify that 
\begin{equation}
-\left(\partial_{U}^{2}\widetilde{\rho}\right)+\left(\partial_{U}\widetilde{\rho}\right)^{2}=u^{\pp2}\left(-\partial_{u}^{2}\rho+\left(\partial_{u}\rho\right)^{2}\right)-\frac{1}{2}\{u,U\},
\end{equation}
\begin{equation}
\partial_{U}\widetilde{\rho}\partial_{U}\phi+\widetilde{\rho}\left(\partial_{U}\phi\right)^{2}=u^{\pp2}\left(\partial_{u}\rho\partial_{u}\phi+\rho\left(\partial_{u}\phi\right)^{2}\right)+\frac{1}{2}\left(\log u^{\pp}\left(\partial_{U}\phi\right)^{2}+\frac{u^{\pp\pp}}{u^{\pp}}\partial_{U}\phi\right),
\end{equation}
where we have defined $u^{\pp}=\dd u/\dd U$. Plugging back to the component value \er{tuu}, we get
\begin{equation}
\bra T_{UU}^{\rm geo} \ket=u^{\pp2} \bra T_{uu}^{\rm geo} \ket+\frac{\hbar}{24\pi}\{u,U\}+\frac{\hbar}{4\pi}\left(\log u^{\pp}\left(\partial_{U}\phi\right)^{2}+\frac{u^{\pp\pp}}{u^{\pp}}\partial_{U}\phi\right).
\end{equation}
Note that apart from the covariant term, the non-covariant part consists of a Schwarzian derivative resulting from the change of the conformal factor $\rho$ together with a dilatonic part resulting from the coupling between the dilaton field $\phi$ and $\rho$. The non-covariant part remains local. 

Now to see how covariance of $\bra T_{ab}\ket$
is maintained at the quantum level, we need to work out the transformation
law for $\langle :T_{ab}: \rangle$. In the following, we will give a brief quantum mechanical derivation in agreement with \cite{Fabbri:2003vy}
by assuming the correct OPE relation of matter fields.

The matter part of the classical action in our setup is
\begin{equation}
S_{m}=-\frac{1}{2}\int\sqrt{-g}e^{-2\phi}\left(\nn f\right)^{2},
\end{equation}
where the equation of motion in conformal gauge can be written as
\begin{equation}
\partial_{v}\left(e^{-2\phi}\partial_{u}f\right)+\partial_{u}\left(e^{-2\phi}\partial_{u}f\right)=0.
\end{equation}
This can be used to derive the following Ward identity
\begin{equation}
\begin{split}
0 & =\del_{\alpha}\bra f\left(x_{1}\right)\ket=\del_{\alpha}\left(\int Dff\left(x_{1}\right)e^{iS_{m}/\hbar+\dots}\right)\\
 & =\alpha\left(x_{1}\right)+\frac{-i}{\hbar}\int\dd^{2}x\partial_{u}\alpha\bra e^{2\phi}\partial_{v}f\left(x\right)f\left(x_{1}\right)\ket+\frac{-i}{\hbar}\int\dd^{2}\partial_{v}\alpha\bra e^{2\phi}\partial_{u}f\left(x\right)f\left(x_{1}\right)\ket\\
 & =\alpha\left(x_{1}\right)+\frac{-i}{\hbar}\int\dd^{2}x\left(\partial_{u}\left(\alpha\bra e^{2\phi}\partial_{v}f\left(x\right)f\left(x_{1}\right)\right)+\alpha\left(\bra e^{2\phi}\partial_{u}f\left(x\right)f\left(x_{1}\right)\ket\right)\right),
\end{split}
\end{equation}
from which we deduce the following OPE relation of the matter fields 
\begin{equation}
e^{-2\phi\left(x_{1}\right)}f\left(x\right)f\left(x_{1}\right)\simeq\frac{1}{4\pi}\log|x-x_{1}|^{2}+{\rm reg}.
\end{equation}
Up to regular part we can actually apply the following symmetric OPE
\begin{equation}
f\left(x\right)f\left(x_{1}\right)\simeq\frac{1}{4\pi}e^{\phi\left(x\right)+\phi\left(x_{1}\right)}\log|x-x_{1}|^{2}+{\rm reg},
\end{equation}
and define the normal-ordered part of the stress tensor as an operator following the point-splitting regularization, which subtracts the divergence in the OPE
\begin{equation}
\langle :T_{uu}\left(x_{1}\right): \rangle\equiv\lim_{x_{2}\to x_{1}}e^{-\phi\left(x_{1}\right)-\phi\left(x_{2}\right)}\left(\partial_{u}\partial_{u^{\pp}}\left(f\left(x_{1}\right)f\left(x_{2}\right)-\bra f\left(x_{1}\right)f\left(x_{2}\right)\ket\right)\right).
\end{equation}
Now we are ready to consider the corresponding transformation law. Under the transformation $x\to X(x)$, where $x=(u,v)$ and $X=(U,V)=(U(u),V(v))$, we find the normal-ordered part transforms as
\bea
\langle :T_{UU}\left(X_{1}\right): \rangle&= & u_{1}^{\pp2}\langle :T_{uu}\left(x_{1}\right): \rangle+\lim_{x_{2}\to x_{1}}e^{-\phi\left(x_{1}\right)-\phi\left(x_{2}\right)}u^{\pp}\left(x_{1}\right)u^{\pp}\left(x_{2}\right)\partial_{u_{1}}\partial_{u_{2}}\bra f\left(x_{1}\right)f\left(x_{2}\right)\ket 
\no\\
&\quad& -\lim_{x_{2}\to x_{1}}e^{-\phi\left(X_{1}\right)-\phi\left(X_{2}\right)}\partial_{U_{1}}\partial_{U_{2}}\bra f\left(U_{1}\right)f\left(U_{2}\right)\ket.
\eea
The non-covariant part results from the subtraction of the singular part in the OPE. Now by expanding $U_2 - U_1$ as power series in $u_2-u_1$, and then take the limit $u_2 \to u_1$, we find the non-covariant part in $\langle :T_{uu}: \rangle$ 
\begin{equation}
\langle :T_{UU}\left(X\right): \rangle=u^{\pp2}\langle :T_{uu}\left(x\right):\rangle-\frac{\hbar}{24\pi}\{u,U\}-\frac{\hbar}{4\pi}\left(\frac{u_{1}^{\pp\pp}}{u_{1}^{\pp}}\partial_{U_{1}}\phi+\log|u_{1}^{\pp}|^{2}\left(\partial_{U}\phi\right)^{2}\right),
\end{equation}
which precisely cancels the one in $\bra T^{\rm geo}_{UU} \ket$. A similar analysis holds exactly for the $VV$-component.

Therefore, if we define the total expectation value of the stress
tensor to be the sum of $\bra T^{\rm geo}_{ab} \ket$ and the normal-ordered part $\langle :T_{ab}: \rangle$,
then clearly general covariance is maintained.

\section{A Non-Perturbative Analysis of Back-Reaction From Vacuum Polarization} \la{sB}

In this appendix, we study the back-reaction problem from the stress tensor we derived for Boulware state in Sec.~\ref{s3.1}. Conventional understanding about the Boulware state is that it is unphysical due to the divergence of the stress tensor in the free-falling frame, which can be easily seen in the Kruskal coordinates due to the diverging blue-shift factor $\frac{du}{dU} \propto \frac{1}{r-r_0}$ at the horizon \cite{Christensen}. That is, regularity at both the future and past horizons imposes the following conditions
\be 
\lim_{r \to r_0} \frac{|\langle B | T_{uu} | B \rangle|}{(1-\frac{r_0}{r})^2} = \lim_{r \to r_0} \frac{|\langle B | T_{vv} | B \rangle|}{(1-\frac{r_0}{r})^2} <\infty , \quad \lim_{r \to r_0} \frac{|\langle B | T_{uv} | B \rangle|}{(1-\frac{r_0}{r})} <\infty.
\ee
We can see clearly that even by considering the non-minimal dilaton coupled matter, the stress tensor we derived in \er{BoulwareT} does not obey these conditions. Hence, the view that the Boulware state is describing the exterior spacetime with $r> r_0$ seems well justified. That is, the physical portion of the state should not contain the horizon.

However, recent studies \cite{Ho:2017joh, Ho:2018fwq, Ho:2019pjr} based on a non-perturbative analysis indicates that the Boulware state is no longer unphysical once we include back-reaction, and it should be the correct state if we are looking at black hole formed from gravitational collapse (the scenario was first considered in \cite{Fabbri:2005zn, Fabbri:2005nt}, see also \cite{Arrechea:2019jgx, Beltran-Palau:2022nec}). In fact, there are a number of compelling reasons to examine this question more closely by taking into account the back-reaction of these vacuum polarization modes: 
\begin{itemize}
\item Boulware state can be defined via a natural boundary condition, namely
the vanishing Hawking flux in the asymptotic boundary. In this context, the boundary fluctuations are small, and the gravitational
effect is localized in the bulk. The effective action is suitable in the scenario where the back-reaction problem and the definition of the state do not depend significantly on UV physics.
\item The main reason that the Boulware state is thought to be unphysical is the divergence of the stress tensor at the horizon in the classical black hole background. The divergence persists perturbatively in $G \hbar$ in the back-reaction sourced by the quantum stress tensor. This argument is used to exclude the horizon as a physical portion of the Boulware state.

However, the above claim is circular in the sense that it assumes the existence of a horizon that persists in the back-reacted geometry. In fact, as demonstrated in \cite{Ho:2017joh, Ho:2018fwq, Ho:2019pjr}, depending on the fine structure of the stress tensor, the back-reacted geometry might create structures without a horizon while still leading to a well-defined state.

\item In other words, the divergence of the stress tensor
at the horizon can be alternatively viewed as a breakdown of the
perturbative analysis in $G \hbar$. This implies that we need a more careful non-perturbative analysis at the horizon scale. We further demonstrate that as one goes deeper into the bulk, the back-reaction will become more important and the geometry deviates from a vacuum black hole to a static quantum star sourced by the stress tensor. The resulting geometry is well-defined with no singularity, and can be viewed as a quantum state of the theory defined by the corresponding (asymptotic) boundary conditions. 

\item In terms of the gravitational path integral, this means the
true saddle that we are considering with vanishing Hawking flux might
not be the classical saddle that we assumed to have a horizon. The saddle point is modified by the quantum effects since the back-reaction makes a significant difference in the geometry of the
saddle. We should move to the correct saddle by including the quantum corrections.
\end{itemize}

To begin with, let us assume the following ansatz for the back-reacted geometry
\be \la{ansatz}
\begin{split}
    ds^2&=-F(r)e^{2 \epsilon \varphi(r)}dt^2 +\frac{dr^2}{F(r)} \\
    &=F(r)e^{2 \epsilon \varphi(r)} \bigg( -dt^2 +\frac{dr^2}{F(r)e^{2 \epsilon \varphi(r)}} \bigg) \\
    &=F(r)e^{2 \epsilon \varphi(r)} (-dt^2 +dr^{\ast 2}).
\end{split}
\ee
where 
\be
F(r)=1-\frac{r_0}{r}+\frac{\epsilon m(r)}{r}  \equiv F_0(r)+\frac{\epsilon m(r)}{r},
\ee
with $\epsilon=\frac{G_N \hbar}{24 \pi}$. We have also defined the tortoise coordinate $r^\ast$
\be
r^\ast \equiv \int \frac{1}{F(r) e^{\epsilon \varphi}}dr.
\ee
Now we introduce the Eddington-Finkelstein coordinates 
\be
u=t-r^\ast,\quad v=t+r^\ast.
\ee
We have
\be
ds^2=-F(r)e^{2 \epsilon \varphi(r)} du dv.
\ee
While the ansatz for the metric already implies the effect of back-reaction sourced by $\epsilon$
is small, we shall see concretely in the following how contradiction from the back-reaction equations arises and the assumption about the existence of a horizon fails when we go nearer to the would-be horizon. 

With the metric ansatz, the semi-classical Einstein equations sourced by the quantum stress tensor read
\be \la{eom-B1}
- \epsilon F_0 (r) m'(r)= 2G_N \langle T_{tt} \rangle,
\ee
\be
2 \epsilon r F_0(r) \varphi'(r)= 2G_N \bigg( F_0(r) \langle T_{rr} \rangle +\frac{\langle T_{tt} \rangle}{F_0(r)} \bigg).
\ee
Starting from the stress tensors we derived in \er{BoulwareT} and \er{BoulwareTuv}, we work out
\begin{equation}
\begin{aligned}\bra T_{rr}\ket= & \frac{\hbar}{12\pi}\left(1-\frac{r_0}{r}\right)^{-2}\left[\frac{3r^2_0}{8r^{4}}-\frac{r_0}{2r^{3}}\right]+\frac{\hbar}{8\pi r^{2}}\ln\left(1-\frac{r_0}{r}\right)\\
 & -\frac{\hbar r_0}{12 \pi r^{3}}\left(1-\frac{r_0}{r}\right)^{-1},
\end{aligned}
\end{equation}

\begin{equation} 
\begin{aligned}\bra T_{tt}\ket= & \frac{\hbar}{12\pi}\left[\frac{3r^2_0}{8r^{4}}-\frac{r_0}{2r^{3}}\right]+\frac{\hbar}{8\pi}\left(1-\frac{r_0}{r}\right)^{2}\frac{1}{r^{2}}\ln\left(1-\frac{r_0}{r}\right)\\
 & +\frac{\hbar r_0}{12 \pi r^3}\left(1-\frac{r_0}{r}\right).
\end{aligned}
\end{equation}
We then find that there is already an inconsistency in the equation
of motion: when $r=r_0$, the LHS of the first equation in \er{eom-B1}
vanishes, while $\bra T_{tt}\ket$ does not, this means $m^{\pp}\left(r\right)$
has to be divergent in $r=r_0$. Indeed, by solving the solutions for the functions $m(r)$ and $\varphi(r)$, one can explicitly check the two functions blow up at the horizon, and we expect such behavior to persist beyond one-loop order. This is due to the fact that the
back-reaction is strong and the ansatz \er{ansatz} that is perturbative in $\epsilon$  is no
longer applicable in the near-horizon regime. 

In other words, \er{ansatz} assumes that we get a smooth geometry with a horizon when we 
turn off the back-reaction in the $\epsilon\to 0$ limit. However, since the stress tensor becomes divergent in the near-horizon regime, where the quantum fluctuation can no longer be viewed as a small perturbation, we expect a dramatic change in the near-horizon structure. Therefore, we need to analyze the near-horizon geometry starting from a generic ansatz that works for arbitrary two-dimensional geometry.  

In the following, we shall apply the ansatz
\begin{equation}
\begin{split}
ds^{2} & =-C(r)dt^{2}+\frac{C(r)}{H^{2}(r)}dr^{2}\\
 & =-e^{2\rho\left(r\right)}\dd u\dd v,
\end{split}
\end{equation}
where we will express everything in terms of a generic conformal factor $\rho(r)$. In the following, we discuss the conditions
for the existence of a horizon when back-reaction is included. We have $C\left(r\right)=e^{2\rho\left(r\right)}$,
and suppose the horizon exists and is specified by $r=r_{H}$,
which leads to the fact that 
\begin{equation}
C\left(r_{H}\right)=0, \quad \rho\left(r_{H}\right)=-\infty, \quad \rho^{\pp}\left(r_{H}\right)=\infty.
\end{equation}
This is essentially the reason why the perturbative analysis in $\epsilon$
breaks down: when we work in the regime where $\left(\epsilon\rho^{\pp2}\right)\simeq O\left(1\right)$,
it can no longer be viewed as perturbation around classical geometry,
instead, it changes the ``classical background'' significantly.
Therefore, we solve the semi-classical Einstein
equation sourced by $\bra T_{ab}\ket$ in the following 
\begin{equation}
e^{-2 \phi} \{2 \nabla_a \nabla_b \phi-2 \nabla_a \phi \nabla_b \phi+g_{ab}[3 (\nabla \phi)^2-2 \Box \phi] \} -g_{ab}=2G_N\left\langle T_{ab}\right\rangle.
\end{equation}
The $tt$-component gives 
\begin{equation}\la{ttt}
\begin{aligned}e^{2\rho}-2rH(r)H^{\prime}(r)+H^{2}(r)\left(2r\rho^{\prime}(r)-1\right)= & 2\epsilon\left[\frac{H(r)}{r}H^{\prime}(r)\left(r\rho^{\prime}(r)+6\right)\right.\\
 & +\frac{H^{2}(r)}{r^{2}}\left(r^{2}\rho^{\prime\prime}(r)-r^{2}\rho^{2}(r)\right.\\
 & \left.\left.-6r\rho^{\prime}(r)+6\rho(r)-6\right)\right],
\end{aligned}
\end{equation}
while the $rr$-component gives
\begin{equation} \la{trr}
1-\frac{e^{2\rho(r)}}{H^{2}(r)}+2r\rho^{\prime}(r)=2\epsilon\left[\frac{6\rho(r)}{r^{2}}-\frac{6\rho^{\prime}(r)}{r}-\rho^{\prime2}(r)\right].
\end{equation}
From \er{trr} we can solve $H\left(r\right)$
in terms of $\rho\left(r\right)$ as 
\begin{equation}
H(r)=\pm\frac{e^{\rho(r)}r}{\sqrt{D(r)}},
\end{equation}
where we have introduced 
\begin{equation}
D(r)=r^{2}+2r^{3}\rho^{\prime}(r)+2\epsilon\left(r^{2}\rho^{\prime2}(r)+6r\rho^{\prime}(r)-6\rho(r)\right).
\end{equation}
Substituting either the positive or negative root of $H\left(r\right)$ back into \er{ttt}, and eliminating some overall factors, we find the equation of motion in terms of $\rho(r)$
\begin{equation} \la{numer}
\begin{aligned}0= & 144\epsilon^{2}\rho^{2}(r)-12r\epsilon\rho(r)\left[\rho^{\prime}(r)\left(3r^{2}+17\epsilon+3r\epsilon\rho^{\prime}(r)\right)-r\epsilon\rho^{\prime\prime}(r)\right]\\
 & +r^{2}\left[6\epsilon+2\rho^{\prime}(r)\left(r^{2}+6\epsilon+r\epsilon\rho^{\prime}(r)\right)\left(r+\rho^{\prime}(r)\left(r^{2}+6\epsilon+r\epsilon\rho^{\prime}(r)\right)\right)\right.\\
 & \left.+\left(r^{4}+11r^{2}\epsilon+36\epsilon^{2}+r\epsilon\left(r^{2}+6\epsilon\right)\rho^{\prime}(r)\right)\rho^{\prime\prime}(r)\right].
\end{aligned}
\end{equation}
This differential equation is extremely complicated and cannot be solved exactly. However, the method of dominant balance suffices for us to analyze the dominant solutions in different regimes. 

As a consistency check, let us first assume the existence of a horizon at $r=r_{H}$, and work in
the regime where $\epsilon^{-1/2}\simeq\rho^{\pp2}\simeq\rho^{\pp\pp}$. We are still in the perturbative framework in terms of $\epsilon$ and we can omit terms that are higher order in $\epsilon$, that is
\be
2r^5 \rho'+2r^{6}\rho^{\pp2}+r^{6}\rho^{\pp\pp}+O(\epsilon)=0.
\ee
The solution can therefore be chosen as
\be \label{sol-B1}
2 \rho(r)= \ln{\bigg(1-\frac{r_H}{r} \bigg)} + {\rm const},
\ee
which reproduces the Schwarzschild form. The scale $r_H$ that appears in the above solution can be defined in the asymptotic boundary as the ADM mass of the whole spacetime. However, the na\"ive solution above breaks down as we get nearer to
$r_{H}$, such that $\rho^{\pp\pp},\rho^{\pp2}\gg\epsilon^{-1}$.
In this case, the following two terms dominate
\begin{equation}
2r^{4}\epsilon^{2}\rho^{\pp4}+r^{5}\epsilon\rho^{\pp}\rho^{\pp\pp}+\dots=0,
\end{equation}
where terms in the dots are with fewer derivatives in $\rho$. We find
then in this case $\rho$ caps off in a very sharp region, with 
\begin{equation}\label{sol-B2}
\rho\left(r\right)=\frac{r_H}{2} \sqrt{\frac{\pi}{\epsilon}}{\rm Erfi}\left(\sqrt{\ln\left(r/r_{H}\right)}\right),
\end{equation}
where $r_{H}$ is an integration constant where we have defined it to
be the position where $\rho\left(r_{H}\right)=0$. Namely, at $r=r_{H}$
the geometry actually caps off and forms a smooth cone. There is no
horizon at all, and the physical portion of spacetime only contains
the part $r\geq r_{H}$. 

We further comment that for the aforementioned analysis to work, we should
keep $r$ very close to $r_{H}$ such that 
\begin{equation}
\frac{r-r_{H}}{r_{H}}\ll\frac{\epsilon}{r_{H}^{2}}.
\end{equation}
Within this small region, the back-reaction of the Boulware modes is so strong that the geometry deviates significantly from the one with a horizon. Instead, it forms a static and spherically symmetric quantum star and ends at a definite value $r=r_H$ of the radius. We interpret the point with $r=r_H$ as the center of the star, which is similar to the origin of the polar coordinates. In this way, we do not need to impose extra boundary conditions there. We present a numerical verification of the claim in Figure~\ref{fig1-B} and Figure~\ref{fig2-B} below.\footnote{Earlier works such as \cite{Ho:2017joh} described the resulting geometry as a wormhole-like structure with $r_H$ being the effective radius of the throat. This scenario is related to our discussion by identifying the two sides of a wormhole together with the asymptotic boundary. }

\begin{figure} 
\centering 
    \includegraphics[width=300pt]{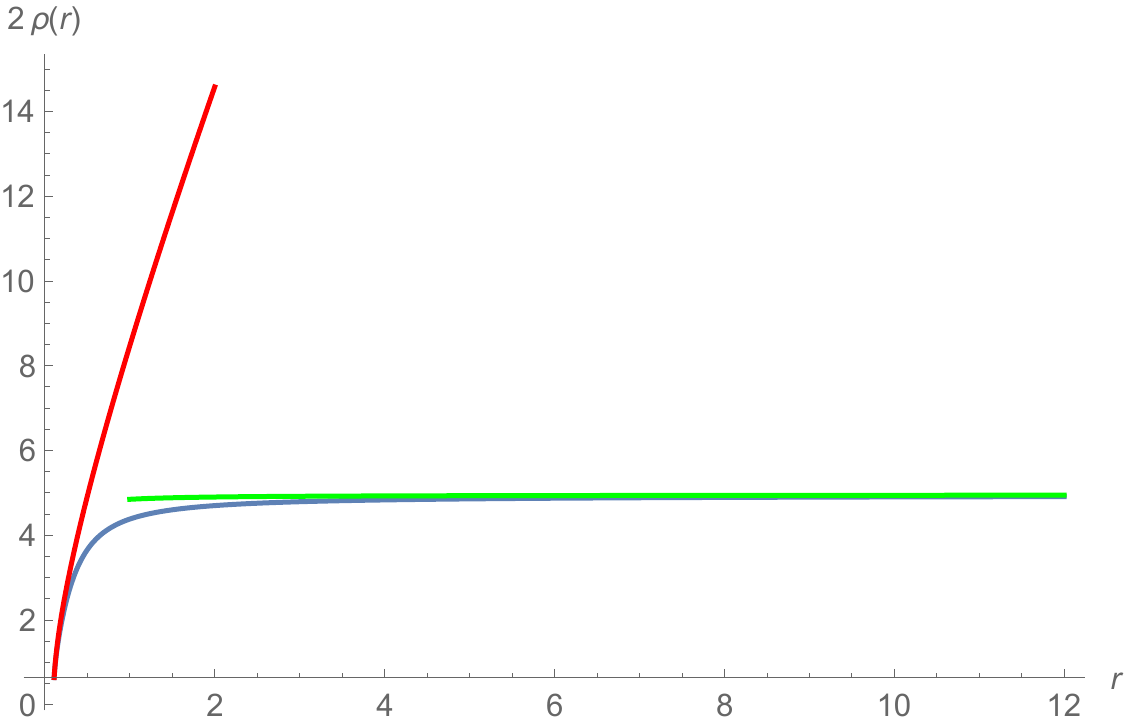}
    \caption{A plot of the numerical solution of \er{numer} (blue), Schwarzschild black hole solution \er{sol-B1} (green) and near $r_H$ solution \er{sol-B2} (red). We have set $\epsilon=0.01, r_H=0.1$ and the constant in \er{sol-B1} to be 4.95.  It is clear that the numerical solution matches \er{sol-B2} in a narrow near $r_H$ regime very well, which justifies our approximation. It asymptotes to a Schwarzschild black hole when the scale $r$ is significantly larger than $\epsilon^{(1/2)}$ .}
    \label{fig1-B}
\end{figure}

\begin{figure}
\centering
    \includegraphics[width=200pt]{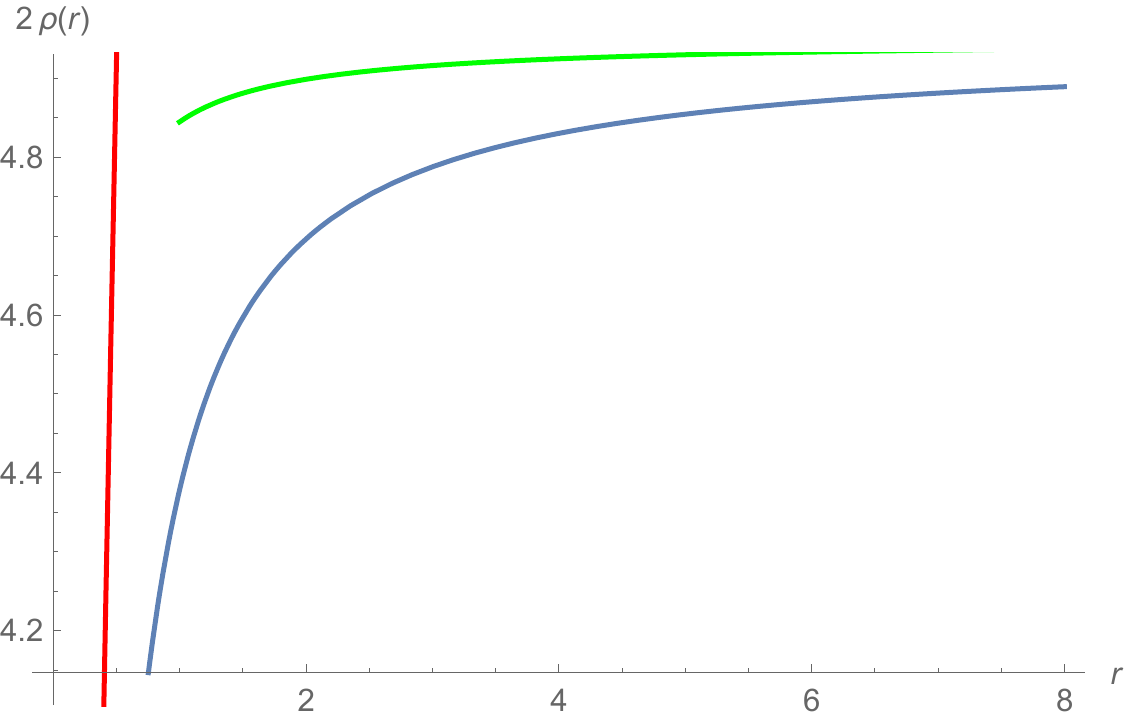}
    \includegraphics[width=200pt]{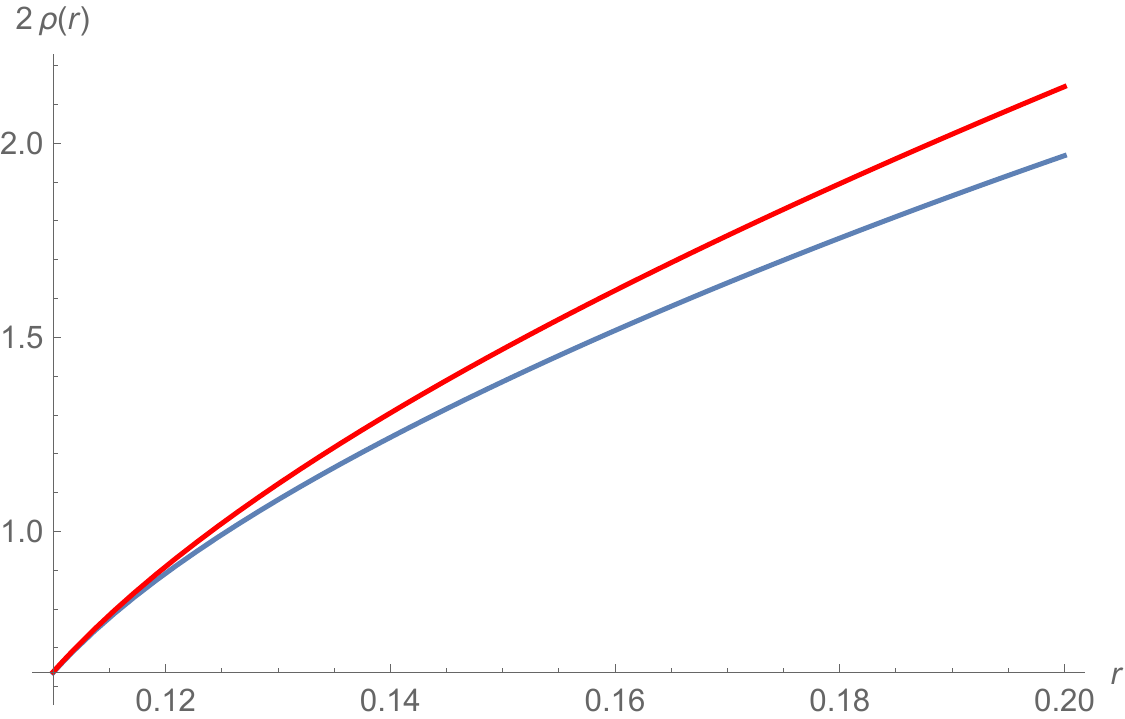}
    \caption{The left graph zooms into the regime where the numerical solution deviates from the two approximate solutions. It is a regime deep inside the bulk but still away from $r_H$. The geometry suffers from the back-reaction sourced by the stress tensor of quantum matter and fails to form a horizon. The right graph zooms into the regime near $r_H$, where the stress tensor component is significant compared with the curvature. We see in this case the numerical solution matches the near $r_H$ solution \er{sol-B2} well. }
    \label{fig2-B}
\end{figure}

Now let us imagine the journey of an infalling observer starting at spatial infinity. When $r\gg r_H$, she feels to be in the vacuum state exterior to a black hole. Then as she moves deeper and closer to $r_H$, but with $r-r_H \gg \epsilon^{1/2}$, the quantum fluctuation becomes significant, which locates at what she thinks to be the putative horizon. She would not encounter anything unusual and will reach the other side of the star if she manages to survive and pass through $r=r_H$.

We end this section by commenting on the fact that the Einstein tensor for the solution discussed here satisfies $G^t_t = G^r_r=O(1)+O(\sqrt{r-r_H})$ near $r\simeq r_H$. This means the prescribed solution and the definition of the Boulware state do not require Planck scale physics and is consistent with the low-energy effective dynamics of gravity specified by the semi-classical Einstein equations.

\section{Island Computation in the Hartle-Hawking State} \la{sC}

In this appendix, we detail the computation of the Page curve following the island prescription
for the fine-grained entropy, namely 
\be
S_\text{gen} (R)=\text{min}_{I} \bigg\{ \text{ext}_{I} \bigg[ \frac{\text{Area}(\partial I)}{4 G_N}+S_{\text{matter}}(I \cup R) \bigg] \bigg\}. 
\ee

We will first consider the no-island case where we shall reproduce
Hawking\textquoteright s prediction on a monotonically increasing
entropy. Note that in this case, the area term is absent, therefore we
identify the matter entropy part as our fine-grained entropy, which
is given by 
\begin{equation}
S_{{\rm matter}}=\frac{1}{12}\ln\frac{\left(V_{R}-V_{L}\right)^{2}\left(U_{R}-U_{L}\right)^{2}}{\delta^{4}e^{-2\rho_{R}}e^{-2\rho_{L}}}.
\end{equation}
Here we denote $L/R$ as the left/right asymptotically flat region
of spacetime. We define the coordinates on the cut-off surface
to be $\left(t,b^{\ast}\right)$. $\delta$ is the UV cut-off. Plugging
the definition of the coordinates
\begin{equation}
\begin{gathered}V_{R}=\frac{1}{\kappa}e^{\kappa\left(t+b^{\ast}\right)},\quad U_{R}=-\frac{1}{\kappa}e^{-\kappa\left(t-b^{\ast}\right)},\\
V_{L}=-\frac{1}{\kappa}e^{\kappa\left(-t+b^{\ast}\right)},\quad U_{L}=\frac{1}{\kappa}e^{-\kappa\left(-t-b^{\ast}\right)}.
\end{gathered}
\end{equation}
where the conformal factor is the same in both regions as $b$ is
the same, hence we know $\rho_{L}=\rho_{R}$. We find 
\begin{equation}
S_{\text{matter }}(t)=\frac{1}{6}\ln\frac{4F(b)\cosh^{2}(\kappa t)}{(\kappa\delta)^{2}e^{-2\epsilon\varphi(b)}}.
\end{equation}
We regularize the UV divergence by demanding that initially the entropy
is zero, namely $S_{{\rm mattter}}\left(0\right)=0$, and this fixes
\be
\delta^2 = \frac{4F(b) e^{2 \epsilon  \varphi(b)}}{\kappa^2}.
\ee
Putting the definition of $\delta$ back into the formula, we get
\begin{equation}
S_{{\rm matter}}=\frac{1}{3}\ln\left(\cosh\left(\ka t\right)\right).
\end{equation}
At early times of the evaporation, namely $\ka t\ll1$, the entropy
behaves as 
\begin{equation}
S_{{\rm matter}}\simeq\frac{1}{6}\left(\ka t\right)^{2},
\end{equation}
while the more interesting case is at late times when $\ka t\gg1$, we find
\begin{equation}
S_{{\rm matter}}\simeq\frac{1}{3}\ka t+{\rm const}.
\end{equation}
The fine-grained entropy of the radiation increases monotonically in time, which is in agreement with Hawking's original calculation corresponding to the no-island case. The above result is general for the two-dimensional dilaton gravity model, with the only difference being that now we have a quantum corrected $\kappa$. We shall see in the following how the island prescription
restores unitarity. 

\begin{figure} 
\centering
    \includegraphics[width=0.90\textwidth]{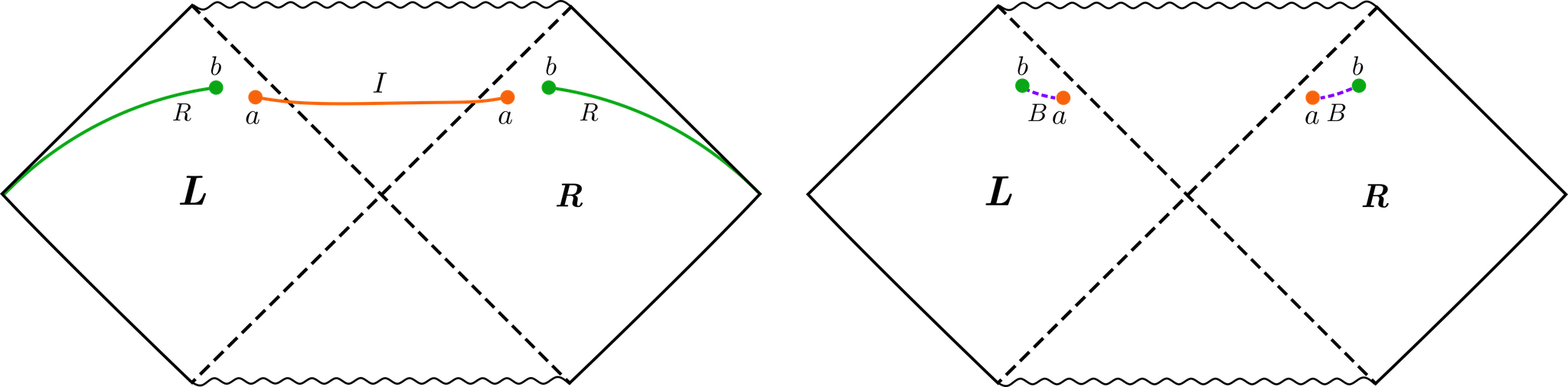}
    \caption{The Penrose diagrams of the two-sided black hole. The left and right asymptotically flat regions are denoted as $\textbf{\textit{L}}$ and $\textbf{\textit{R}}$. We take the states associated with the Hawking radiation to be represented by $R$ where the cut-off surfaces are written simply as $b$. Similarly, for the island region $I$ and the associated quantum extremal surfaces $\pa I=a$. By complementarity, the calculation of the matter entropy involves two disjoint intervals $B$ on the two sides.}
    \label{twosided}
\end{figure}

With island, we need to consider the entropy formula for two disjoint intervals with (See Figure~\ref{twosided})
\be \la{twoint}
S_{\text{matter}} = \frac{1}{6 } \ln{\frac{d^2_{12} d^2_{23} d^2_{14} d^2_{34}}{\delta^4 d^2_{24} d^2_{13} e^{-\rho_1}e^{-\rho_2} e^{- \rho_3} e^{- \rho_4}}},
\ee
where $d^2_{ij}=(V_i -V_j) (U_i-U_j)$. Suppose the position of the island is given by $(t',a^\ast)$ on the right patch (and similarly $(-t', a^\ast)$ on the left patch), we will be able to find the exact position by varying with respect to $a^\ast$. Now we have the following relations
\be
V_{Rb}=\frac{1}{\kappa} e^{\kappa (t+b^\ast)}, \quad U_{Rb}=- \frac{1}{\kappa}e^{-\kappa (t-b^\ast)},
\ee
\be
V_{Lb}=-\frac{1}{\kappa} e^{\kappa (-t+b^\ast)}, \quad U_{Lb}=\frac{1}{\kappa}e^{-\kappa (-t-b^\ast)},
\ee
\be
V_{Ra}=\frac{1}{\kappa} e^{\kappa (t+a^\ast)}, \quad U_{Ra}=-\frac{1}{\kappa}e^{-\kappa (t-a^\ast)},
\ee
\be
V_{La}=-\frac{1}{\kappa} e^{\kappa (-t'+a^\ast)}, \quad U_{La}=\frac{1}{\kappa}e^{-\kappa (-t'-a^\ast)},
\ee
By using the fact that at late times $t\simeq t^{\pp}$
\begin{equation}
\frac{d_{23}^{2}d_{14}^{2}}{d_{24}^{2}d_{13}^{2}}\rightarrow1,\quad d_{12}=d_{34},\quad\rho_{1}=\rho_{4},\quad\rho_{2}=\rho_{3},
\end{equation}
the matter entropy term becomes 
\begin{equation}
S_{\text{matter }}=\frac{1}{3}\left(\rho_{a}+\rho_{b}\right)+\frac{2}{3}\ln\left(e^{\kappa b^{\ast}}-e^{\kappa a^{\ast}}\right)-\frac{2}{3}\ln\kappa\delta.
\end{equation}
Adding to the area term (with a factor of $2$ because we have two asymptotically flat regions), we
write the fine-grained entropy as 
\begin{equation}
S_{\text{gen }}=\frac{2\pi a^{2}}{G_{N}\hbar}+\frac{1}{3}\left(\rho_{a}+\rho_{b}\right)+\frac{2}{3}\ln\left(e^{\kappa b^{\ast}}-e^{\kappa a^{\ast}}\right)-\frac{2}{3}\ln\kappa\delta.
\end{equation}
We extremize the entropy with respect to $a^{\ast}$
\begin{equation}
\partial_{a^{\ast}}S_{\text{gen }}=\frac{4\pi a}{G_{N}\hbar}\frac{da}{da^{\ast}}+\frac{1}{3}\frac{d\rho(a)}{da}\frac{da}{da^{\ast}}-\frac{2}{3}\frac{\kappa}{e^{\kappa\left(b^{\ast}-a^{\ast}\right)}-1}=0.
\end{equation}
By definition, we know $\dd a/\dd a^{\ast}=F\left(a\right)e^{\epsilon\vp\left(a\right)}$,
and by noting that $\epsilon = \frac{G_N \hbar}{24 \pi}$, we get 
\begin{equation} \la{islandHH}
\left[a+2\epsilon\rho^{\pp}\left(a\right)\right]F(a)e^{\epsilon\varphi(a)}=4\epsilon\frac{\kappa}{e^{\kappa\left(b^{\ast}-a^{\ast}\right)}-1}.
\end{equation}
In the case of an eternal black hole, we expect to find the quantum extremal surface to be near but outside the horizon \cite{Almheiri:2019yqk}. Without loss of generality, we consider the near-horizon expansion where we take the island position to be
\be
a =r_H+x, \quad x \ll r_H.
\ee
The expansion on the LHS of \er{islandHH} becomes
\be \la{LHS}
\text{LHS} \simeq \bigg( r_H+2 \epsilon\frac{d \rho}{dr}\bigg|_H \bigg) \bigg(\frac{d F}{dr}e^{\epsilon \varphi(r_H)} \bigg) \bigg|_{H} x= 2 \kappa x ( r_H+2 \epsilon \rho'|_H ),
\ee
by dropping terms with $F(r_H)=0$. We have also used the definition of the surface gravity where $(\frac{dh}{dr} e^{\epsilon \varphi(r)})|_H=2 \kappa$. For the RHS of \er{islandHH}, we consider the cut-off surface to be far away from the horizon, hence only $a^\ast=r^\ast(a)$ is relevant in the expansion. Note that $r^\ast (r_H) \to -\infty$, we have the following two choices for $e^{\kappa a^\ast}$
\begin{itemize}
    \item The leading order piece of $e^{\kappa a^\ast}$ is an $O(1)$ constant, which means that the island position is at a small fixed location away from the horizon. In this case, we do not need to expand on the RHS of \er{islandHH} and we expect the correction coming from $x$ is $O(\epsilon)$. This can be verified by substituting \er{LHS} into \er{islandHH}
    \be
    x = \frac{2 \epsilon}{(r_H+2 \epsilon \rho'|_H)[e^{\kappa (b^\ast-a^\ast)}-1]} \approx \frac{2 \epsilon}{r_0 [e^{\kappa (b^\ast-a^\ast)}-1]}+O(\epsilon^2).
    \ee
    \item The leading order piece of $e^{\kappa a^\ast}$ is $O(x)$, which means the island is extremely close to the horizon and they are nearly identical. In this case, the correction coming from $x$ will be of $O(\epsilon^2)$, in agreement with \cite{Hashimoto:2020cas, Matsuo:2020ypv, Djordjevic:2022qdk}.  Let us consider the expansion
    \bea \la{exp}
e^{2 \kappa a^\ast} &\approx& e^{2 \kappa r^\ast (r_H)}+2 \kappa \bigg(e^{2 \kappa r^\ast} \frac{dr^\ast}{dr} \bigg) \bigg|_{H }x 
\\
&=& 2 \kappa x e^{1+\epsilon \alpha (r_H)} \frac{e^{-\epsilon \varphi(r_H)}}{r_H h'(r_H)}=\frac{x}{r_H} e^{1+\epsilon \alpha(r_H)}.
\eea
Note that we have used the following near-horizon expansion for $r^\ast (r)$ \cite{Djordjevic:2022qdk} 
\be \la{tort}
r^\ast (r) \approx \frac{1}{2 \kappa} \bigg[ \frac{r}{r_H}+\ln{\bigg(\frac{r}{r_H}-1 \bigg)} +\epsilon \alpha (r) \bigg],
\ee
where $\alpha (r)$ denotes the terms that do not diverge at the horizon. We can work out this relation easily by considering the near-horizon expansion of the tortoise coordinate
\bea
r^\ast &\equiv&\int \frac{e^{-\epsilon \varphi(r)}}{h(r)}dr
\no\\
&\approx& \int \frac{e^{- \epsilon \varphi(r_H)}-e^{-\epsilon \varphi(r_H)}(r-r_H)\epsilon \varphi'(r_H) +\cdots }{h(r_H)+h'(r_H)(r-r_H)+\frac{1}{2} h''(r_H) (r-r_H)^2+\cdots} dr
\no \\
&=&\frac{1}{2 \kappa}\int \frac{1}{r-r_H} \bigg[ \frac{1-(r-r_H) \epsilon \varphi'(r_H)+\cdots}{1+\frac{1}{2}\frac{h''(r_H)}{h'(r_H)}(r-r_H)+\cdots} \bigg] dr
\no\\
&=&\frac{1}{2 \kappa} \int \bigg[\frac{1}{r-r_H}-\bigg(\frac{1}{2} \frac{h''}{h'}+\epsilon \varphi' \bigg)\bigg|_H +\frac{\epsilon}{2} \bigg( \frac{h''}{h'} \varphi'\bigg) \bigg|_H (r-r_H)\bigg]dr
\no\\
&=&\frac{1}{2 \kappa} \bigg[\ln{\bigg(\frac{r}{r_H}-1\bigg)}-\frac{1}{2}\frac{h''(r_H)}{h'(r_H)} r -\epsilon \varphi'(r_H) r 
\no\\
&\quad& +\frac{\epsilon}{2} \frac{h''(r_H)}{h'(r_H)} \varphi'(r_H) \bigg(\frac{r^2}{2}-rr_H \bigg) +C \bigg],
\eea
where we have used $h(r_H)=0$ and $2 \kappa = h'(r_H) e^{\epsilon \varphi(r_H)}$. With the following relation
\be
-\frac{1}{2}\frac{h''(r_H)}{h'(r_H)} r \approx \frac{r}{r_H}-\frac{\epsilon r r_H}{2 r_0} m''(r_H),
\ee
we confirm \er{tort}. Hence the RHS of \er{islandHH} can be expanded as
\bea
\text{RHS} &\approx& 4 \epsilon \kappa  e^{-\kappa b^\ast} e^{\kappa a^\ast}(1+e^{-\kappa b^\ast} e^{\kappa a^\ast})
\\
&=& 4 \epsilon \kappa  \bigg[\sqrt{\frac{x}{r_H}} e^{\frac{1}{2}(1+\epsilon \alpha(r_H))-\kappa b^\ast}+\frac{x}{r_H}e^{1+\epsilon \alpha(r_H)-2 \kappa b^\ast} \bigg].
\eea
We solve for $x$ as
\bea
x&=& \frac{1}{r_H} \frac{ (\frac{2 \epsilon }{r_H} )^2 e^{1-2 \kappa b^\ast+\epsilon \alpha(r_H)}}{[1+ \frac{2 \epsilon }{r_H} (\rho'|_H -\frac{1}{r_H}e^{1-2\kappa b^\ast+\epsilon \alpha(r_H)} ) ]^2}
\no\\
&\approx&\frac{4 \epsilon^2}{r^3_h}e^{1-2 \kappa b^\ast}+O(\epsilon^3).
\eea
\end{itemize}
Note that $x$ is positive in both cases, which means the island is indeed outside the horizon. This confirms our initial assumption. We
then conclude this section by comparing the entropy in the no-island and island phases. We have in the island case the $S_{\text{gen}}$ being
\bea \la{SgenHH}
S_{\text{gen}}(a)&=&S_{\text{gen}}(r_H)+S_{\text{gen}}'(r_H)x + O(x^2)
\no\\
&\approx&S_{\text{gen}}(r_H)+\frac{4 \pi r_H}{G_N \hbar}x,
\eea
where
\bea
S_{\text{gen}}(r_H)&=&\frac{2 \pi r^2_H}{G_N \hbar} +\frac{1}{3}(\rho_H+\rho_b)+\frac{2}{3} \ln{\frac{e^{\kappa b^\ast}}{\kappa \delta}}
\\
&=&2 \bigg[ \frac{ \pi r^2_H}{G_H \hbar} +\frac{1}{12} \ln{\frac{e^{4 \kappa b^\ast}}{(\kappa \delta)^4 e^{-2 \rho_H} e^{-2 \rho_b}}\bigg]}.
\eea
We can see from \er{SgenHH} that if $x \sim O(\epsilon)$, the correction can be $O(1)$ in $\epsilon=\frac{G_N \hbar}{24 \pi}$. If $x \sim O(\epsilon^2)$, the correction is essentially negligible. Therefore, if we keep only up to the $O\left(1\right)$ terms of the
entropy, we can approximately think of the island located at the position of the back-reacted horizon. In either case, the fine-grained
entropy at late times is given by 
\be
S_{\text{FG}}=\text{min} \bigg \{\frac{1}{3} \kappa t, S_{\text{gen}}(a) \bigg\}.
\ee
The Page time is then the transition time where 
\be
\frac{1}{3}\kappa t_P \approx S_{\text{gen}}(a) \implies t_P=3 \kappa S_{\text{gen}}(a).
\ee

\section{Island Computation in the Unruh State} \la{sD}

Similarly, we compute the island position where the generalized
entropy is given by 
\begin{equation}
S_{{\rm gen}}=\frac{\pi r_{a}^{2}}{G}+S_{{\rm matter}},
\end{equation}
and $S_{{\rm matter}}$ can be written in $\left(U,v\right)$ coordinate as 
\begin{equation}
S_{\text{matter }}=\frac{c}{6}\log\left(U_{a}-U_{b}\right)\left(v_{a}-v_{b}\right)+\frac{c}{12}\left(\kappa_{a}v_{a}+\epsilon \varphi_a+\kappa_{b}v_{b} +\epsilon \varphi_b\right) +\text{const},
\end{equation}
where in the following, we will denote $r_a=r(v_a), r_{Ha}=r_{H}\left(v_{a}\right),\ka_{a}=\ka\left(v_{a},r_{Ha}\right)$,
and similarly for $b$. Here $a$ corresponds to $\pa I$, and $b$ is the cut-off surface. It is necessary to specify the concrete surface
gravity at $a$ and $b$ because now we are not in a stationary case,
and it is likely that for general extremal island configuration that
$v_{a}\not=v_{b}$.

We are mainly interested in the late-time configuration of the island, and the endpoint of it is expected to be in the near-horizon regime \cite{Penington:2019npb}, where the following approximation is applicable 
\begin{equation} \la{appro} 
r\left(v\right)=r_{H}\left(v\right)-Ue^{\ka v}+\frac{1}{\ka}r_{H}^{\pp}+O\left(\epsilon^{2}\right).
\end{equation}
The physical meaning of $U$ coordinate is the relative deviation for a radial in-falling null geodesic 
from the horizon, and $\ka$ is the surface gravity at $r_{H}$. The above equation \er{appro} should be understood to work perturbatively in $\epsilon$, 
where the horizon position $r_H$ and surface gravity $\ka$ should all be viewed as functions of $\epsilon$. Therefore we expect that the time derivatives of $r_H$ and $\ka$ should belong to $O(\epsilon)$, and \er{appro} is exact only up to $O(\epsilon)$.

Now we start by extremizing the generalized entropy
\begin{equation}
\begin{split} \la{ex-Unruh}
\frac{\partial S_{{\rm gen}}}{\partial v_{a}} & =\frac{ r_{a}}{12\epsilon}\left(r_{Ha}^{\pp}-\ka_{a}U_{a}e^{\ka_{a}v_{a}}\right)+\frac{c}{6}\frac{1}{v_{a}-v_{b}}+\frac{c} {12}\ka_{a}+O\left(\epsilon\right)=0,\\
\frac{\partial S_{{\rm gen}}}{\partial U_{a}} & =\frac{ r_{a}}{12\epsilon}\left(-e^{\ka_{a}v_{a}}\right)+\frac{c}{6}\frac{1}{U_{a}-U_{b}}+O\left(\epsilon\right)=0,
\end{split}
\end{equation}
where we only keep terms up to $O\left(1\right)$ in $\epsilon$,
and used the fact that $r_{Ha}^{\pp},\ka_{a}^{\pp}\simeq O\left(\epsilon\right)$,
where the prime denotes the derivative with respect to $v$. Combining the two
equations we find a useful relation between $U_{a}$ and $U_{b}$
\begin{equation} \la{sol-Ub}
U_{a}=U_{b}\frac{ r_{a}r_{Ha}^{\pp}\left(v_{a}-v_{b}\right)+c\epsilon\left(2+\ka_{a}\left(v_{a}-v_{b}\right)\right)}{ r_{a}r_{Ha}^{\pp}\left(v_{a}-v_{b}\right)+c\epsilon\left(2-\ka_{a}\left(v_{a}-v_{b}\right)\right)}.
\end{equation}
Since the leading order of the fraction above is $O(1)$ in $\epsilon$, the equation indicates that $U_a$ and $U_b$ are at the same order in $\epsilon$.  Plugging \er{sol-Ub} back into the extremal equation \er{ex-Unruh} of $S_{\rm{gen}}$, we find the solution of $U_a$ reads
\begin{equation} \la{sol-Ua}
U_{a}=\frac{\epsilon ce^{-\ka_{a}v_{a}}\left(2+\ka_{a}\left(v_{a}-v_{b}\right)\right)}{\ka_{a}r_{Ha}\left(v_{a}-v_{b}\right)}+\frac{r_{Ha}^{\pp}}{\ka_{a}}e^{-\ka_{a}v_{a}}+O\left(\epsilon^{2}\right),
\end{equation}
and the result is manifestly at $O\left(\epsilon\right)$, which is consistent with the near-horizon approximation \er{appro}.

Notice that we have assumed that the endpoint of the cut-off surface $b=(U_b, v_b)$ locates outside the horizon. Let us verify this by plugging $U_{a}$ in \er{sol-Ua} back into \er{sol-Ub}, which gives
\begin{equation}\label{rb}
    \begin{split}
       U_{b}&=\frac{\epsilon ce^{-\kappa_{a}v_{a}}\left(2-\kappa_{a}\left(v_{a}-v_{b}\right)\right)}{\kappa_{a}r_{Ha}\left(v_{a}-v_{b}\right)}+\frac{r_{Ha}^{\prime}}{\kappa_{a}}e^{-\kappa_{a}v_{a}}+O\left(\epsilon^{2}\right), \\
        r_{b} & =r_{Hb}-\frac{\epsilon c\left(2-\ka_{a}\left(v_{a}-v_{b}\right)\right)}{\ka_{a}r_{Ha}\left(v_{a}-v_{b}\right)}e^{\ka_{b}v_{b}-\ka_{a}v_{a}}\\
 & +\left(\frac{r_{Hb}^{\pp}}{\ka_{b}}-\frac{r_{Ha}^{\pp}}{\ka_{a}}e^{\ka_{b}v_{b}-\ka_{a}v_{a}}\right)+O\left(\epsilon^{2}\right).
    \end{split}
\end{equation}
where we have used the definition \er{appro} in deriving the radial position of $b$. For an evaporating black hole with a shrinking horizon, one always has
$r_{H}^{\pp}<0$ and $\ka^{\pp}>0$. Then it is clear to deduce from
\er{rb} that the endpoint of the cut-off surface is outside the horizon
as long as $v_{a}-v_{b}<0$. The condition $v_{a}-v_{b}<0$ is imposed because a timelike surface is not included in the extremization procedure. The fact that $b$ is always outside the horizon is consistent with our setup.

Finally, let us discuss the location of the island by considering the difference
$r_{a}-r_{Ha}$ with $U_a$ given by \er{sol-Ua}, The result is 
\begin{equation}\label{island-a}
r_{a}= r_{Ha}-\frac{\epsilon c\left(2+\ka_{a}\left(v_{a}-v_{b}\right)\right)}{\ka_{a}r_{Ha}\left(v_{a}-v_{b}\right)}+O\left(\epsilon^{2}\right),
\end{equation}
An interesting point of the result is that there is no explicit dependence on the kinematics of the horizon. That is, terms with explicit dependence on time derivative of $r_{H}$ all cancel among themselves. The contribution from back-reaction can all be absorbed into quantum corrections of the surface gravity $\ka_a$. This implies that we do not need to specify a concrete form of evolution of the horizon when analyzing the position of the island. One can easily deduce from \er{island-a} that for $|v_{a}-v_{b}|>2/\ka_{a}$, then $r_{a}<r_{Ha}$ where the island sits inside the horizon, and 
for $|v_{a}-v_{b}|< 2/\ka_{a}$, it extends outside the horizon.

\end{appendix}

 \bibliographystyle{JHEP}
 \bibliography{bibliography}

\end{document}